\documentclass[format=acmsmall, review=false, screen=true]{acmart}

\usepackage{booktabs} % For formal tables

\usepackage{url}  %Required
\usepackage{graphicx}  %Required
\usepackage{amsmath}
\usepackage{adjustbox}
\usepackage{subfigure}
\usepackage{blindtext}
\usepackage{enumerate}
\usepackage{bm}
\usepackage{array}
\usepackage{longtable}

\usepackage[ruled]{algorithm2e} % For algorithms

\SetAlFnt{\small}
\SetAlCapFnt{\small}
\SetAlCapNameFnt{\small}
\SetAlCapHSkip{0pt}
\IncMargin{-\parindent}
\newcommand{\etal}{\textit{et al.~}}

\newcolumntype{L}[1]{>{\raggedright\arraybackslash}p{#1}}
\newcolumntype{C}[1]{>{\centering\arraybackslash}p{#1}}
\newcolumntype{R}[1]{>{\raggedleft\arraybackslash}p{#1}}

\AtBeginDocument{%
  \providecommand\BibTeX{{%
    \normalfont B\kern-0.5em{\scshape i\kern-0.25em b}\kern-0.8em\TeX}}}

\begin{document}

%
% The "title" command has an optional parameter, allowing the author to define a "short title" to be used in page headers.
\title[Affective Computing for Large-Scale Heterogeneous Multimedia Data: A Survey]{Affective Computing for Large-Scale Heterogeneous Multimedia Data: A Survey}

\author{Sicheng Zhao}
\affiliation{%
  \institution{University of California, Berkeley}
  \city{Berkeley}
  \postcode{94720}
  \country{USA}}
\email{schzhao@gmail.com}
\author{Shangfei Wang}
\authornote{Corresponding authors: Shangfei Wang, Mohammad Soleymani.}
\affiliation{%
  \institution{University of Science and Technology of China}
  \city{Hefei}
  \postcode{230027}
  \country{China}}
\email{sfwang@ustc.edu.cn}
\author{Mohammad Soleymani}
\authornotemark[1]
\affiliation{%
  \institution{University of Southern California}
  \city{Playa Vista}
  \postcode{90094}
  \country{USA}}
\email{soleymani@ict.usc.edu}
\author{Dhiraj Joshi}
\affiliation{%
  \institution{IBM Research AI}
  \city{Yorktown Heights}
  \postcode{10598}
  \country{USA}}
\email{djoshi@us.ibm.com}
\author{Qiang Ji}
\affiliation{%
  \institution{Rensselaer Polytechnic Institute}
  \city{Troy}
  \postcode{12180}
  \country{USA}}
\email{qji@ecse.rpi.edu}

% By default, the full list of authors will be used in the page headers. Often, this list is too long, and will overlap
% other information printed in the page headers. This command allows the author to define a more concise list
% of authors' names for this purpose.
\renewcommand{\shortauthors}{Zhao et al.}

%
% The abstract is a short summary of the work to be presented in the article.
\begin{abstract}
The wide popularity of digital photography and social networks has generated a rapidly growing volume of multimedia data (\textit{i.e.}, image, music, and video), resulting in a great demand for managing, retrieving, and understanding these data. Affective computing (AC) of these data can help to understand human behaviors and enable wide applications. In this article, we survey the state-of-the-art AC technologies comprehensively for large-scale heterogeneous multimedia data. We begin this survey by introducing the typical emotion representation models from psychology that are widely employed in AC. We briefly describe the available datasets for evaluating AC algorithms. We then summarize and compare the representative methods on AC of different multimedia types, \textit{i.e.}, images, music, videos, and multimodal data, with the focus on both handcrafted features-based methods and deep learning methods. Finally, we discuss some challenges and future directions for multimedia affective computing.
\end{abstract}

\copyrightyear{2019}

\setcopyright{acmcopyright}
\acmJournal{TOMM}
\acmYear{2019} \acmVolume{1} \acmNumber{1} \acmArticle{1} \acmMonth{1} \acmPrice{15.00}\acmDOI{10.1145/3363560}

\begin{CCSXML}
<ccs2012>
<concept>
<concept_id>10002944.10011122.10002945</concept_id>
<concept_desc>General and reference~Surveys and overviews</concept_desc>
<concept_significance>500</concept_significance>
</concept>
<concept>
<concept_id>10002951.10003317.10003347.10003353</concept_id>
<concept_desc>Information systems~Sentiment analysis</concept_desc>
<concept_significance>500</concept_significance>
</concept>
<concept>
<concept_id>10003120.10003121</concept_id>
<concept_desc>Human-centered computing~Human computer interaction (HCI)</concept_desc>
<concept_significance>500</concept_significance>
</concept>
</ccs2012>
\end{CCSXML}

\ccsdesc[500]{General and reference~Surveys and overviews}
\ccsdesc[500]{Information systems~Sentiment analysis}
\ccsdesc[500]{Human-centered computing~Human computer interaction (HCI)}

\keywords{Affective computing, emotion recognition, sentiment analysis, large-scale multimedia}

%
% A "teaser" image appears between the author and affiliation information and the body
% of the document, and typically spans the page.
%%  \includegraphics[width=\textwidth]{sampleteaser}
%%  \caption{Seattle Mariners at Spring Training, 2010.}
%%  \Description{Enjoying the baseball game from the third-base seats. Ichiro Suzuki preparing to bat.}

%
% This command processes the author and affiliation and title information and builds
% the first part of the formatted document.
\maketitle

\begin{figure}[t]
   \centering
   \includegraphics[width=0.8\linewidth]{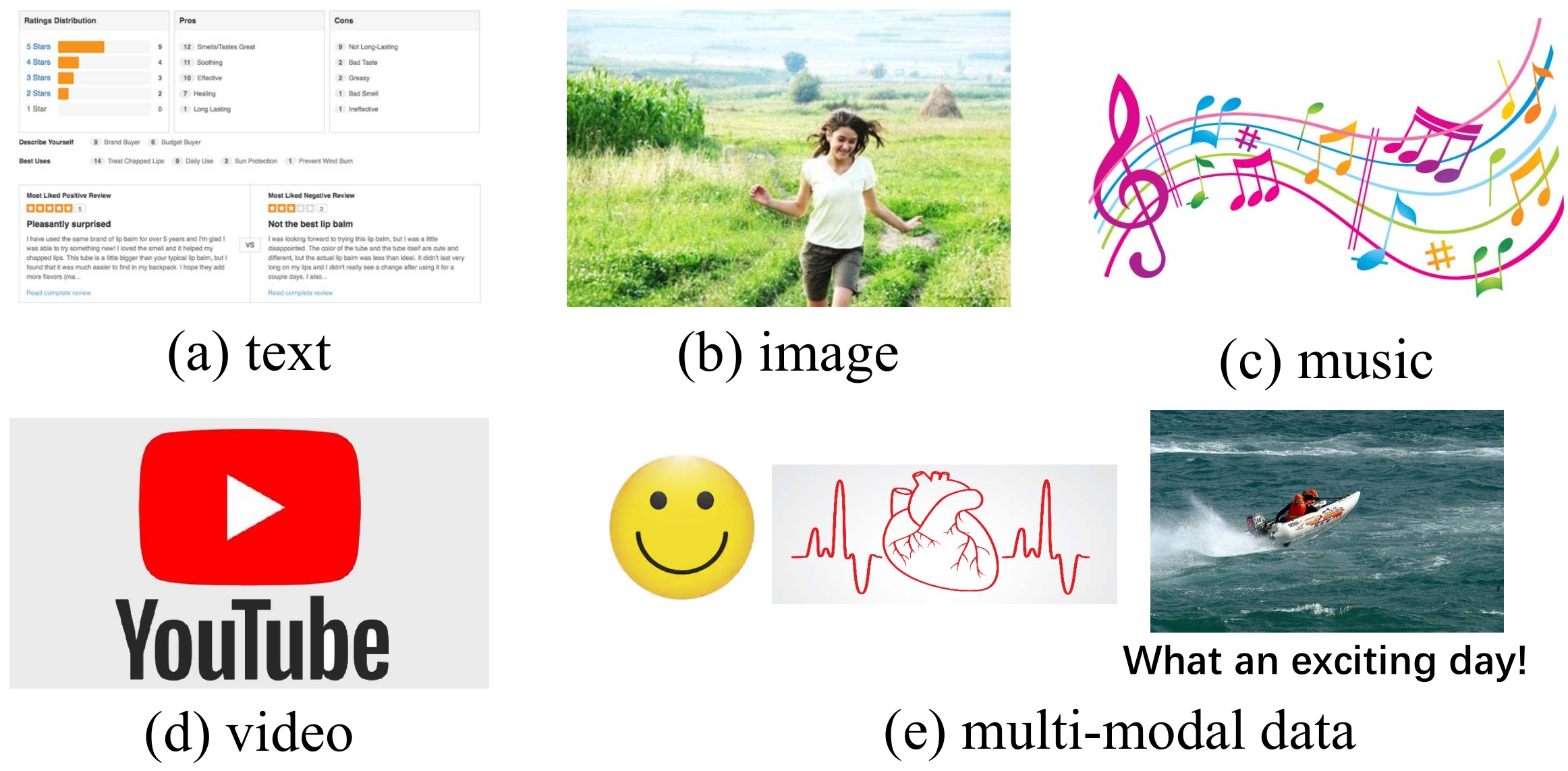}
   \caption{Different multimedia data that are widely used to express emotions.}
   \label{fig:multimedia}
\end{figure}

\section{Introduction}
\label{sec:Introduction}
Users are increasingly recording their daily activities, sharing interesting experiences, and expressing personal viewpoints using mobile devices on social networks, such as Twitter\footnote{\url{https://twitter.com}}, Facebook\footnote{\url{https://www.facebook.com}}, and Weibo\footnote{\url{https://www.weibo.com}}, \textit{etc}. As a result, a rapidly growing volume of multimedia data (\textit{i.e.}, image, music, and video) has been generated, as shown in Figure~\ref{fig:multimedia}, which results in a great demand for the management, retrieval, and understanding of these data. Most existing work on multimedia analysis focus on the cognitive aspects, \textit{i.e.}, understanding the objective content, such as object detection in images~\cite{han2018advanced}, speaker recognition in speech~\cite{hansen2015speaker}, and action recognition in videos~\cite{herath2017going}. Since what people feel have a direct influence on their decision making, affective computing (AC) of these multimedia data is of significant importance and has attracted increasing attention \cite{chen2014object,zhang2016exploring,yao2019attention,zhan2019zero,zhao2019pdanet}. For example, companies would like to know how customers evaluate their products and can thus improve their services~\cite{jansen2009twitter}; depression and anxiety detection from social media can help understand psychological distress and thus potentially prevent suicidal actions~\cite{shen2017depression}.

While the sentiment analysis in text~\cite{pang2008opinion} has long been a standard task, AC from other modalities, such as image and video, has just begun to be considered recently. In this article, we aim to review the existing AC technologies comprehensively for large-scale heterogeneous multimedia data, including image, music, video, and multimodal data.

Affective computing of multimedia (ACM) aims to recognize the emotions that are expected to be evoked in viewers by a given stimuli. Similar to other supervised learning tasks, ACM is typically composed of three steps: data collection and annotation, feature extraction, and mapping learning between features and emotions~\cite{zhao2017approximating}. One main challenge for ACM is the affective gap, \textit{i.e.}, ``the lack of coincidence between the features and the expected affective state in which the user is brought by perceiving the signal''~\cite{hanjalic2006extracting}. In the early stage, various hand-crafted features were designed to bridge this gap with traditional machine learning algorithms, while more recently researchers have focused on end-to-end deep learning from raw multimedia data to recognize emotions. Existing ACM methods mainly assign the dominant (average) emotion category (DEC) to an input stimuli, based on the assumption that different viewers have similar reactions to the same stimuli. We can usually formulate this task as a single-label learning problem.

However, emotions are influenced by subjective and contextual factors, such as the educational background, cultural diversity, and social interaction~\cite{peng2015mixed,zhao2016predicting,yang2017learning}. As a result, different viewers may react differently to the same stimuli, which creates the subjective perception challenge. Therefore, the perception inconsistency makes it insufficient to simply predict the DEC for the highly subjective variable. As stated in~\cite{zhao2016predicting}, we can perform two kinds of ACM tasks to deal with the subjectivity challenge: predicting personalized emotion perception for each viewer and assigning multiple emotion labels for each stimuli. For the latter one, we can either assign multiple labels to each stimuli with equal importance using multi-label learning methods, or predict the emotion distributions which tries to learn the degrees of each emotion~\cite{yang2017learning}.

In this article, we concentrate on surveying the existing methods on ACM and analyzing potential research trends. Section~\ref{sec:EmotionModels} introduces the widely-used emotion representation models from psychology. Section~\ref{sec:Datasets} summarizes the existing available datasets for evaluating ACM tasks. Section~\ref{sec:Image}, Section~\ref{sec:Music}, Section~\ref{sec:Video}, and  Section~\ref{sec:multimodal} survey the representative methods on AC of images, music, videos, and multimodal data, respectively, including both handcrafted features-based methods and deep learning methods. Section~\ref{sec:FutureDirections} provides some suggestions for future research, followed by conclusion in Section~\ref{sec:Conclusion}.

To the best of our knowledge, this article is among the first that provide a comprehensive survey of affective computing of multimedia data from different modalities. Previous surveys mainly focus on a single modality, such as images~\cite{zhao2018affective,joshi:emotion-survey}, speech~\cite{el2011survey}, music~\cite{Kim2010,yang2012machine}, video~\cite{wang2015video,Baveye2018}, and multimodal data~\cite{soleymani2017survey}. From this survey, readers can more easily compare the correlations and differences among different AC settings. We believe that this will be instrumental in generating novel research ideas.

\section{Emotion Models from Psychology}
\label{sec:EmotionModels}

There are two dominant emotion representation models deployed by psychologists: categorical emotions (CE), and dimensional emotion space (DES). CE models classify emotions into a few basic categories, such as \emph{happiness} and \emph{anger}, \emph{etc}. Some commonly used models include Ekman's six basic emotions \cite{ekman1992argument} and Mikels's eight emotions~\cite{mikels2005emotional}. When classifying emotions into \emph{positive} and \emph{negative}  (polarity)~\cite{zhao2018personality,zhao2019personalized}, sometimes including \emph{neutral}, ``emotion'' is called ``sentiment''. However, sentiment is usually defined as an atitude held toward an object \cite{soleymani2017survey}. Emotions are usually represented by DES models as continuous coordinate points in a 3D or 2D Cartesian space, such as valence-arousal-dominance (VAD) \cite{schlosberg1954three} and activity-temperature-weight~\cite{lee2011fuzzy}. VAD is the most widely used DES model, where valence represents the pleasantness ranging from positive to negative, arousal represents the intensity of emotion ranging from excited to calm, and dominance represents the degree of control ranging from controlled to in control. Dominance is difficult to measure and is often omitted, leading to the commonly used two dimensional VA space~\cite{hanjalic2006extracting}.

The relationship between CE and DES and the transformation from one to the other are studied in~\cite{sun2009improved}. For example, positive valence relates to a happy state, while negative valence relates to a sad or angry state. CE models are easier for users to understand and label, but the limited set of categories may not well reflect the subtlety and complexity of emotions. DES can better describe detailed emotions with subtle differences flexibly, but it is difficult for uses to distinguish the absolute continuous values, which may also be problematic. 
CE and DES are mainly employed in classification and regression tasks, respectively, with discrete and continuous emotion labels. If we discretize DES into several constant values, we can also use it for classification~\cite{lee2011fuzzy}. Ranking based labeling can be applied to ease DEC comprehension difficulties in raters.

Although less explored in this context, one of the most well-known theories that explains the development of emotional experience is appraisal theory. According to this theory, cognitive evaluation or appraisal of a situation or content in case of multimedia results in emergence of emotions~\cite{ortony88emotion,citeulike:3014462}. According to Ortony, Clore and Collins (OCC)~\cite{ortony88emotion}, emotions are experienced following a scenario comprising a series of phases. First, there is a perception of an event, object or an action. Then, there is an evaluation of events, objects or action according to personal wishes or norms. Finally, perception and evaluation result in a specific emotion or emotions arising. Certain appraisal dimensions such as novelty and complexity can be labeled and detected from content. For example, \citeauthor{Soleymani2015} automatically recognized image novelty and complexity that are related to interestingness. There are also domain specific emotion taxonomy and scales. Geneva Emotional Music Scale \cite{Zentner2008} is a music specific emotion model for describing emotions induced by music. It consists of a hierarchical structure with 45 emotions, nine emotional categories and three superfactors that can describe emotion in music.

\begin{table}[!t]
\centering\scriptsize
\caption{Representative emotion models employed in ACM.}
\begin{tabular}
%{C{1cm}<{\centering}  C{0.8cm}<{\centering}  C{0.81cm}<{\centering}  p{8cm}<{\centering}}
{cccc}
\hline
\textbf{Model} & \textbf{Ref} & \textbf{Type} & \multicolumn{1}{c}{\textbf{Emotion states/dimensions}} \\
\hline
Ekman & \cite{ekman1992argument}  & CE & happiness, sadness, anger, disgust, fear, surprise\\
Mikels & \cite{mikels2005emotional} & CE & amusement, anger, awe, contentment, disgust, excitement, fear, sadness\\
Plutchik & \cite{plutchik1980emotion} & CE & ($\times$ 3 scales) anger, anticipation, disgust, joy, sadness, surprise, fear, trust\\
Clusters & \cite{hu2007exploring} & CE &  29 discrete labels are consistently grouped into 5 clusters at a similar distance level \\
Sentiment & & CE & positive, negative, (or neutral)\\
VA(D) & \cite{schlosberg1954three} & DES & valence-arousal(-dominance)\\
ATW & \cite{lee2011fuzzy} &DES & activity-temperature-weight\\
\hline
\end{tabular}
\label{tab:EmotionModels}
\end{table}

Another relevant concept worth mentioning is that emotion in response to multimedia can be expected, induced or perceived emotion. 
Expected emotion is the emotion that the multimedia creator intends to make people feel, perceived emotion is what people perceive as being expressed, while induced/felt emotion is the actual emotion that is felt by a viewer. Discussing the difference or correlation of various emotion models is out of the scope of this article. The typical emotion models that have been widely used in ACM are listed in in Table~\ref{tab:EmotionModels}.

\section{Datasets}
\label{sec:Datasets}

\subsection{Datasets for AC of Images}
\label{sec:ImageDatasets}

The early datasets for AC of images mainly come from the psychology community with small-scale images. The International Affective Picture System (\textbf{IAPS}) is an image set that is widely used in psychology to evoke emotions~\cite{lang1997international}. Each image that depicts complex scenes is associated with the mean and standard deviation (STD) of VAD ratings in a 9-point scale by about 100 college students. The \textbf{IAPSa} dataset is selected from IAPS with 246 images~\cite{mikels2005emotional}, which are labeled by 20 undergraduate students. The \textbf{Abstract} dataset consists of 279 abstract paintings without contextual content. Approximately 230 people peer rated these paintings. The Artistic dataset (\textbf{ArtPhoto}) includes 806 artistic photographs from a photo sharing site~\cite{machajdik2010affective} with emotions determined by the artist uploading the photos. The Geneva affective picture database (\textbf{GAPED}) is composed of 520 negative, 121 positive, and 89 neutral images~\cite{dan2011geneva}. Besides, these images are also rated with valence and arousal values, ranging from 0 to 100 points. There are 500 abstract paintings in both \textbf{MART} and \textbf{devArt} datasets, which are collected from the Museum of Modern and Contemporary Art of Trento and Rovereto~\cite{alameda2016recognizing}, and the ``DeviantArt'' online social network~\cite{alameda2016recognizing}, respectively.

\begin{table*}[!t]
\centering\scriptsize
\caption{Released and freely available datasets for AC of images, where `Ref' is short for Reference, `\# Images' and `\# Annotators' respectively represent the total number of images and annotators (f: female, m: male), `Labeling' represents the method to obtain labels, such as human annotation (annotation) and keyword searching (keyword), and `Labels' means the detailed labels in the dataset, such as dominant emotion category (dominant), average dimension values (average), personalized emotion (personalized), and emotion distribution (distribution).}
\begin{tabular}
{cccccccc}
\hline
\textbf{Dataset} & \textbf{Ref} & \textbf{\# Images} & \textbf{Type} & \textbf{\# Annotators}& \textbf{Emotion model} & \textbf{Labeling} & \textbf{Labels}\\
\hline
IAPS & \cite{lang1997international}  & 1,182 & natural & $\approx$100 (half f) & VAD & annotation  & average  \\
IAPSa & \cite{mikels2005emotional} & 246 & natural & 20 (10f,10m) & Mikels  &  annotation  & dominant \\
Abstract & \cite{machajdik2010affective} & 279 & abstract & $\approx$230 & Mikels   &  annotation  & dominant \\
ArtPhoto & \cite{machajdik2010affective} & 806 & artistic & -- & Mikels &  keyword &  dominant \\
GAPED & \cite{dan2011geneva} & 730 & natural & 60  & Sentiment, VA  & annotation  & dominant, average \\
MART & \cite{alameda2016recognizing} & 500 & abstract & 25 (11f,14m) & Sentiment   & annotation  &  dominant\\
devArt & \cite{alameda2016recognizing} & 500 & abstract & 60 (27f,33m) & Sentiment   & annotation  &  dominant\\
Tweet & \cite{borth2013large} & 603 & social & 9 & Sentiment  & annotation  & dominant \\
FlickrCC & \cite{borth2013large} & $\approx$500,000 & social & -- & Plutchik   & keyword  & dominant \\
Flickr & \cite{yang2014your} & 301,903 & social & 6,735 & Ekman  & keyword  & dominant  \\
Emotion6 & \cite{peng2015mixed} & 1,980 & social & 432 & Ekman+neutral  & annotation  &  distribution \\

FI & \cite{you2016building} & 23,308 & social & 225 & Mikels  & annotation  &  dominant \\
IESN & \cite{zhao2016predicting} & 1,012,901 & social & 118,035 & Mikels, VAD  &  keyword & personalized \\
FlickrLDL & \cite{yang2017learning} & 10,700 & social & 11 & Mikels  & annotation &  distribution \\
TwitterLDL & \cite{yang2017learning} & 10,045 & social & 8 & Mikels  &  annotation &  distribution \\
\hline
\end{tabular}
\label{tab:ImageDataset}
\end{table*}

Recent datasets, especially the large-scale ones, are constructed using images from social networks. The Tweet dataset (\textbf{Tweet}) consists of 470 and 113 tweets for positive and negative sentiments, respectively~\cite{borth2013large}. The \textbf{FlickrCC} dataset includes about 500k Flickr creative common (CC) images which are generated based on 1,553 adjective noun pairs (ANPs)~\cite{borth2013large}. The images are mapped to the Plutchnik's Wheel of Emotions with 8 basic emotions, each with 3 scales. The \textbf{Flickr} dataset contains about 300k images~\cite{yang2014your} with the emotion category defined by the synonym word list which has the most same words as the adjective words of an image's tags and comments. The \textbf{FI} dataset consists of 23,308 images which are collected from Flicker and Instagram by searching the emotion keywords~\cite{you2016building} and labeled by 225 Amazon Mechanical Turk (MTurk) workers. The number of images in each Mikels emotion category is larger than 1,000. The \textbf{Emotion6} dataset~\cite{peng2015mixed} consists of 1,980 images collected from Flickr with 330 images for each Ekman's emotion category. Each image was scored by 15 MTurk workers to obtain the discrete emotion distribution information. The \textbf{IESN} dataset that is constructed for personalized emotion prediction \cite{zhao2016predicting} contains about 1M images from Flickr. Lexicon-based methods and VAD averaging~\cite{warriner2013norms} are used to segment the text of metadata from uploaders for expected emotions and comments from viewers for personalized emotions. There are 7,723 active users with more than 50 involved images.  We can also easily obtain the DEC and emotion distribution for each image. \textbf{FlickrLDL} and \textbf{TwitterLDL} datasets~\cite{yang2017learning} are constructed for discrete emotion distribution learning. The former one is a subset of FlickrCC, which are labeled by 11 viewers. The latter one consists of 10,045 images which are collected by searching various sentiment key words from Twitter and labeled by 8 viewers. These datasets are summarized in Table~\ref{tab:ImageDataset}.

\subsection{Datasets for AC of Music}
\label{sec:MusicDatasets}
A notable benchmark for music recognition is music mood classification (AMC) task, organized by annual Music Information Retrieval Evaluation eXchange\footnote{\url{http://www.music-ir.org/mirex/wiki/}} (MIREX)~\cite{mirex07}. In MIREX mood classification task, initially 600 songs were shared with the participants. Starting from 2013, 1,438 30 seconds excerpts from Korean pop songs have been added to MIREX. MIREX benchmark aims to automatically classify songs into five emotion clusters derived from cluster analysis of online tags. MIREX mood challenge emotional representation has been debated in the literature due to its data-driven origin rather than psychology of emotion. For example, in~\cite{laurier07mirex}, semantic and acoustic overlaps have been found between clusters. MIREX mood challenge considers only one label for the whole song and disregards the dynamic time evolving nature of music.

Computer Audition Lab 500 (CAL500) is a dataset of 500 popular songs which is labeled by multiple tags including emotions~\cite{turnbull2007towards}. The dataset is labeled in the lab by 66 labelers. Soundtracks datasets~\cite{Eerola2011} for music and emotion is developed by~\citeauthor{Eerola2011} and contains instrumental music from soundtrack of 60 movies. The expert annotators selected songs based on five basic discrete categories (\textit{anger}, \textit{fear}, \textit{sadness}, \textit{happiness}, and \textit{tenderness}) and dimensional VA representation of emotions. Although not developed with music content analysis in mind, the Database for Emotion Analysis using Physiological Signals or DEAP dataset~\cite{koelstra2012tac} also includes valence, arousal and dominance ratings for 120 one-minute music video clips of western pop music. Each video clip is annotated by by 14--16 participants who were asked to report their felt valence, arousal, and dominance on a 9-point scale. AMG1608~\cite{chen15icassp} is another music dataset that contains arousal and valence ratings for 1,608 Western songs in different genres and is annotated through MTurk.

Music datasets with emotion labels usually consider one emotion label per song (static). MoodSwings dataset~\cite{speck11ismir} was the first to annotate music dynamically over time. MoodSwings was developed by Schmidt \etal and includes 240 15s excerpts of western pop songs  with per-second valence and arousal labels,  collected on MTurk. The MediaEval ``Emotion in Music'' challenge was organized in years 2013--2015 in MediaEval Multimedia Evaluation initiative\footnote{http://www.multimediaeval.org}. MediaEval is a community-driven benchmarking campaign dedicated to evaluating algorithms for social and human-centered multimedia access and retrieval~\cite{mediaeval}. Unlike MIREX, ``Emotion in Music'' task focused on dynamic emotion recognition in music tracking arousal and valence over time~\cite{Soleymani1000songs,deam}. The data from  MediaEval tasks were compiled in MediaEval Database for Emotional Analysis in Music (DEAM) which is the largest available dataset with dynamic annotations, at 2Hz,  with valence and arousal annotations for 1,802 songs and song excerpts licensed under Creative Commons license. PMEmo is a dataset of 794 songs with dynamic and static arousal and valence annotations in addition to electrodermal responses from ten participants~\cite{Zhang2018PMEmo}.

These datasets are summarized in Table~\ref{tab:MusicDataset}. For a more detailed review of available music datasets with emotional labels, we refer the readers to~\cite{panda_thesis}.

\begin{table*}[!t]
\centering\scriptsize
\caption{Released and freely available datasets for music emotion recognition, where `\# Songs' and `\# Annotators' respectively represent the total number of songs and annotators per song, `Labeling' represents the method to obtain labels, such as human annotation (annotation), and `Labels' means the detailed labels in the dataset, such as dominant emotion category (dominant), average dimension values (average), personalized emotion (personalized), and emotion distribution (distribution).}
\begin{tabular}
{cccccccc}
\hline
\textbf{Dataset} & \textbf{Ref} & \textbf{\# Songs} & \textbf{Type} & \textbf{\# Annotators}& \textbf{Emotion model} & \textbf{Labeling} & \textbf{Labels}\\
\hline
MIREX mood& \cite{mirex07} & 2,038 & western and kpop & 2--3 & Clusters & annotation & dominant, distribution\\
CAL500 & \cite{turnbull2007towards}& 500 & western & >2 & -- & annotation & dominant\\
Soundtracks& \cite{Eerola2011} & 110  & instrumental &110 & self-defined, VA & annotation & distribution \\
MoodSwings& \cite{speck11ismir} & 240 & western & Unknown & VA & annotation & distribution\\
DEAP & \cite{koelstra2012tac} & 120 & western & 14--16 & VAD & annotation & average \\
AMG1608& \cite{chen15icassp} & 1,608 & western & 15 & VA & annotation & distribution\\
DEAM & \cite{deam} & 1,802 & diverse & 5--10 & VA & annotation & average, distribution \\
PMEmo & \cite{Zhang2018PMEmo} & 794 & western & 10 & VA  & annotation & distribution \\
\hline
\end{tabular}
\label{tab:MusicDataset}
\end{table*}

\subsection{Datasets for AC of Videos}

The target of video affective content computing is to recognize the emotions evoked by videos. In this field, it is necessary to construct a large benchmark dataset with precise emotional tags. However, the majority of existing research evaluate their proposed methods on their own collected datasets. The scarce video resources in those self-collected datasets, combined with the copyright restrictions result in limited accessibility for other researchers to reproduce existing work. Therefore, it is beneficial to summarize some publicly available datasets in this field. In general, publicly available datasets can be classified into two types: datasets consisting of videos only, such as movie clips or user generated videos, and datasets including both videos and audience's information.

\subsubsection{Datasets consisting of videos only}

The LIRIS-ACCEDE dataset~\cite{dellandrea2019datasets} is one of the largest datasets in this area. Because it is collected under Creative Commons licenses, there are no copyright issues. The LIRIS-ACCEDE dataset is a living database in development. In order to fulfill requirements of different tasks, new data, features and tags are included. The LIRIS-ACCEDE dataset includes the Discrete LIRIS-ACCEDE collection and the Continuous LIRIS-ACCEDE collection in 2015 and was used for the MediaEval Emotional Impact of Movies tasks from 2015 to 2018.

The Discrete LIRIS-ACCEDE collection~\cite{baveye2015liris} includes 9,800 clips, which is derived from 40 feature films and 120 short films. Specifically, the majority of the 160 films are collected from the video platform VODO. The duration of all 160 films is about 73.6 hours in total. All of the 9,800 video clips last about 27 hour in total and the duration of each clip is between 8 to 12 seconds, which is long enough for viewers to feel emotions.
In this collection, all the 9,800 video clips are labeled by values of valence and arousal.

The Continuous LIRIS-ACCEDE collection~\cite{baveye2015deep} differs from the Discrete LIRIS-ACCEDE collection in annotation type. Roughly speaking, the annotations for movie clips in the Discrete  LIRIS-ACCEDE collection are global. It means that a whole 8 to 12 second video clip is represented by a single value of valence and arousal. This annotation type limits the possiblity for tracking emotions. To address this issue,  30 longer films are selected from the 160 films mentioned above. The total duration of all the selected films is about 7.4 hours. There are  emotional annotations according to valence and arousal of each second of the films in the collection.

The MediaEval Affective Impact of Movies collections between 2015 and 2018 are used for the MediaEval affective Impact of Movies tasks in each corresponding year. Specifically, the MediaEval 2015 Affective Impact of Movies~\cite{sjoberg2015mediaeval} includes two sub-tasks: affect detection and violence detection. The Discrete LIRIS-ACCEDE collection was used as the development set. And 1,100 additional video clips were extracted from 39 new movies and included. Indeed, all the new collected data were shared under Creative Commons licenses. In addition, three values were used to label the 10,900 video clips: a binary signal representing the presence of violence, a class tag of the excerpt for felt arousal and an annotation for felt valence.

The MediaEval 2016 Affective Impact of Movies Task~\cite{dellandrea2016mediaeval} also includes two sub-tasks: Global emotion prediction and Continuous emotion prediction. The Discrete LIRIS-ACCEDE collection and the Continuous LIRS-ACCEDE collection were used as the development sets for the first and second sub-tasks, respectively. In addition, 49 new movies were chosen as the test sets. 1,200 short video clips from the new movies were extracted for the first task, and 10 long movies were selected for the second task. For the first sub-task, the tags include scores of valence and arousal for each whole movie clip. And for the second sub-task, scores of valence and arousal for each second of the movies are evaluated.

\label{sec:VideoDatasets}

The MediaEval 2017 Affective Impact of Movies Task~\cite{DBLP:conf/mediaeval/DellandreaH0BS17} is focused on long movies for two sub-tasks: valence/arousal prediction and Fear prediction. The Continuous LIRIS-ACCEDE collection was selected as the development set, and an additional 14 new movies were collected as the test set. The annotations contain a valence value and an arousal value. In addition, there are a binary value to represent whether the segment is supposed to induce fear or not for each 10-second segment.

The MediaEval 2018 Affective Impact of Movies task~\cite{DBLP:conf/mediaeval/DellandreaH0BXS18} is also dedicated to valence/arousal prediction and fear prediction. The Continuous LIRIS-ACCEDE collection and the test set of the MediaEval 2017 Emotional Impact of Movies task were used as the development set. In addition, 12 other movies selected from the set of the 160 movies mentioned in the Discrete LIRIS-ACCEDE part were used as test set. Specifically, for the first sub-task, there are annotations containing valence and arousal values for each second of the movies. And the beginning and ending times of each sequence in movies that induce fear are recorded for the second sub-task. 

The VideoEmotion dataset~\cite{jiang2014predicting} is a well-designed user-generated video collection. It contains 1,101 videos downloaded from web platforms, such as YouTube and Flickr. The annotations of the videos in this dataset are based on Plutchik's wheel of emotions~\cite{plutchik1980emotion}.

Both the YF-E6 Dataset and the VideoStory-P14 Dataset are introduced in~\cite{xu2016heterogeneous}. In order to collect the YF-E6 emotion dataset, six basic emotion types are used as keywords to search videos on YouTube and Flickr. There are 3,000 videos collected in the YF-E6 dataset totally. Then there were 10 annotators performing the labeling tasks. Only when all tags for a video clip were more than 50 percent consistent, the video clip was added to the dataset. Finally, the dataset includes 1,637 videos labeled with six basic emotion types. The VideoStory-P14 Dataset is based on the VideoStory dataset. Similar to the VideoEmotion  Dataset, the keywords in Plutchik's Wheel of Emotions were used for the search process of the construction of the VideoStory dataset. Finally, there are 626 videos in the videoStory-P14 dataset with each having a unique emotion tag.

\subsubsection{Datasets including both videos and audience's reactions}

The DEAP dataset~\cite{koelstra2012tac} includes the EEG and peripheral physiological signals that are collected from 32 participants during watching 40 one-minute long excerpts of music videos. In addition, frontal face videos collected from 22 of the 32 participants are gathered. Annotators labeled each video according to the level of like/dislike, familiarity, arousal, valence, and dominance. Though the DEAP dataset is publicly available, it should be noted that it does not include the actual videos because of the licensing issues, but the links of videos are provided.

The MAHNOB-HCI~\cite{soleymani2012multimodal} is a multimodal dataset including multi-class information recorded in response to video affective stimuli. Particularly, speeches, face videos, and eye gazes are recorded. In addition, two experiments were conducted to record both peripheral and central nervous system physiological signals from 27 subjects. In the first experiment, subjects were assigned to report their emotional responses to 20 affective induced videos, including the level of arousal, valence and dominance, and predictability as well as emotion categories. In the second experiment, the participants evaluated whether they agreed with the displayed labels after watching short videos and images. The dataset is available for academic use through a web-interface.

\begin{table*}[!t]
\centering\scriptsize
    \caption{Released and freely available datasets for video emotion recognition, where `\#Clips' and `Hours' respectively represent the total number and hours of video clips, `Type' means the genre of the videos in the dataset, `Emotion model' represents the labeling type, `Labeling' represents the method to obtain labels, such as human annotation (annotation) and keyword searching (keyword), and `Labels' means the detailed labels in the dataset, such as dominant emotion category (dominant), average dimension values (average), personalized emotion (personalized), and emotion distribution (distribution).}
\resizebox{\textwidth}{!}{%
\begin{tabular}
{C{1.8 cm} cccccccc}
\hline
\textbf{Dataset} & \textbf{Ref} & \textbf{\#Clips}& \textbf{Hours} & \textbf{Type} & \textbf{\# Annotators} & \textbf{Emotion model} & \textbf{Labeling} & \textbf{Labels} \\
\hline
Discrete LIRIS-ACCEDE & \cite{baveye2015liris} & 9,800 & 26.9 & film & - & VA & annotation & dominant \\
Continuous LIRIS-ACCEDE & \cite{baveye2015deep} & 30 & 7.4 & film & 10 (7f,3m) & VA & annotation & average \\
MediaEval 2015 & \cite{sjoberg2015mediaeval} & 1,100  & - & film & - & 3 discrete VA values & annotation & dominant \\
MediaEval 2016 & \cite{dellandrea2016mediaeval} & 1,210 & - & film & - & VA & annotation & distribution, average \\
MediaEval 2017 & \cite{DBLP:conf/mediaeval/DellandreaH0BS17} & 14 & 8 & film & - & VA, fear & annotation & average \\
MediaEval 2018 & \cite{DBLP:conf/mediaeval/DellandreaH0BXS18} & 12 & 9 & film & - & VA, fear & annotation & average \\
VideoEmotion & \cite{jiang2014predicting} & 1,101 & 32.7 & user-generated & 10 (5f,5m) & Plutchik & annotation & dominant \\
YF-E6 & \cite{xu2016heterogeneous} & 1,637 & 50.9 & user-generated & 10(5f,5m) & Emkan & annotation & dominant \\
VideoStory-P14 & \cite{xu2016heterogeneous} & 626 & - & user-generated & - & Plutchik & keyword & dominant \\
DEAP & \cite{koelstra2012tac} & 120 & 2 & music video & - & VAD & annotation & personalized  \\
MAHNOB-HCI & \cite{soleymani2012multimodal} & - & - & multiple types & - & VAD, Ekman+neutral & annotation & personalized \\
DECAF & \cite{abadi2015decaf} & 76 & - & music video/movies & - & VAD & annotation & personalized \\
AMIGOS & \cite{DBLP:journals/corr/CorreaASP17} & 20 & - & movies collection & - & VAD, Ekman & annotation & personalized \\
ASCERTAIN & \cite{subramanian2018ascertain} & 36 & - & movies collection & 58 (21f,37m) & VA & annotation & personalized \\
\hline
\end{tabular}
}
\label{tab:VideoDataset}
\end{table*}

The DECAF dataset~\cite{abadi2015decaf} consists of  Infra-red facial video signals, Electrocardiogram (ECG), Magnetoencephalogram (MEG), horizontal Electrooculogram (hEOG) and Trapezius Electromyogram (tEMG), recorded from 30 participants watching 36 movie clips and 40 one-minute music videos, which are derived from the DEAP dataset~\cite{koelstra2012tac}. The subjective feedback is based on valence, arousal, and dominance space. In addition, time-continuous emotion annotations for movie clips are also included in the dataset.

The AMIGOS dataset~\cite{DBLP:journals/corr/CorreaASP17} includes multi-class affective data, individual and groups of viewers' responses to both short and long videos. The EEG, ECG, GSR, frontal, and full body video were recorded in two experimental settings, \textit{i.e.}, 40 participants watching 16 short emotional clips and 4 long clips. The duration of each selected short videos is between 51 and 150 seconds, and the duration of each long excerpt is about 20 minutes. Finally, participants annotated the affective level of valence, arousal, control, familiarity, liking, and basic emotions.

Big-five personality scales and affective self-ratings of 58 users together with their EEG, ECG, GSR, and facial activity data were included in the ASCERTAIN dataset~\cite{subramanian2018ascertain} . The number of videos used as the stimulus is 36 and the length of each video clip is between 51 and 128 seconds. It is the first physiological dataset that is useful for both affective and personality recognition.

The publicly available datasets for video affective content analysis are summarized in Table~\ref{tab:VideoDataset}.

\subsection{Datasets for AC of Multimodal Data}
\label{sec:multimodalDatasets}
In addition to audiovisual content and viewers' reactions, other modalities, such as language, also contain significant information for affective understanding of multimedia content. 

Visual sentiment is the sentiment associated with the concepts depicted in images. Two datasets were developed through mining images associated with the adjective-noun pair (ANP) representations that have affective significance~\cite{borth2013sentibank}. ANPs in \cite{borth2013sentibank} were generated by first using seed terms from Plutchik's Wheel of Emotion~\cite{plutchik1980emotion} to query Flickr\footnote{\url{https://www.flickr.com}} and YouTube\footnote{\url{https://www.youtube.com/}}. After mining the tags associated with visual content on YouTube and Flickr, adjective and noun candidates were identified through part-of-speech tagging. Then adjective and nouns were paired to create ANP candidates which were  filtered by sentiment strength, named entities, and popularity. The  Visual Sentiment Ontology (VSO), \cite{borth2013sentibank}\footnote{\url{https://visual-sentiment-ontology.appspot.com}}, is the results of this process. Sentibank resulted in the creation of a set of photo-tweet sentiment dataset, with both visual and textual data with polarity labels, collected on Amazon Mechanical Turk\footnote{\url{http://www.ee.columbia.edu/ln/dvmm/vso/download/twitter_dataset.html}}. This work was later extended to form a multilingual ANP set and its dataset, in~\cite{Jou:MVSO,dalmia2016columbia}\footnote{\url{http://mvso.cs.columbia.edu}}, containing 15,630 ANPs from 12 major languages and 7.37M images~\cite{senticart:icmr16}. My Reaction When (MRW) dataset contains 50,107 video-sentence pairs crawled from social media, depicting physical or emotional reactions to the situations described in sentences~\cite{Song_2019_CVPR}. The GIFs are sourced from Giphy\footnote{\url{https://giphy.com/}}. Even though there is no emotional labels, the language and visual associations are mainly based on sentiment which makes this dataset an interesting resource for affective content analysis.

CMU Multimodal Opinion Sentiment and Emotion Intensity (CMU-MOSEI) is a collection of multiple datasets for multimodal sentiment analysis and emotion recognition. This collection includes more than 23,500 sentence utterance videos from more than 1,000 people from YouTube~\cite{zadeh2018multi}\footnote{\url{https://github.com/A2Zadeh/CMU-MultimodalSDK}}. All the videos are transcribed and aligned with audiovisual modalities. 
A multimodal multi-party dataset for emotion recognition in conversation (MELD) was primarily developed for emotion recognition in multiparty interaction purposes~\cite{poria-etal-2019-meld}. MELD contains visual, audio, and textual modalities and includes 13,000 utterances from 1,433 dialogues from the TV-series Friends, with each utterance labeled with emotion and sentiment.

\begin{table*}[!t]
\centering\scriptsize
    \caption{Released and freely available datasets for multimodal multimodal emotion recognition. Disc. for MELD corresponds to six Ekman emotions in addition to neutral. `Labeling' represents the method to obtain labels, such as human annotation (annotation), self-reported felt emotion and keyword searching (keyword), `Labels' means the detailed labels in the dataset, such as dominant emotion category (dominant), average dimension values (average), personalized emotion (personalized), and emotion distribution (distribution).} 
\resizebox{\textwidth}{!}{%
\begin{tabular}
%{p{1.2cm} c p{1cm} p{2.1cm} l l p{2cm} l}
{c c c c c c c c}
\hline
\textbf{Dataset} & \textbf{Ref} & \textbf{\#Samples}& \textbf{Modalities} & \textbf{Type} & \textbf{Emotion model} & \textbf{Labeling}&\textbf{Labels}\\
\hline
SentiBank tweet & \cite{borth2013sentibank} & 603  & images, text & images & Sentiment&annotation& dominant\\
MVSO & \cite{Jou:MVSO} & 7.36M  & image, metadata & photos & Sentiment &automatic & average\\
CMU-MOSEI & \cite{zadeh2018multi} &  23,500  & video, audio, text & YouTube videos & Sentiment &annotation& average\\
MELD & \cite{poria-etal-2019-meld} &  13,000  & video, audio, text & TV series & Sentiment, Disc. &annotation& dominant\\
COGNIMUSE & \cite{Zlatintsi2017} &  3.5h  & video, audio, text & movies & VA&annotation, self-report& dominant\\
VR & \cite{Li_VR} &  73  & video, audio & VR videos & VA&self-report&average \\
\hline
\end{tabular}
}
\label{tab:MMDataset}
\end{table*}
%Ekman+(neutral,non-neutral)

COGNINMUSE is a collection of videos annotated with sensory and semantic saliency, events, cross-media semantics, and emotions~\cite{Zlatintsi2017}. A subset of 3.5h extracted from movies, including textual modality, are annotated on arousal and valence. 
\citeauthor{Li_VR} collected a dataset of 360 degrees virtual reality videos that can elicit different emotions~\cite{Li_VR}. Even though the dataset consists of 73 short videos, on average 183s long, it is one of the first datasets of its kind whose content understanding stays limited. These multimodal datasets are summarized in Table~\ref{tab:MMDataset}.

\section{Affective Computing of Images}
\label{sec:Image}

In the early stages, AC researchers mainly worked on designing handcrafted features to bridge the affective gap. Recently, with the advent of deep learning especially convolutional neural networks (CNNs), current methods have shifted to an end-to-end deep representation learning. Motivated by the fact that the perception of image emotions may be dependent on different types of features~\cite{zhao2014affective}, some methods employ fusion strategies to jointly consider multiple features. In this section, we summarize and compare these methods. Please note that here we classify the directly extracted CNN features based on pre-trained deep models into handcrafted features category.

\subsection{Handcrafted Features-Based Methods for AC of Images}
\label{sec:ImageHandcrafted}

\textbf{Low-level Features} are difficult to be understood by viewers. These features are often directly derived from other computer vision tasks. Some widely extracted features include GIST, HOG2x2, self-similarity and geometric context color histogram features as in~\cite{patterson2012sun}, because of their individual power and distinct description of visual phenomena in a scene perspective.

Compared with the above generic features, some specific features derived from art theory and psychology have been designed. For example, \citeauthor{machajdik2010affective}~\shortcite{machajdik2010affective} extracted elements-of-art features, including \emph{color} and \emph{texture}. The MPEG-7 visual descriptors are employed in~\cite{lee2011fuzzy}, which include four color-related ideas and two texture-related ideas. How shape features in natural images influence emotions is investigated in~\cite{lu2012shape} by modeling the concepts of roundness-angularity and simplicity-complexity. \citeauthor{sartori2015s}~\shortcite{sartori2015s} designed two kinds of visual features to represent different color combinations based on Itten's color wheel.

\textbf{Mid-level Features} contain more semantics, are more easily interpreted by viewers than low-level features, and thus are more relevant to emotions. \citeauthor{patterson2012sun}~\shortcite{patterson2012sun} proposed to detect 102 attributes in 5 different categories, including materials, surface properties, functions or affordances, spatial envelop attributes, and object presence. Besides these attributes, eigenfaces that may contribute to facial images are also incorporated in~\cite{yuan2013sentribute}. More recently, in~\cite{rao2016multi}, SIFT features are first extracted as basic features, which are fed into bag-of-visual-words (BoVW) to represent the multi-scale blocks. Another mid-level representation is the latent topic distribution estimated by probabilistic latent semantic analysis.

Harmonious composition is essential in an artwork. Several compositional features, such as low depth of field, are designed to analyze such characteristics of an image~\cite{machajdik2010affective}. Based on the fact that figure-ground relationships, color patterns, shapes and their diverse combinations are often jointly employed by artists to express emotions in their artworks, \citeauthor{wang2013interpretable}~\shortcite{wang2013interpretable} proposed to extract interpretable aesthetic features.
Inspired by princiles-of-art, \citeauthor{zhao2014exploring}~\shortcite{zhao2014exploring} designed corresponding mid-level features, including \emph{balance}, \emph{emphasis}, \emph{harmony}, \emph{variety}, \emph{gradation}, and \emph{movement}. For example, Itten's color contrasts and the rate of focused attention are employed to measure \textit{emphasis}.

\begin{table*}
\centering\scriptsize
\caption{Summary of the hand-crafted features at different levels for AC of images. `\# Feat' indicates the dimension of each feature.}
\resizebox{\textwidth}{!}{
\begin{tabular}{ccc p{8cm} c}
\hline
\textbf{Feature} & \textbf{Ref} & \textbf{Level} & \multicolumn{1}{c}{\textbf{Short description}} & \textbf{\# Feat} \\
\hline
LOW\_C & \cite{patterson2012sun} & low  & GIST, HOG2x2, self-similarity and geometric context color histogram features & 17,032\\
Elements & \cite{machajdik2010affective} & low  & color: mean saturation, brightness and hue, emotional coordinates, colorfulness, color names, Itten contrast, Wang's semantic descriptions of colors, area statistics; texture: Tamura, Wavelet and gray-level co-occurrence matrix & 97\\
MPEG-7 & \cite{lee2011fuzzy} & low  & color: layout, structure, scalable color, dominant color; texture: edge histogram, texture browsing &  $\approx$200 \\
Shape & \cite{lu2012shape} & low & line segments, continuous lines, angles, curves &  219\\
IttenColor & \cite{sartori2015s} & low & color co-occurrence features and patch-based color-combination features & 16,485\\
\hline
Attributes & \cite{patterson2012sun} & mid & scene attributes & 102 \\
Sentributes & \cite{yuan2013sentribute} & mid & scene attributes, eigenfaces & 109 \\
Composition & \cite{machajdik2010affective} & mid & level of detail, low depth of field, dynamics, rule of thirds & 45 \\
Aesthetics & \cite{wang2013interpretable} & mid & figure-ground relationship, color pattern, shape, composition & 13 \\
Principles & \cite{zhao2014exploring} & mid & principles-of-art: balance, contrast, harmony, variety, gradation, movement &  165\\
BoVW & \cite{rao2016multi} & mid & bag-of-visual-words on SIFT, latent topics & 330 \\
\hline
FS & \cite{machajdik2010affective} & high & number of faces and skin pixels, size of the biggest face, amount of skin w.r.t. the size of faces & 4 \\
ANP & \cite{borth2013large} & high & semantic concepts based on adjective noun pairs & 1,200 \\
Expressions & \cite{yang2010exploring} & high & automatically assessed facial expressions (anger, contempt, disgust, fear, happiness, sadness, surprise, neutral) & 8 \\
\hline
\end{tabular}
}
\label{tab:ImageHandCraftedFeatures}
\end{table*}

\textbf{High-level Features} that represent the semantic content contained in images can be easily understood by viewers. We can also well recognize the conveyed emotions in images through these semantics. In the early years, simple semantic content including faces and skins contained in images are extracted in~\cite{machajdik2010affective}. For the images that contain faces, facial expressions may directly determine the emotions. \citeauthor{yang2010exploring}~\shortcite{yang2010exploring} extracted 8 kinds of facial expressions as high-level features. They built compositional features of local Haar appearances by a minimum error based optimization strategy, which are embedded into an improved AdaBoost algorithm. For the images detected without faces, the experessions are simply set as \emph{neutral}. Finally, they generated a 8 dimensional vector with each element representing the number of corresponding facial expressions.

\begin{table*}[!t]
\centering\scriptsize
\caption{Representative work on AC of images using hand-crafted features, where `Fusion' indicates the fusion strategy of different features, `cla, reg, ret, cla\_p, dis\_d, dis\_c' in the Task column are short for classification, regression, retrieval, personalized classification, discrete distribution learning, continuous distribution learning (the same below), respectively, `Result' is the reported best accuracy for classification, mean squared error for regression, discounted cumulative gain for retrieval, F1 for personalized classification, and KL divergence for distribution learning (the first line~\cite{zhao2015predicting} is the result on sum of squared difference) on the corresponding datasets.}
\begin{tabular}
{l p{2.5cm} l p{1.5cm} p{2.5cm} l R{2.5cm}} %Bjoern: "leftified" :)
\hline
\textbf{Ref} & \textbf{Feature} & \textbf{Fusion} & \textbf{Learning} & \textbf{Dataset} & \textbf{Task} & \textbf{Result} \\
\hline
\cite{machajdik2010affective} & Elements, Composition, FS & early & NB  & IAPSa, Abstract, ArtPhoto & cla & 0.471, 0.357, 0.495\\
\cite{lee2011fuzzy} & MPEG-7 & -- & KNN  & unreleased &  cla  & 0.827\\
\cite{lu2012shape} & Shape, Elements  & early & SVM, SVR &  IAPSa; IAPS &  cla; reg & 0.314; V-1.350, A-0.912\\%shape:0.299, V: 1.708, A: 0.943\\
\cite{li2012context} & Segmented objects  & -- & SL &  IAPS, ArtPhoto &  cla & 0.612, 0.610\\
\cite{yuan2013sentribute} & Sentributes  & -- & SVM, LR & Tweet & cla  & 0.824\\
\cite{wang2013interpretable} & Aesthetics & -- & NB & Abstract, ArtPhoto & cla  & 0.726, 0.631\\
\cite{zhao2014exploring} & Principles  & -- & SVM, SVR & IAPSa, Abstract, ArtPhoto; IAPS &  cla; reg & 0.635, 0.605, 0.669; V-1.270, A-0.820\\
\cite{zhao2014affective} & LOW\_C, Elements, Attributes, Principles, ANP, Expressions & graph & MGL & IAPSa, Abstract, ArtPhoto, GAPED, Tweet &  ret & 0.773, 0.735, 0.658, 0.811, 0.701\\
\cite{sartori2015s} & IttenColor & -- & SL & MART, devArt & cla  & 0.751, 0.745\\
\cite{rao2016multi} & BoVW & -- & MIL & IAPSa, Abstract, ArtPhoto & cla  & 0.699, 0.636, 0.707\\
\cite{alameda2016recognizing} & IttenColor  & -- & MC & MART, devArt & cla  & 0.728, 0.761\\
\hline
\cite{zhao2016predicting} & GIST, Elements, Attributes, Principles, ANP, Expressions & graph & RMTHG & IESN &  cla\_p & 0.582\\
\hline
\cite{zhao2015predicting} & GIST, Elements, Principles & - & SSL  & Abstract & dis\_d & 0.134\\
\cite{zhao2017approximating} & GIST, Elements, Attributes, Principles, ANP, deep features from AlexNet  & weighted & WMMSSL  & Abstract, Emotion6, IESN & dis\_d & 0.482, 0.479, 0.478 \\
\cite{yang2017learning} & ANP, VGG16  & - & ACPNN &  Abstract, Emotion6, FlickrLDL, TwitterLDL & dis\_d & 0.480, 0,506, 0,469, 0.555\\
\cite{zhao2017learning} & GIST, Elements, Attributes, Principles, ANP, AlexNet  & weighted & WMMCPNN &  Abstract, Emotion6, IESN  & dis\_d & 0.461, 0.464, 0.470\\
\cite{zhao2017continuous} & GIST, Elements, Attributes, Principles, ANP, AlexNet  & -- & MTSSR &  IESN  & dis\_c & 0.436\\
\hline
\end{tabular}
\label{tab:ImageHandCraftedMethods}
\end{table*}

More recently, the semantic concepts are described by adjective noun pairs (ANPs)~\cite{borth2013large,Chen2014DeepSentiBank}, which are detected by SentiBank~\cite{borth2013large} or DeepSentiBank~\cite{Chen2014DeepSentiBank}. The advantages of ANP are that it turns a neutral noun into an ANP with strong emotions and makes the concepts more detectable, compared to nouns and adjectives, individually. A 1,200 dimensional vector representing the probability of the ANPs can form a feature vector.

Table~\ref{tab:ImageHandCraftedFeatures} summarizes the above-mentioned hand-crafted features at different levels for AC of images. Some recent methods also extracted CNN features from pre-trained deep models, such as AlexNet~\cite{zhao2017approximating,zhao2017learning} and VGGNet~\cite{yang2017learning}.

To map the extracted handcrafted features to emotions,
\textbf{Machine Learning Methods} are commonly employed. Some typical learning models include Naive Bayes (NB)~\cite{machajdik2010affective,wang2013interpretable}, support vector machine (SVM)~\cite{lu2012shape,yuan2013sentribute,zhao2014exploring}, $K$ nearest neighbor (KNN)~\cite{lee2011fuzzy}, sparse learning (SL)~\cite{li2012context,sartori2015s}, logistic regression (LR)~\cite{yuan2013sentribute}, multiple instance learning (MIL)~\cite{rao2016multi}, and matrix completion (MC)~\cite{alameda2016recognizing} for emotion classification , support vector regression (SVR)~\cite{lu2012shape,zhao2014exploring} for emotion regression, and multi-graph learning (MGL)~\cite{zhao2014affective} for emotion retrieval.

Instead of assigning the DEC to an image, some recent methods began to focus on the perception subjectivity challenge, \textit{i.e.}, predicting personalized emotions for each viewer or learning emotion distributions for each image. The personalized emotion perceptions of a specified user after viewing an image is predicted in~\cite{zhao2016predicting,zhao2018predicting}, associated with online social networks. They considered different types of factors that may contribute to emotion recognition, including the images' visual content, the social context related to the corresponding users, the emotions' temporal evolution, and the images' location information. To jointly model these factors, they proposed rolling multi-task hypergraph learning (RMTHG), which can also easily hanlde the data incompleteness issue.

Generally, the distribution learning task can be formulated as a regression problem, which slightly differs for different distribution categories (\textit{i.e.}, discrete or continuous). For example, if emotion is represented by CE, the regression problem targets predicting the discrete probability of each emotion category with the sum equal to 1; if we represent emotion based on DES, the regression problem is typically transformed to the prediction of the parameters of specified continuous probability distributions. For the latter one, we usually need to firstly determine the form of continuous distributions, such as exponential distribution and Gaussian distribution. Some representative learning methods for emotion distribution learning of discrete emotions include shared sparse learning (SSL)~\cite{zhao2015predicting}, weighted multimodal SSL (WMMSSL) \cite{zhao2017approximating,zhao2018discrete}, augmented conditional probability neural network (ACPNN)~\cite{yang2017learning}, and weighted multi-model CPNN (WMMCPNN)~\cite{zhao2017learning}. Both SSL and WMMSSL can only model one test image each time, which is computationally inefficient. After the parameters are learned, ACPNN and WMMCPNN can easily predict the emotion distributions of a test image. Based on the assumption that the VA emotion labels can be well modeled by a mixture of 2 bidimensional Gaussian mixture models (GMMs), \citeauthor{zhao2017continuous}~\shortcite{zhao2017continuous} proposed to learn continuous emotion distributions in VA space by multi-task shared sparse regression (MTSSR). Specifically, the parameters of GMMs are regressed, including the mean vector and covariance matrix of the 2 Gaussian components as well as the mixing coefficients.

Table~\ref{tab:ImageHandCraftedMethods} summarizes some representative work based on hand-crafted features. Generally, high-level features (such as ANP) can achieve better recognition performance for images with rich semantics, mid-level features (such as Principles) are more effective for artistic photos, while low-level features (such as Elements) perform better for abstract paintings.

\subsection{Deep Learning-Based Methods for AC of Images}
\label{sec:ImageDeep}
To deal with the situation where images are weakly labeled, a potentially cleaner subset of the training instances are selected progressively~\cite{you2015robust}. First, they trained an initial CNN model based on the training data. Second, they selected the training samples with distinct sentiment scores between the two classes with a high probability based on the prediction score of the trained model on the training data itself. Finally, the pre-trained AlexNet on ImageNet is fine-tuned to classify emotions into 8 categories by changing the last layer of the CNN from 1000 to 8~\cite{you2016building}. Besides using the fully connected layer as classifier, they also trained an SVM classifier based on the extracted features from the second to the last layer of the pre-trained AlexNet model.

Multi-level deep representations (MldrNet) are learned in~\cite{rao2016learning} for image emotion classification. They segmented the input image into 3 levels of patches, which are input to 3 different CNN models, including Alexnet, aesthetics CNN (ACNN), and texture CNN (TCNN). The fused features are fed into multiple instance learning (MIL) to obtain the emotion labels. \citeauthor{zhu2017dependency}~\shortcite{zhu2017dependency} proposed to integrate the different levels of features by a Bidirectional GRU model (BiGRU) to exploit their dependencies based on MldrNet. They generated two features from the Bi-GRU model and concatenated them as the final feature representations. To enforce the feature vectors extracted from each pair of images from the same category to be close enough, and those from different categories to be far away, they proposed to jointly optimize a contrastive loss together with the traditional cross-entropy loss.

More recently, \citeauthor{yang2018retrieving}~\shortcite{yang2018retrieving} employed deep metric learning to explore the correlation of emotional labels with the same polarity, and proposed a multi-task deep framework to optimize both retrieval and classification tasks. By considering the relations among emotional categories in the Mikels' wheel, they jointly optimized a novel sentiment constraint with the cross-entropy loss. Extending triplet constraints to a hierarchical structure, the sentiment constraint employs a sentiment vector based on the texture information from the convolutional layer to measure the difference between affective images. In~\cite{yang2018weakly,she2019wscnet}, \citeauthor{yang2018retrieving} proposed a weakley supervised coupled convolutional neural network to exploit the discriminability of localized regions for emotion classification. Based on the image-level labels, a sentiment map is firstly detected in one branch with the cross spatial pooling strategy. And then the holistic and localized information are jointly combined in the other branch to conduct a classification task. The detected sentiment map can easily explain which regions of an image determine the emotions.

The above deep methods mainly focused on the dominant emotion prediction. There are also some work on emotion distribution learning based on deep models. The very first work is a mixed bag of emotions, which trains a deep CNN regressor (CNNR) for each emotion category in Emotion6~\cite{peng2015mixed} based on the AlexNet architecture. They changed the number of output nodes to 1 to predict a real value for each emotion category and replaced the Softmax loss with Euclidean loss. To ensure the sum of different probabilities to be 1, they normalized the predicted probabilities of all emotion categories. However, CNNR has some limitations. First, the predicted probability cannot be guaranteed to be non-negative. Second, the probability correlations among different emotions are ignored, since the regressor for each emotion category is trained independently. In~\cite{yang2017joint}, \citeauthor{yang2017joint} designed a multi-task deep framework based on VGG16 by jointly optimizing the cross-entropy loss for emotion classification and Kullback-Leibler (KL) divergence loss for emotion distribution learning. To match the single emotion dataset to emotion distribution learning settings, they transformed each single label to emotion distribution with emotion distances computed on Mikels' wheel~\cite{zhao2016predicting,zhao2018predicting}. By extending the size of training samples, this method achieves the state-of-the-art performance for discrete emotion distribution learning.

\begin{table*}[!t]
\centering\scriptsize
\caption{Representative work on deep learning based AC of images, where `Pre' indicates whether the network is pre-trained using ImageNet, `\# Feat' indicates the dimension of last feature mapping layer before the emotion output layer, `Cla' indicates the classifier used after the last feature mapping with default Softmax, `Loss' indicates the loss objectives (besides the common cross-entropy loss for classification), and `Result' is the reported best accuracy for classification, discounted cumulative gain for retrieval, and KL divergence for distribution learning on the corresponding datasets.}
\resizebox{\textwidth}{!}{%
\begin{tabular}
{c C{1.5cm} c C{1cm} cc C{3cm}  C{0.4cm} C{2cm}} %Bjoern: changed alignment
\hline
\textbf{Ref} & \textbf{Base net} & \textbf{Pre} & \textbf{\#Feat} & \textbf{Cla} & \textbf{Loss} & \textbf{Dataset} & \textbf{Task} & \textbf{Result} \\
\hline
\cite{you2015robust} & self-defined & no  & 24 & -- & -- & FlickrCC & cla & 0.781\\
\cite{you2016building} & AlexNet & yes  & 4,096 & SVM & -- & FI, IAPSa, Abstract, ArtPhoto &  cla  & 0.583, 0.872, 0.776, 0.737\\
\cite{rao2016learning} & AlexNet, ACNN, TCNN & yes & 4,096, 256, 4,096 & MIL & -- &  \tiny{FI, IAPSa, Abstract, ArtPhoto, MART} &  cla & \tiny{0.652, 0.889, 0.825, 0.834, 0.764}\\
\cite{zhu2017dependency} & self-defined & no & 512  & -- & contrastive &  FI, IAPSa, ArtPhoto &  cla & 0.730, 0.902, 0.855\\
\cite{yang2018retrieving} & GoogleNet-Inception & yes & 1,024  & -- & sentiment & FI, IAPSa, Abstract, ArtPhoto & cla; ret  & 0.676, 0.442, 0.382, 0.400; 0.780, 0.819, 0.788, 0.704\\
\cite{yang2018weakly,she2019wscnet} & ResNet-101 & yes & 2,048  & -- & -- & FI, Tweet & cla  & 0.701, 0.814\\
\hline
\cite{peng2015mixed} & AlexNet & yes & 4,096 & -- & Euclidean & Emotion6 & dis\_d  & 0.480\\
\cite{yang2017joint} & VGG16 & yes & 4,096 & -- & KL & Emotion6, FlickrLDL, TwitterLDL & dis\_d & 0.420, 0,530, 0,530\\
\hline
\end{tabular}
}
\label{tab:ImageDeepMethods}
\end{table*}

The representative deep learning based methods are summarized in Table~\ref{tab:ImageDeepMethods}. The deep representation features generally perform better than the hand-crafted ones, which are intuitively designed for specific domains based on several small-scale datasets. However, how the deep features correlate to specific emotions is unclear.

\section{Affective Computing of Music}
\label{sec:Music}
Music emotion recognition (MER) strives to identify emotion expressed by music and subsequently predict listener's felt emotion from acoustic content and music metadata, \textit{e.g.}, lyrics, genre, \textit{etc.} Emotional understanding of music have applications in music recommendation and is particularly useful for producing music retrieval. An analysis of search queries from creative professionals showed that 80\% contain emotional terms, showing emotions prominence in that field~\cite{inskip2012}. A growing number of work have tried to address emotional understanding of music from acoustic content and metadata (see~\cite{yang2012machine,Kim2010} for earlier reviews on this topic). 

Earlier work on emotion recognition from music relied on extracting acoustic features similar to the ones used in speech analysis, such as audio energy and formants. Acoustic features describe attributes related to musical dimensions. Musical dimensions include melody, harmony, rhythm, dynamics, timbre (tone color), expressive techniques, musical texture, and musical form~\cite{panda_thesis}, as shown in Table~\ref{tab:MusicFeatures}. Some also add energy as a musical feature which is important for MER~\cite{yang2012machine}. 
Melody is a linear succession of tones and can be captured by features representing key, pitch and tonality. Among others, chroma is often used to represent melodic features~\cite{yang2012machine}. 
Harmony is how the combination of various pitches are processed during hearing. Understanding harmony involves chords or multiple notes played together. Examples of acoustic features capturing harmony include chromagram, key, mode, and chords~\cite{aljanaki2016emotion}.
Rhythm consists of repeated patterns of musical sounds, \textit{i.e.}, notes and pulses that can be describes in terms of tempo and meter. Higher tempo songs often induce higher arousal and fluent rhythm is associated with higher valence and firm rhythm is associated with sad songs~\cite{yang2012machine}. Mid-level acoustic features, such as onset rate, tempo and beat histogram, can represent rhythmic characteristics of music. 
Dynamics of music involve the variation in softness or loudness of notes which include change of loudness (contrast) and emphasis on individual sounds (accent)~\cite{panda_thesis}. Dynamics of music can be captured by changes in acoustic features related to energy such as root mean square (RMS) energy.
Timbre is the perceived sound quality of musical notes. Timbre is what differentiates different voices and instruments playing the same sound. Acoustic features capturing timbre, such as MFCC and spectrum shape, describe sound quality~\cite{Yang2018}.  Acoustic features describing timbre include MFCC, spectral features (centroid, contract, flatness), and zero crossing rate~\cite{aljanaki2016emotion}.
Expressive techniques are the way a musical piece is played including tempo and articulation~\cite{panda_thesis}. Acoustic features, such as tempo, attack slope, and time, can be used to describe this dimension.
Musical texture is how rhythmic, melodic, and harmonic features are combined in music production~\cite{panda_thesis}. It is related to the range of tones played at the same time. Musical form describes how a song is structured, such as introduction verse and chorus~\cite{panda_thesis}. 
Energy whose dynamics are described in music dynamic features is strongly associated with arousal perception.

\begin{table*}[!t]
\centering\scriptsize
    \caption{Musical dimensions and acoustic features describing them.}
\begin{tabular}
{ l p{10cm} }
\hline
\textbf{Musical dimension} & \textbf{Acoustic features} \\
\hline
Melody & Pitch \\%\hline
Harmony & chromagram, chromagram peak, key, mode, key clarity, harmonic, change, chords \\%\hline
Rhythm & tempo, beat histograms, rhythm regularity, rhythm strength, onset rate\\%\hline
Dynamics and loudness & RMS energy, loudness, timpral width\\%\hline
Timbre & MFCC, spectral shapres (centroid, shape, spread, skewness, kurtosis, contrast and flatness),  brightness, rolloff frequency, zero crossing rate, spectral contrast, auditory modulation features, inharmonicity, roughness, dissonance, odd to even harmonic ratio \cite{aljanaki2016emotion}\\%\hline
Musical form & Similarity Matrix (similarity between all possible frames) \cite{panda_thesis}\\%\hline
Texture & attack slope, attack time\\
\hline
\end{tabular}
\label{tab:MusicFeatures}
\end{table*}

There are a number of toolboxes available for extracting acoustic features from music that can be used for music emotion recognition. 
Music Analysis, Retrieval and Synthesis for Audio Signals (Marsyas)~\cite{tzanetakis2000marsyas} is an open source framework developed in C++ that supports extracting a large range of acoustic features with music information retrieval applications in mind, including time-domain zero-crossings, spectral centroid, rolloff, flux, and Mel-Frequency Cepstral Coefficients (MFCC) \textit{etc.}
MIRToolbox~\cite{mirtoolbox} is an open source toolbox implemented in MATLAB for music information retrieval applications. MIRToolbox offers the ability to extract a comprehensive set of acoustic features at  different levels including features related to tonality, rhythm, and structures.
Speech and music interpretation by large-space extraction or OpenSMILE~\cite{eyben2010,eyben2013} is an open source software developed in C++ with the ability to extract a large number of acoustic features for speech and music analysis in real-time. 
LibROSA~\cite{mcfee2015librosa} is a Python package for music and audio analysis. It is mainly developed with music information retrieval application in mind and supports importing from different audio sources and extracting musical features such as onsets chroma and tempo in addition to the low-level acoustic features. 
ESSENTIA~\cite{ESSENTIA} is an open source library developed in C++ with Python interface that is developed for audio analysis. ESSENTIA contains an extensive collection of algorithms supporting audio input/output functionality, standard digital signal processing blocks, statistical characterization of data, and a large set of spectral, temporal, tonal and high-level music descriptors. 

Music emotion recognition either attempts to classify songs or excerpt into categories (classification) or estimate their expressed emotions on continuous dimensions (regression). The choice of machine learning model in music emotion recognition depends on the emotional representation used. Mood clusters~\cite{mirex07}, dimensional representations such as arousal, tension and valence as well as music specific emotion representation can be used. An analysis of the methods proposed for MediaEval ``Music in Emotion'' task submissions revealed that using deep learning accounted for the superior performance for emotion recognition much more than the choice features~\cite{deam}. Recent methods for emotion recognition in music rely on deep learning and often use spectrogram features that are converted to images~\cite{aljanaki2018data}. \citeauthor{aljanaki2018data} proposed learning musically meaningful mid-level perceptual features that can describe emotions in music~\cite{aljanaki2018data}. They demonstrated that perceptual features such as  melodiousness, modality, rhythmic complexity and dissonance can describe a large portion of emotional variance in music both in dimensional representation and MIREX clusters. They also trained a deep convolutional neural network to recognize these mid-level attributes. There have been also work attempting to use lyrics in addition to acoustic content for recognizing emotion in music~\cite{7536113}. However, lyrics are copyrighted and not easily available which hinders further work in this direction.

\section{Affective Computing of Videos}
% \label{sec:Feature}
\label{sec:Video}

Currently, the features used in affective video content analysis are mainly from two categories~\cite{shukla2018multimodal, shukla2017evaluating}. One is considering the stimulus of video content and extracting the features reflecting the emotions conveyed by the video content itself. And the other is extracting features from the viewers. Features extracted from the video content are content-based features, and features formed from the signals of the viewers' responses are viewer-related features.

\subsection{Content-related Features}
% \label{sec:LowlevelFeatures}
Generally speaking, the video content comprise of a series of ordered frames as well as corresponding audio signals. Therefore, it is natural to extract features from these two modalities. The audiovisual features can further be divided into low-level and mid-level according to their ability to describe the semantics of video content.

\subsubsection{Low-level features}
Commonly, the low-level features are directly computed from the raw visual and audio content, and usually carry no semantic information. As for visual content, color, lighting, and tempo are important elements that can endow the video with strong emotional rendering and further give viewers direct visual stimuli. In many cases, computations are conducted over each frame of the video, and the average values of the computational results of the overall video are considered as visual features. Specifically, the color-related features often contain the histogram and variance of color~\cite{shukla2018multimodal,chen2018identifying,zhu2019hybrid}, the proportions of color~\cite{nemati2016incorporating,Niu2017TemporalFV}, the number of white frame and fades~\cite{zhu2019hybrid}, the grayness~\cite{chen2018identifying}, darkness ratio, color energy~\cite{zhao2013flexible,Wang2015Emotion}, brightness ratio and saturation~\cite{niu2017novel,Niu2017TemporalFV}, \textit{etc}. In addition, the differences of dark and light can be reflected by the lighting key, which is used to evoke emotions in video and draw the attention of viewers by creating an emotional atmosphere~\cite{nemati2016incorporating}. As for the tempo-related features, properties of shot can reinforce the expression of video, such as shot change rate and shot length variance~\cite{niu2016novel,niu2017novel,chen2018identifying} according to movie grammar. To better take advantage of the temporal information of the video, the motion vectors have been computed as features in~\cite{zhong2019video}. Since the optical flow can characterize the influence of camera motions, the histogram of optimal flow matrix (HOF) has been computed as features in~\cite{yi2018multi}. Additionally, \citeauthor{yi2018multi}~\shortcite{yi2018multi} traced motion key points at multiple spatial scales and  computed the mean motion magnitude of each frame as features.

To represent audio content, pitch, zero crossing rates (ZCR), Mel frequency cepstrum coefficients (MFCC), and energy are the most popular features  \cite{zhao2013flexible,Wang2015Emotion,han2015arousal,nemati2016incorporating,acar2017comprehensive}. In particular, the MFCC~\cite{hu2016multi,zhu2016video,mcduff2017large,ben2018deep,yi2018multi,zhu2019hybrid} and its $\Delta{MFCC}$ are used to characterize emotions in video clips frequently; while the derivatives and statistics (min, max,mean) of MFCC or $\Delta{MFCC}$ are also explored widely. As for pitch, ~\cite{niu2016novel} shows that pitch of sound is associated closely with some emotions, such as anger with higher pitch and sadness with lower standard deviation of pitch. Similar situation can also occur in the energy~\cite{niu2016novel,Niu2017TemporalFV}. For example, the total energy of anger or happiness is higher than the counterpart of unexciting emotions. ZCR~\cite{Niu2017TemporalFV} is used to separate different types of audio signals, such as music, environmental sound and speech of human. Besides these frequent related features, audio flatness~\cite{zhu2019hybrid}, spectral flux~\cite{zhu2019hybrid}, delta spectrum magnitude, harmony~\cite{Niu2017TemporalFV,sivaprasad2018multimodal,zhu2019hybrid}, band energy ratio, spectral centroid~\cite{hu2016multi,zhu2019hybrid}, and spectral contrast~\cite{Niu2017TemporalFV} are also utilized.

Evidently, the aforementioned features are mostly handcrafted. With the emergence of  deep learning, features can be automatically learned through deep neural networks. Some pre-trained convolutional neural networks (CNNs) are used to learn static representations from every frame or some selected key frames, while  a Long-short term memory (LSTM) is exploited to capture dynamic  representations existing in videos.

For instance, in~\cite{xu2016heterogeneous}, an AlexNet with seven fully-connected layers trained on 2600 ImageNet classes is used to learn features. A Convolutional Auto-Encoder (CAE) is designed to ensure the CNNs can extract the visual features effectively in~\cite{pang2015mutlimodal}. \citeauthor{ben2018deep}~\cite{ben2018deep} first used the pre-trained ResNet-152 to extract feature vectors. And then, these vectors are fed into an LSTM according to their temporal order to extract high-order representations. Pre-trained model, SoundNet, is utilized to learn audio features. Because the expressive emotions of video are induced and communicated by the protagonist in video in many cases, the features of protagonist are extracted from the key frame by a pre-trained CNN and used in video affective analysis in~\cite{zhu2016video,zhu2019hybrid}. In addition to the protagonist, other objects in each frame of video also give insights into emotional expression of video. For example, in~\cite{shukla2018looking}, \citeauthor{shukla2018looking} removed the non-gaze regions from video frames (Eye ROI) and built the coarse grained scene structure remaining gist information by Guassian filter with variance. After the operations above in~\cite{shukla2018looking}, the next video affective analysis may pay more attention to important information and reduce unnecessary noise.

\subsubsection{Mid-level features}
Unlike low-level features, mid-level features often contain semantic information. For example, EmoBase10 feature depicting audio cues is computed in~\cite{yi2018multi}. \citeauthor{hu2016multi}~\cite{hu2016multi} proposed a method of combining the audio and visual features to model contextual structures of the key frames selected from video. This can produce a kind of so-called multi-instance sparse coding (MI-SC) for next analysis. In addition, the lexical features~\cite{muszynski2019recognizing} are extracted from the dialogues of speakers by using a natural language toolkit. These features can reflect the emotional changes in videos and can also represent a certain emotional expression in overall videos. \citeauthor{muszynski2019recognizing}~\cite{muszynski2019recognizing} used aesthetic movie highlights related to occurrences of meaningful movie scenes to define some experts. These features produced by experts are more knowledgeable and abstract for video affective analysis, especially movies. HHTC features, which are computed on the basis of combination of Huang Transform in visual-audio and cross-correlation features, are proposed in~\cite{niu2017novel}.

\subsection{Viewer-related Features}
\label{sec:Viewer-relatedFeatures}
Besides the content-related features, viewers' facial expressions and changes of physiological signals evoked by content of videos are the most common sources for extracting viewer-related features. \citeauthor{mcduff2017large}~\cite{mcduff2017large} coded the facial actions of viewers for further affective video analysis. Among various physiological signals, electrocardiography (ECG), galvanic skin response (GSR), electroencephalography (EEG) are the mostly frequently ones and their statistical measures, such as mean, median, spectral power bands, \textit{etc}., are often recommended as features. \citeauthor{Wang2015Emotion}~\cite{Wang2015Emotion} used EEG signals to construct a new EEG feature with the assistance of the relationship among video content by exploiting canonical correlation analysis (CCA).
In~\cite{gui2018implicit}, some viewer-related features are extracted from the whole pupil dilation ratio time-series without the differences among pupil diameter in human eyes, such as its average and derivation for global features as well as the four spectral power bands for local features.

In addition to viewers' responses mentioned above, the comments or other textual information produced by viewers can also reflect their attitudes or emotional reactions toward the videos. In the light of this, it is reasonable to consider users' textual comments or other textual information to extract features. In~\cite{nemati2016incorporating}, the ``sentiment analysis'' module using Unigrams and Bigrams is built to learn comment-related features of the collected data according to the YouTube link provided by the DEAP dataset.

\subsection{Machine Learning Methods}    
After feature extracting,  a classifier or a regressor is used to obtain emotional analysis results. For classification, there are several  frequently used classifiers, including support vector machines (SVM)~\cite{nemati2017evidential,acar2017comprehensive,yi2018multi,wang2017content,gupta2016quality,shukla2017evaluating}, Naive Bayes (NB)~\cite{gupta2016quality,nemati2017evidential}, Linear Discriminant Analysis~\cite{shukla2018looking}, logistic regression (LR)~\cite{xu5731visual}, and ensemble learning~\cite{acar2017comprehensive}, \textit{etc}.

Recent work show that the SVM-based methods are very popular for affective video content analysis due to its simplicity, max-margin training property, and use of kernels ~\cite{wang2015video}.  For example,  \citeauthor{yi2018multi}' work ~\cite{yi2018multi} demonstrated that linear SVM is more suitable for classification than RBM, MLP, and LR . In~\cite{shukla2017evaluating}, LDA, linear SVM (LSVM), and Radial Basis SVM (RSVM) classifiers are employed in emotion recognition experiments, and the RSVM obtained the best F1 scores. In~\cite{gupta2016quality}, both Navie Bayes and SVM are used as classifiers in unimodal and multimodal conditions. In the unimodal experimental condition, NB is not better than SVM. And the fusion results showed that SVM is much better than NB in multimodal situations. However, SVM also has its shortages, such as the difficulty of selecting suitable kernel functions. Indeed, SVM is not always the best choice. In~\cite{acar2017comprehensive}, the results demonstrated that ensemble learning outperforms SVM in terms of classification accuracy. Ensemble learning has acquired a lot of attention in many fields because of its accuracy, simplicity, and robustness. In addition, in~\cite{xu5731visual}, LR is adopted as the classifier for its effectiveness and simplicity. In fact, LR is used frequently in many transfer learning tasks.

However, all the classifiers mentioned above are not able to capture the temporal information. Some other methods try to use temporal information. For example, \citeauthor{gui2018implicit}~\cite{gui2018implicit} combined SVM and LSTM to predict emotion labels. Specifically, global features and sequence features are proposed to represent the pupillary response signals. Then a SVM classifier is trained with the global features and a LSTM classifier is trained with the sequence features. Finally, a decision fusion strategy is proposed to combine these two classifiers.

\begin{table*}[!t]
\centering\scriptsize
\caption{Representative work on AC of videos using kinds of features, where $P_{a}$, $P_{v}$, $MSE_{a}$, $MSE_{v}$, $Acc_{a}$, $Acc_{v}$, $Acc$ and $MAP$ indicates the Pearson correlation coefficients of arousal and valence, the mean sum error of arousal and valence, the accuracy of arousal and valence, the average accuracy and mean average precision respectively. `statistics' means (min, max, mean).}
\resizebox{\textwidth}{!}{%
\begin{tabular}
{p{0.4 cm} p{4 cm} p{0.7 cm} p{2 cm} p{2 cm} p{0.3 cm} R{2cm}} %Bjoern: changed alignment
\hline
\textbf{Ref} & \textbf{Feature} & \textbf{Fusion} & \textbf{Learning} & \textbf{Dataset} & \textbf{Task} & \textbf{Result} \\
\hline
\cite{sivaprasad2018multimodal} & Mel frequency spectral; MFCC, Chroma and their derivatives and statistics; Audio compressibility; Harmonicity; Shot frequency; HOF and statistics; Histogram of 3D HSV and statistics; Video compressibility; Histogram of facial area & decision & LSTM & Dataset described by Malandrakis \cite{5946961} & reg & $P_{a}$:$0.84 \pm 0.06$ $MSE_{a}$:$0.08 \pm 0.04$ $P_{v}$:$0.50 \pm 0.14$ $MSE_{v}$:$0.21 \pm 0.06$\\

\cite{gui2018implicit} & Average, standard deviation and four spectral power bands of pupil dilation ratio time-series & decision & SVM, LSTM  & MAHNOB-HCI & cla & $Acc_{a}$:0.730, $Acc_{v}$:0.780\\

\cite{ben2018deep} & Audio and visual deep features from pretrained model & feature & CNN, LSTM, SVM & PMIT & cla & $MAP$:0.0.2122\\

\cite{acar2017comprehensive} & MFCC; Color values; HoG; Dense trajectory descriptor; CNN-learned features & decision & CNN, SVM, Ensemble & DEAP & cla & $Acc$: 0.81, 0.49 \\

\cite{mo2018novel} & HHTC features & feature & SVR & Discrete LIRIS-ACCEDE & reg & $MSE_{a}$:0.294, $MSE_{v}$: 0.290 \\

\cite{gupta2016quality} & Statistical measures (such as mean, median, skewness kurtosis) for EEG data, power spectral features, ECG, GSR, Face/Head-pose  & decision & SVM/NB &  music excerpts~\cite{morreale2013robin} & cla & F1 (v: 0.59, 0.58, a: 0.60, 0.57) \\

\cite{guo2019affective} & Time-span visual and visual features & feature & CNN Opensmile toolbox &  music excerpts~\cite{morreale2013robin} & cla & $MSE_{a}$:0.082 $MSE_{v}:0.071$\\

\cite{han2015arousal} & tempo; pitch; zero cross; roll off; MFCCs; Saturation; Color heat; Shot length feature; General preferences; Visual excitement; Motion feature; fMRI feature & feature & DBM SVM & TRECVID & cla & - \\

\cite{zhu2019hybrid} & Colorfulness; MFCC; CNN-learned features from the keyframes containing protagonist & decision & CNN, SVM, SVR & LIRIS-ACCEDE, PMSZU & cla/reg & - \\

\cite{zhong2019video} & Multi-frame motion vectors & decision & CNN & SumMe, TVsum, Continuous LIRIS-ACCEDE & reg & - \\

\cite{hu2016multi} &  The median of the L values in Luv space; means and variances of components in HSV space; texture feature; mean and standard deviation of motions between frames in a short; MFCC; Spectral power; mean and variance of the spectral centroids; Time domain zero crossings rate; Multi-instance sparse coding & feature & SVM & Musk1, Musk2, Elephant, Fox, Tiger & cla & $Acc$: 0.911, 0.906, 0.885, 0.627, 0.868\\

\cite{nemati2016incorporating} & Lighting key; Color; Motion vectors; ZCR; energy; MFCC; pitch; Textual features & decision & SMO, Navie Bayes & DEAP & cla & F1:0.849 0.811 $Acc$: 0.911 0.883 \\

\cite{chen2018identifying} & Key lighting; Grayness; Fast motion; Shot chanage rate; Shot length variation; MFCC; CNN-learned features; power spectral density; EEG; ECG; respiation; galvanic skin resistance & feature & SVM & DEAP & cla & $Acc_v$:0.7 0.7 0.7125, $Acc_a$:0.6876 0.7 0.8 F1 (A:0.664 0.687 0.789) \\

\cite{nemati2017evidential} & MFCC; ZCR; energy; pitch, color histograms; lighting key; motion vector & decision & SVM, Navie Bayes & DEAP & cla & F1: 0.869, 0.846 $Acc$: 0.925, 0.897 \\

\cite{shukla2017evaluating} & CNN feature, low-level audio visual features, EEG & decision & LDA, LSVM, RSVM & Dataset introduced by the authors & cla & - \\

\cite{yi2018multi} & MKT; ConvNets feature; EmoLarge; IS13; MFCC; EmoBase10; DSIFT; HSH & decision & SVM, LR, RBM, MLP & MediaEval 2015, 2016 Affective Impact of Movies & cla & $Acc_a$:0.574, $Acc_v$:0.462\\
%, $MSE_a$:0.117, $MSE_v$:0.198 \\

\cite{shukla2018looking} & CNN feature & - & SVM, LDA & dataset in~\cite{shukla2017affect} & cla & - \\

\cite{baveye2015deep} & CNN feature & - & SVR & LIRIS-ACCEDE & reg &  $MSE_{a}$: 0.021,  $MSE_{v}$: 0.027\\

\hline
\end{tabular}
}
\label{tab:VideoHandCraftedMethods}
\end{table*}

A regressor is needed when mapping the extracted features to the continuous dimensional emotion space. Recently, one of the most popular regression method is support vector regression (SVR)~\cite{mo2018novel,baveye2015deep,zhu2019hybrid}. For example,  in~\cite{baveye2015deep}, video features like audio, color, aesthetic are fed into SVR in the SVR-Standard experiment. And in the SVR-Transfer learning experiment, the pre-trained CNN is treated as a feature extractor. The CNN's outputs are used as the input to the SVR. The experimental results showed that the SVR-Transfer learning outperforms other methods. Indeed, the various kernel functions in SVR  provide a stronger adaptability.

\subsection{Data Fusion}
In total, there are two fusion strategies for multimodal information: feature-level fusion and decision-level fusion. Feature-level fusion means that the multimodal features are combined and  then used as the input of a classifier or a regressor. Decision-level fusion fuses several results of different classifiers, and the final results are computed according to the fusion methods.

One way of feature-level fusion is implemented by feature accumulation or concatenation~\cite{ben2018deep,zhu2019hybrid,yi2018multi}. In~\cite{ben2018deep}, two feature vectors for visual and audio data are averaged as the global genre representations. In~\cite{zhu2019hybrid}, multi-class features are concatenated to generate a high dimensional joint representation. Some machine learning methods are also employed to learn joint features~\cite{guo2019affective,han2015arousal,pandeya2019music,xing2019exploiting}. In~\cite{guo2019affective}, a two-branch network is used to combine the visual and audio features. The outputs of the two-branch network are then fed into a classifier, and the experiment results showed that the joint features outperform other methods. In~\cite{han2015arousal}, the low-level audio-visual features and fMRI-derived features are fed into multimodal DBM to learn joint representations. The target of this method is to learn the relation between audio-visual features and fMRI-derived features. In~\cite{xing2019exploiting}, PCA is used to learn the multimodal joint features. In~\cite{Wang2015Emotion}, canonical correlation analysis (CCA) is used to construct a new video feature space with the help of EEG features and a new EEG feature space with the assistance of video content, so only one modality is needed to predict emotion during the testing process.

By combining the results of different classifiers, decision-level fusion strategy is able to achieve better results~\cite{gui2018implicit,nemati2016incorporating,acar2017comprehensive,gupta2016quality,nemati2017evidential,sivaprasad2018multimodal}. In~\cite{acar2017comprehensive}, linear fusion and SVM-based fusion techniques are explored to combine outputs of several classifiers. Specifically, the output of each classifier has its own weight in linear fusion. The final result is the weighted sum of all outputs. In SVM-based fusion, the outputs of unimodal classifiers are concatenated together. And then the higher level representations for each video clip are fed into a fusion SVM to predict the emotion. Based on these results, linear fusion is better than SVM-based fusion. In~\cite{gupta2016quality,nemati2017evidential}, linear fusion is also used to fuse the outputs of multiple classifiers. The differences among these linear fusion methods depend on the distribution of weights.

\subsection{Deep Learning Methods}
\label{sec:Deep learning method}

In tradition, video emotional recognition includes two steps, \textit{i.e.}, feature extraction step and regression or classification step. Because of the lack of consensus on the most relevant emotional features, we may not be able to extract the best features for the problem at hand. As a result, this two-step mode has hampered the development of affective video content analysis. In order to solve this problem, some methods based on end-to-end training frameworks are proposed. \citeauthor{khorrami2016deep} \cite{khorrami2016deep} combined CNN and RNN to recognize the emotional information of videos. According to their method, a CNN is trained using frame facial images sampled from videos to extract features. Then the features are fed into a RNN to perform continuous emotion recognition. In~\cite{huang2018end}, a single network using ConvLSTM is proposed, where videos are input to the network and the predicted emotional information is output directly. In fact, due to the complexity of CNNs and RNNs, the training of these frameworks needs large amounts of data. However, in video affective content analysis, the samples in existing datasets are usually limited. This is the reason why end-to-end methods are still less common compared to the traditional two step methods, despite of their influential potentials.

\section{Affective Computing of Multimodal Data}
\label{sec:multimodal}
In this section, we survey the work that analyze multimodal data beyond audiovisual content. Most of the existing work on affective understanding of multimedia rely on one modality, even when additional modalities are available, for example in videos~\cite{jiang2014predicting}. Earlier work on emotional understanding of multimedia used hand crafted features from different modalities that are fused at feature or decision levels~\cite{hanjalic2005affective,Arifin2008,Benini:2011,soleymani2009,Teixeira2012}. The more recent work mainly use deep learning models~\cite{jiang2014predicting,Pang2015}. 

%here some work on text analysis and its features
%lexically-based LIWC
% bag-of-words
%owrd embedding
Language is a commonly used modality in addition to vision and audio. There is a large body of work on text-based sentiment analysis~\cite{pang2008opinion}. Sentiment analysis from text is well-established and is deployed at scale in industry at a broad set of applications involving opinion mining~\cite{soleymani2017survey}. With the shift toward an increasingly multimodal social web, multimodal sentiment analysis is becoming more relevant. For example, vloggers post their opinions on YouTube, and photos commonly accompany user posts on Instagram and Twitter. 
Analyzing text for emotion recognition requires representing terms by features. Lexically-based approaches are one of the most popular methods for text-based emotion recognition. They involve using knowledge of words' affect for estimating document or content's affect. Linguistic Inquiry and Word Count (LIWC) is a well-known lexical tool that matches the terms in a document with its dictionary and generates  scores along different dimensions including affective and cognitive constructs such as ``present focus'' and ``positive emotion''~\cite{liwc2007}. The terms in each category or selected by experts is extensively validated on different content. AffectNet is another notable lexical resource which includes a semantic netowrk of 10,000 items with representations  for ``pleasantness'', ``attention'', ``sensitivity'', and ``aptitude''~\cite{affectnet}. The continuous representations can be mapped to 24 distinct emotions. DepecheMood is a lexicon created through a data-driven method mining a news website annotated with its particular set of discrete emotions, namely, ``afraid'', ``amusemed'', ``anger'', ``annoyed'', ``don't care'', ``happy'', and ``inspired''~\cite{staiano2014depeche}. DepecheMood is extended to DepecheMood++ by including Italian~\cite{depechemoodpp}.

The more recent development in text-based affective analysis is models powered by deep learning. Leveraging large scale data, deep neural networks are able to learn representations that are relevant for affective analysis in language. Word embeddings are one of the most common representations used to represent language. Word embeddings, such as Word2Vec~\cite{word2vec} or GloVe~\cite{pennington2014glove}, learn language context of the word by learning a representation (a vector), that can capture semantic and syntactic similarities. More recently, representation learning models that can encode the whole sequence of terms (sentences, documents) showed impressive performance in different language understanding tasks, including sentiment and emotional analysis. 
Bidirectional Encoder Representations from Transformers (BERT)~\cite{devlin-etal-2019-bert} is a method for learning a language model that can be trained on large amount of data in an unsupervised manner. This pre-trained model is very effective in representing a sequence of terms as a fixed-length representation (vector). BERT architecture is a multi-layer bidirectional Transformer network that encodes the whole sequence at once. BERT representation achieves state-of-the-art results in multiple natural language understanding tasks. 

The audiovisual features that are used for multimodal understanding of affect are similar to the ones discussed in previous sections. The main technique between miltimodal models lies in methods for multimodal fusion. Multimodal methods involve extracting features from multiple modalities, \textit{e.g.}, audiovisual, and training joint or separate machine learning models for fusion~\cite{tadas_mm_survey}. Multimodal fusion can be done in model-based and model-agnostic ways. The model-agnostic fusion methods do not rely on a specific classification or regression method and include feature-level, decision-level, or hybrid fusion techniques.  Model-based methods address multimodal fusion in model construction. Examples of model-based fusion methods include Multiple Kernel Learning (MKL)~\cite{Gonen2011}, graphical models, such as Conditional Random Fields~\cite{Baltrusaitis2013} and neural networks~\cite{Rajagopalan2016,Nicolaou2011}.

\citeauthor{Pang2015}~\cite{Pang2015} used Deep Boltzmann Machine (DBM) to learn a joint representation across text, vision, and audio to recognize expected emotions from social media videos. Each modality is separately encoded with stacking multiple Restricted Boltzmann Machines (RBM) and pathways are merged to a joint representation layer. The model was evaluated for recognizing eight emotion categories for 1,101 videos from~\cite{jiang2014predicting}. \citeauthor{muszynski2019recognizing}~\cite{muszynski2019recognizing} studied perceived vs induced emotion in movies. To this end, they collected additional labels on a subset of LIRIS-ACCEDE dataset~\cite{baveye2015liris}. They found that perceived and induced emotions do not always agree. Using multimodal Deep Belief Networks (DBN), they could demonstrate that fusion of  electrodermal responses with audiovisual content features improves the overall accuracy for emotion recognition~\cite{muszynski2019recognizing}.

In~\cite{sivaprasad2018multimodal}, authors performed regression to estimate intended arousal and valence levels (as judged by experts in~\cite{5946961}). LSTM recurrent neural networks are used for unimodal regressions and fused via early and late fusion for audiovisual estimation with late fusion achieving the best results. \citeauthor{Tarvainen2018}~\cite{Tarvainen2018} performed an in-depth analysis on how emotions are constructed in movies. They identified scene type as a major factor in emotions in movies. They then used content features to recognize emotions along three dimensions of hedonic tone (valence), energetic arousal (awake--tired) and tense arousal (tense--calm).

Bilinear fusion is a method that is proposed to model inter- and intra- modality interaction among modalities by performing outer product between unimodal embeddings~\cite{Lin_2015_ICCV}. \citeauthor{zadeh2017tensor}~\cite{zadeh2017tensor} extended this to a Tensor Fusion Network to model intra-modality and inter-modality dynamics in multimodal sentiment analysis. The tensor fusion network includes modality embedding sub-networks, a tensor fusion layer modeling the unimodal, bimodal and trimodal interactions using a three-fold Cartesian product from modality embeddings along with a final sentiment inference sub-network conditioned on the tensor fusion layer. The main drawback of such methods is the increase in the dimensionality of the resulting multimodal representation.

\section{Future Directions}
\label{sec:FutureDirections}
Although remarkable progress has been made on affective computing of multimedia (ACM) data, there are still several open issues and directions that can boost the performance of ACM.

\textbf{Multimedia Content Understanding.} As emotions may be directly evoked by the multimedia content in viewers, accurately understanding what is contained in multimedia data can significantly improve the performance of ACM. Sometimes it is even necessary to analyze the subtle details. For example, we may feel ``amused'' on a video with a laughing baby; but if the laugh is from a negative character, it is more possible for us to feel ``angry''. In such cases, besides the common property, such as ``laugh'', we may need to further recognize the identity, such as ``a lovely baby'' and ``an evil antagonist''.

\textbf{Multimedia Summarization.} Emotions can play a vital role in selection of multimedia for creation of summaries or highlights. This is an important application in entertainment and sports industries (\textit{e.g.} movie trailers, sports highlights). There has been some recent work in this direction where affect information from audio visual cues has led to the successful creation of video summaries~\cite{Smith:augmenting, merler:highlights}. In particular, work reported in~\cite{Smith:augmenting} used audiovisual emotions in part to create an AI trailer for a $20^{th}$ Century Fox film in 2016. Similarly, AI Highlights described in~\cite{merler:highlights} hinged on audiovisual emotional cues and have successfully been employed to create the official highlights at Wimbledon and US Open since 2017. This is a very promising direction for affective multimedia computing which can have a direct impact on real world media applications.

\textbf{Contextual Knowledge Modeling.} The contextual information of a viewer watching some multimedia is very important. Similar multimedia data under different contexts may evoke totally different emotions. For example, we may feel ``happy" when listening a song about love in a wedding; but if the same song is played when two lovers are departing, it is more likely that we feel ``sad". The prior knowledge of viewers or multimedia data may also influence the emotion perceptions. An optimistic viewer and a pessimistic viewer may have totally different emotions about the same multimedia data.

\textbf{Group Emotion Clustering.} It is too generic to simply recognize the dominant emotion, while it is too specific to predict personalized emotion. It would make more sense to model emotions for groups or cliques of viewers with similar interests and backgrounds. Clustering different viewers into corresponding groups possibly based on the user profiles may provide a feasible solution to this problem.

\textbf{New AC Setting Adaptation.} Because of the domain shift~\cite{torralba2011unbiased}, the deep learning models trained on one labeled source domain may not work well on the other unlabeled or sparsely labeled target domain, which results in the models' low transferability to new domains. Exploring domain adaptation techniques that fit well on the AC tasks is worth investigating. One possible solution is to translate the source data to an intermediate domain that are indistinguishable from the target data while preserving the source labels~\cite{zhao2018emotiongan,zhao2019cycleemotiongan} using Generative Adversarial Networks~\cite{goodfellow2014generative,zhu2017unpaired}. How do deal with some practical settings, such as multiple labeled source domains and emotion models' homogeneity, is more challenging.

\textbf{Regions-of-Interest Selection.} The contributions of different regions of given multimedia may vary to the emotion recognition. For example, the regions that  contain the most important semantic information in images are more discriminative than background; some video frames are of no use to emotion recognition. Detecting and selecting the regions-of-interest may significantly improve the recognition performance as well as the computation efficiency.

\textbf{Viewer-Multimedia Interaction.}
Instead of direct analysis of the multimedia content or implicit consideration of viewers' physiological signals (such as facial expressions, Electroencephalogram signals, \textit{etc}.), joint modeling of both multimedia content and viewers' responses may better bridge the affective gap and result in superior performances. We should also study how to deal with missing or corrupted data. For example, some physiological signals are unavailable during the data collection stage.

\textbf{Affective Computing Applications.} Although AC is claimed to be important in real-world applications, few practical systems have been developed due to the relatively low performance. With the availability of larger datasets and improvements in self-supervised and semi-supervised learning, we foresee the deployment of ACM in real-world applications. For example, in media analytics, the content understanding methods will identify the emotional preferences of users and emotional nuances of social media content to better target advertising effort; in fashion recommendation, intelligent costumer service, such as customer-multimedia interaction, can provide better experience to customers; in advertisement, generating or curating multimedia that evokes strong emotions can attract more attention. We believe that an emotional artificial intelligence will become a significant component of mainstream multimedia applications.

\textbf{Benchmark Dataset Construction.}
Existing studies on ACM mainly adopt small-scale datasets or construct relatively larger-scale ones using keyword searching strategy without annotation quality guaranteed. To advance the development of ACM, creating a large-scale and high-quality dataset is in urgent need. It has shown that there are three critical factors for dataset construction of ACM, \textit{i.e.}, the context of viewer response, personal variation among viewers, and the effectiveness and efficiency of corpus creation~\cite{soleymani2014corpus}. In order to include a large number of samples, we may exploit online systems and crowdsourcing platforms to recruit large numbers of viewers with a representative spread of backgrounds to annotate multimedia and provide contextual information on their emotional responses. Since emotion is a subjective variable, personalized emotion annotation would make more sense, from which we can obtain the dominant emotion and emotion distribution. Further, accurate understanding of multimedia content can boost the affective computing performance. Inferring emotional labels from social media users' interaction with data, \textit{e.g.}, likes, comments, in addition to their spontaneous responses, \textit{e.g.}, facial expression, where possible, will provide new avenues for enriching affective datasets.

\section{Conclusion}
\label{sec:Conclusion}
In this article, we have surveyed affective computing (AC) methods for heterogeneous multimedia data. For each multimedia type, \textit{i.e.}, image, music, video, and multimodal data, we summarized and compared available datasets, handcrafted features, machine learning methods, deep learning models, and experimental results. We also briefly introduced the commonly employed emotion modelds and outlined potential research directions in this area. Although deep learning-based AC methods have achieved remarkable progress in recent years, an efficient and robust AC method that is able to obtain high accuracy under unconstrained conditions is yet to be achieved. With the advent of deep understanding of emotion evocation in brain science, accurate emotion measurement in psychology, and novel deep learning network architectures in machine learning, affective computing of multimedia data will remain an
active research topic for a long time.

%
% The acknowledgments section is defined using the "acks" environment (and NOT an unnumbered section). This ensures
% the proper identification of the section in the article metadata, and the consistent spelling of the heading.
\begin{acks}
This work was supported by Berkeley DeepDrive, the National Natural Science Foundation of China (Nos. 61701273, 91748129), and the National Key R\&D Program of China (Grant No. 2017YFC011300). The work of MS is supported in part by the U.S. Army. Any opinion, content or information presented does not necessarily reflect the position or the policy of the United States Government, and no official endorsement should be inferred.
\end{acks}

%
% The next two lines define the bibliography style to be used, and the bibliography file.
\bibliographystyle{ACM-Reference-Format}
\bibliography{references}


\begin{thebibliography}{178}

%%% the Web address.

\ifx \showCODEN    \undefined \def \showCODEN     #1{\unskip}     \fi
\ifx \showDOI      \undefined \def \showDOI       #1{#1}\fi
\ifx \showISBNx    \undefined \def \showISBNx     #1{\unskip}     \fi
\ifx \showISBNxiii \undefined \def \showISBNxiii  #1{\unskip}     \fi
\ifx \showISSN     \undefined \def \showISSN      #1{\unskip}     \fi
\ifx \showLCCN     \undefined \def \showLCCN      #1{\unskip}     \fi
\ifx \shownote     \undefined \def \shownote      #1{#1}          \fi
\ifx \showarticletitle \undefined \def \showarticletitle #1{#1}   \fi
\ifx \showURL      \undefined \def \showURL       {\relax}        \fi
% invisible to TeX
\providecommand\bibfield[2]{#2}
\providecommand\bibinfo[2]{#2}
\providecommand\natexlab[1]{#1}
\providecommand\showeprint[2][]{arXiv:#2}

\bibitem[\protect\citeauthoryear{Abadi, Subramanian, Kia, Avesani, Patras, and
  Sebe}{Abadi et~al\mbox{.}}{2015}]%
        {abadi2015decaf}
\bibfield{author}{\bibinfo{person}{Mojtaba~Khomami Abadi},
  \bibinfo{person}{Ramanathan Subramanian}, \bibinfo{person}{Seyed~Mostafa
  Kia}, \bibinfo{person}{Paolo Avesani}, \bibinfo{person}{Ioannis Patras},
  {and} \bibinfo{person}{Nicu Sebe}.} \bibinfo{year}{2015}\natexlab{}.
\newblock \showarticletitle{DECAF: MEG-based multimodal database for decoding
  affective physiological responses}.
\newblock \bibinfo{journal}{\emph{IEEE Transactions on Affective Computing}}
  \bibinfo{volume}{6}, \bibinfo{number}{3} (\bibinfo{year}{2015}),
  \bibinfo{pages}{209--222}.
\newblock


\bibitem[\protect\citeauthoryear{Acar, Hopfgartner, and Albayrak}{Acar
  et~al\mbox{.}}{2017}]%
        {acar2017comprehensive}
\bibfield{author}{\bibinfo{person}{Esra Acar}, \bibinfo{person}{Frank
  Hopfgartner}, {and} \bibinfo{person}{Sahin Albayrak}.}
  \bibinfo{year}{2017}\natexlab{}.
\newblock \showarticletitle{A comprehensive study on mid-level representation
  and ensemble learning for emotional analysis of video material}.
\newblock \bibinfo{journal}{\emph{Multimedia Tools and Applications}}
  \bibinfo{volume}{76}, \bibinfo{number}{9} (\bibinfo{year}{2017}),
  \bibinfo{pages}{11809--11837}.
\newblock


\bibitem[\protect\citeauthoryear{Alameda-Pineda, Ricci, Yan, and
  Sebe}{Alameda-Pineda et~al\mbox{.}}{2016}]%
        {alameda2016recognizing}
\bibfield{author}{\bibinfo{person}{Xavier Alameda-Pineda},
  \bibinfo{person}{Elisa Ricci}, \bibinfo{person}{Yan Yan}, {and}
  \bibinfo{person}{Nicu Sebe}.} \bibinfo{year}{2016}\natexlab{}.
\newblock \showarticletitle{Recognizing emotions from abstract paintings using
  non-linear matrix completion}. In \bibinfo{booktitle}{\emph{IEEE Conference
  on Computer Vision and Pattern Recognition}}. \bibinfo{pages}{5240--5248}.
\newblock


\bibitem[\protect\citeauthoryear{Aljanaki}{Aljanaki}{2016}]%
        {aljanaki2016emotion}
\bibfield{author}{\bibinfo{person}{Anna Aljanaki}.}
  \bibinfo{year}{2016}\natexlab{}.
\newblock \emph{\bibinfo{title}{Emotion in Music: representation and
  computational modeling}}.
\newblock \bibinfo{thesistype}{Ph.D. Dissertation}. \bibinfo{school}{Utrecht
  University}.
\newblock


\bibitem[\protect\citeauthoryear{Aljanaki and Soleymani}{Aljanaki and
  Soleymani}{2018}]%
        {aljanaki2018data}
\bibfield{author}{\bibinfo{person}{Anna Aljanaki} {and}
  \bibinfo{person}{Mohammad Soleymani}.} \bibinfo{year}{2018}\natexlab{}.
\newblock \showarticletitle{A data-driven approach to mid-level perceptual
  musical feature modeling}. In \bibinfo{booktitle}{\emph{International Society
  for Music Information Retrieval Conference}}.
\newblock


\bibitem[\protect\citeauthoryear{Aljanaki, Yang, and Soleymani}{Aljanaki
  et~al\mbox{.}}{2017}]%
        {deam}
\bibfield{author}{\bibinfo{person}{Anna Aljanaki}, \bibinfo{person}{Yi-Hsuan
  Yang}, {and} \bibinfo{person}{Mohammad Soleymani}.}
  \bibinfo{year}{2017}\natexlab{}.
\newblock \showarticletitle{Developing a benchmark for emotional analysis of
  music}.
\newblock \bibinfo{journal}{\emph{PloS One}} \bibinfo{volume}{12},
  \bibinfo{number}{3} (\bibinfo{year}{2017}), \bibinfo{pages}{e0173392}.
\newblock


\bibitem[\protect\citeauthoryear{Araque, Gatti, Staiano, and Guerini}{Araque
  et~al\mbox{.}}{2018}]%
        {depechemoodpp}
\bibfield{author}{\bibinfo{person}{Oscar Araque}, \bibinfo{person}{Lorenzo
  Gatti}, \bibinfo{person}{Jacopo Staiano}, {and} \bibinfo{person}{Marco
  Guerini}.} \bibinfo{year}{2018}\natexlab{}.
\newblock \showarticletitle{DepecheMood++: a Bilingual Emotion Lexicon Built
  Through Simple Yet Powerful Techniques}.
\newblock \bibinfo{journal}{\emph{arXiv preprint arXiv:1810.03660}}
  (\bibinfo{year}{2018}).
\newblock


\bibitem[\protect\citeauthoryear{Arifin and Cheung}{Arifin and Cheung}{2008}]%
        {Arifin2008}
\bibfield{author}{\bibinfo{person}{Sutjipto Arifin} {and}
  \bibinfo{person}{Peter~YK Cheung}.} \bibinfo{year}{2008}\natexlab{}.
\newblock \showarticletitle{Affective level video segmentation by utilizing the
  pleasure-arousal-dominance information}.
\newblock \bibinfo{journal}{\emph{IEEE Transactions on Multimedia}}
  \bibinfo{volume}{10}, \bibinfo{number}{7} (\bibinfo{year}{2008}),
  \bibinfo{pages}{1325--1341}.
\newblock


\bibitem[\protect\citeauthoryear{Baltru{\v{s}}aitis, Ahuja, and
  Morency}{Baltru{\v{s}}aitis et~al\mbox{.}}{2019}]%
        {tadas_mm_survey}
\bibfield{author}{\bibinfo{person}{Tadas Baltru{\v{s}}aitis},
  \bibinfo{person}{Chaitanya Ahuja}, {and} \bibinfo{person}{Louis-Philippe
  Morency}.} \bibinfo{year}{2019}\natexlab{}.
\newblock \showarticletitle{Multimodal Machine Learning: A Survey and
  Taxonomy}.
\newblock \bibinfo{journal}{\emph{IEEE Transactions on Pattern Analysis and
  Machine Intelligence}} \bibinfo{volume}{41}, \bibinfo{number}{2}
  (\bibinfo{year}{2019}), \bibinfo{pages}{423--443}.
\newblock


\bibitem[\protect\citeauthoryear{Baltrusaitis, Banda, and
  Robinson}{Baltrusaitis et~al\mbox{.}}{2013}]%
        {Baltrusaitis2013}
\bibfield{author}{\bibinfo{person}{Tadas Baltrusaitis},
  \bibinfo{person}{Ntombikayise Banda}, {and} \bibinfo{person}{Peter
  Robinson}.} \bibinfo{year}{2013}\natexlab{}.
\newblock \showarticletitle{Dimensional affect recognition using Continuous
  Conditional Random Fields}. In \bibinfo{booktitle}{\emph{IEEE International
  Conference and Workshops on Automatic Face and Gesture Recognition}}.
  \bibinfo{pages}{1--8}.
\newblock


\bibitem[\protect\citeauthoryear{Baveye, Chamaret, Dellandr{\'e}a, and
  Chen}{Baveye et~al\mbox{.}}{2018}]%
        {Baveye2018}
\bibfield{author}{\bibinfo{person}{Yoann Baveye}, \bibinfo{person}{Christel
  Chamaret}, \bibinfo{person}{Emmanuel Dellandr{\'e}a}, {and}
  \bibinfo{person}{Liming Chen}.} \bibinfo{year}{2018}\natexlab{}.
\newblock \showarticletitle{Affective video content analysis: A
  multidisciplinary insight}.
\newblock \bibinfo{journal}{\emph{IEEE Transactions on Affective Computing}}
  \bibinfo{volume}{9}, \bibinfo{number}{4} (\bibinfo{year}{2018}),
  \bibinfo{pages}{396--409}.
\newblock


\bibitem[\protect\citeauthoryear{Baveye, Dellandr{\'e}a, Chamaret, and
  Chen}{Baveye et~al\mbox{.}}{2015a}]%
        {baveye2015deep}
\bibfield{author}{\bibinfo{person}{Yoann Baveye}, \bibinfo{person}{Emmanuel
  Dellandr{\'e}a}, \bibinfo{person}{Christel Chamaret}, {and}
  \bibinfo{person}{Liming Chen}.} \bibinfo{year}{2015}\natexlab{a}.
\newblock \showarticletitle{Deep learning vs. kernel methods: Performance for
  emotion prediction in videos}. In \bibinfo{booktitle}{\emph{International
  Conference on Affective Computing and Intelligent Interaction}}.
  \bibinfo{pages}{77--83}.
\newblock


\bibitem[\protect\citeauthoryear{Baveye, Dellandrea, Chamaret, and Chen}{Baveye
  et~al\mbox{.}}{2015b}]%
        {baveye2015liris}
\bibfield{author}{\bibinfo{person}{Yoann Baveye}, \bibinfo{person}{Emmanuel
  Dellandrea}, \bibinfo{person}{Christel Chamaret}, {and}
  \bibinfo{person}{Liming Chen}.} \bibinfo{year}{2015}\natexlab{b}.
\newblock \showarticletitle{Liris-accede: A video database for affective
  content analysis}.
\newblock \bibinfo{journal}{\emph{IEEE Transactions on Affective Computing}}
  \bibinfo{volume}{6}, \bibinfo{number}{1} (\bibinfo{year}{2015}),
  \bibinfo{pages}{43--55}.
\newblock


\bibitem[\protect\citeauthoryear{Ben-Ahmed and Huet}{Ben-Ahmed and
  Huet}{2018}]%
        {ben2018deep}
\bibfield{author}{\bibinfo{person}{Olfa Ben-Ahmed} {and}
  \bibinfo{person}{Benoit Huet}.} \bibinfo{year}{2018}\natexlab{}.
\newblock \showarticletitle{Deep Multimodal Features for Movie Genre and
  Interestingness Prediction}. In \bibinfo{booktitle}{\emph{International
  Conference on Content-Based Multimedia Indexing}}. \bibinfo{pages}{1--6}.
\newblock


\bibitem[\protect\citeauthoryear{Benini, Canini, and Leonardi}{Benini
  et~al\mbox{.}}{2011}]%
        {Benini:2011}
\bibfield{author}{\bibinfo{person}{Sergio Benini}, \bibinfo{person}{Luca
  Canini}, {and} \bibinfo{person}{Riccardo Leonardi}.}
  \bibinfo{year}{2011}\natexlab{}.
\newblock \showarticletitle{A connotative space for supporting movie affective
  recommendation}.
\newblock \bibinfo{journal}{\emph{IEEE Transactions on Multimedia}}
  \bibinfo{volume}{13}, \bibinfo{number}{6} (\bibinfo{year}{2011}),
  \bibinfo{pages}{1356--1370}.
\newblock


\bibitem[\protect\citeauthoryear{Bogdanov, Wack, G{\'o}mez, Gulati, Herrera,
  Mayor, Roma, Salamon, Zapata, and Serra}{Bogdanov et~al\mbox{.}}{2013}]%
        {ESSENTIA}
\bibfield{author}{\bibinfo{person}{Dmitry Bogdanov}, \bibinfo{person}{Nicolas
  Wack}, \bibinfo{person}{Emilia G{\'o}mez}, \bibinfo{person}{Sankalp Gulati},
  \bibinfo{person}{Perfecto Herrera}, \bibinfo{person}{O. Mayor},
  \bibinfo{person}{Gerard Roma}, \bibinfo{person}{Justin Salamon},
  \bibinfo{person}{J.~R. Zapata}, {and} \bibinfo{person}{Xavier Serra}.}
  \bibinfo{year}{2013}\natexlab{}.
\newblock \showarticletitle{{ESSENTIA}: an Audio Analysis Library for Music
  Information Retrieval}. In \bibinfo{booktitle}{\emph{International Society
  for Music Information Retrieval Conference}}. \bibinfo{pages}{493--498}.
\newblock


\bibitem[\protect\citeauthoryear{Borth, Chen, Ji, and Chang}{Borth
  et~al\mbox{.}}{2013a}]%
        {borth2013sentibank}
\bibfield{author}{\bibinfo{person}{Damian Borth}, \bibinfo{person}{Tao Chen},
  \bibinfo{person}{Rongrong Ji}, {and} \bibinfo{person}{Shih-Fu Chang}.}
  \bibinfo{year}{2013}\natexlab{a}.
\newblock \showarticletitle{Sentibank: large-scale ontology and classifiers for
  detecting sentiment and emotions in visual content}. In
  \bibinfo{booktitle}{\emph{ACM International Conference on Multimedia}}.
  \bibinfo{pages}{459--460}.
\newblock


\bibitem[\protect\citeauthoryear{Borth, Ji, Chen, Breuel, and Chang}{Borth
  et~al\mbox{.}}{2013b}]%
        {borth2013large}
\bibfield{author}{\bibinfo{person}{Damian Borth}, \bibinfo{person}{Rongrong
  Ji}, \bibinfo{person}{Tao Chen}, \bibinfo{person}{Thomas Breuel}, {and}
  \bibinfo{person}{Shih-Fu Chang}.} \bibinfo{year}{2013}\natexlab{b}.
\newblock \showarticletitle{Large-scale visual sentiment ontology and detectors
  using adjective noun pairs}. In \bibinfo{booktitle}{\emph{ACM International
  Conference on Multimedia}}. \bibinfo{pages}{223--232}.
\newblock


\bibitem[\protect\citeauthoryear{Cambria, Mazzocco, Hussain, and Eckl}{Cambria
  et~al\mbox{.}}{2011}]%
        {affectnet}
\bibfield{author}{\bibinfo{person}{Erik Cambria}, \bibinfo{person}{Thomas
  Mazzocco}, \bibinfo{person}{Amir Hussain}, {and} \bibinfo{person}{Chris
  Eckl}.} \bibinfo{year}{2011}\natexlab{}.
\newblock \showarticletitle{Sentic Medoids: Organizing Affective Common Sense
  Knowledge in a Multi-Dimensional Vector Space}. In
  \bibinfo{booktitle}{\emph{Advances in Neural Networks}}.
  \bibinfo{pages}{601--610}.
\newblock


\bibitem[\protect\citeauthoryear{Chen, Cheng, and Guo}{Chen
  et~al\mbox{.}}{2018}]%
        {chen2018identifying}
\bibfield{author}{\bibinfo{person}{Mo Chen}, \bibinfo{person}{Gong Cheng},
  {and} \bibinfo{person}{Lei Guo}.} \bibinfo{year}{2018}\natexlab{}.
\newblock \showarticletitle{Identifying affective levels on music video via
  completing the missing modality}.
\newblock \bibinfo{journal}{\emph{Multimedia Tools and Applications}}
  \bibinfo{volume}{77}, \bibinfo{number}{3} (\bibinfo{year}{2018}),
  \bibinfo{pages}{3287--3302}.
\newblock


\bibitem[\protect\citeauthoryear{Chen, Borth, Darrell, and Chang}{Chen
  et~al\mbox{.}}{2014a}]%
        {Chen2014DeepSentiBank}
\bibfield{author}{\bibinfo{person}{Tao Chen}, \bibinfo{person}{Damian Borth},
  \bibinfo{person}{Trevor Darrell}, {and} \bibinfo{person}{Shih-Fu Chang}.}
  \bibinfo{year}{2014}\natexlab{a}.
\newblock \showarticletitle{DeepSentiBank: Visual Sentiment Concept
  Classification with Deep Convolutional Neural Networks}.
\newblock \bibinfo{journal}{\emph{Computer Science}} (\bibinfo{year}{2014}).
\newblock


\bibitem[\protect\citeauthoryear{Chen, Yu, Chen, Cui, Chen, and Chang}{Chen
  et~al\mbox{.}}{2014b}]%
        {chen2014object}
\bibfield{author}{\bibinfo{person}{Tao Chen}, \bibinfo{person}{Felix~X Yu},
  \bibinfo{person}{Jiawei Chen}, \bibinfo{person}{Yin Cui},
  \bibinfo{person}{Yan-Ying Chen}, {and} \bibinfo{person}{Shih-Fu Chang}.}
  \bibinfo{year}{2014}\natexlab{b}.
\newblock \showarticletitle{Object-based visual sentiment concept analysis and
  application}. In \bibinfo{booktitle}{\emph{ACM International Conference on
  Multimedia}}. \bibinfo{pages}{367--376}.
\newblock


\bibitem[\protect\citeauthoryear{Chen, Yang, Wang, and Chen}{Chen
  et~al\mbox{.}}{2015}]%
        {chen15icassp}
\bibfield{author}{\bibinfo{person}{Yu-An Chen}, \bibinfo{person}{Yi-Hsuan
  Yang}, \bibinfo{person}{Ju-Chiang Wang}, {and} \bibinfo{person}{Homer Chen}.}
  \bibinfo{year}{2015}\natexlab{}.
\newblock \showarticletitle{The AMG1608 dataset for music emotion recognition}.
  In \bibinfo{booktitle}{\emph{IEEE International Conference on Acoustics,
  Speech and Signal Processing}}. \bibinfo{pages}{693--697}.
\newblock


\bibitem[\protect\citeauthoryear{Correa, Abadi, Sebe, and Patras}{Correa
  et~al\mbox{.}}{2017}]%
        {DBLP:journals/corr/CorreaASP17}
\bibfield{author}{\bibinfo{person}{Juan Abdon~Miranda Correa},
  \bibinfo{person}{Mojtaba~Khomami Abadi}, \bibinfo{person}{Nicu Sebe}, {and}
  \bibinfo{person}{Ioannis Patras}.} \bibinfo{year}{2017}\natexlab{}.
\newblock \showarticletitle{{AMIGOS:} {A} dataset for Mood, personality and
  affect research on Individuals and GrOupS}.
\newblock \bibinfo{journal}{\emph{arXiv preprint arXiv:1702.02510}}
  (\bibinfo{year}{2017}).
\newblock


\bibitem[\protect\citeauthoryear{Dalmia, Liu, and Chang}{Dalmia
  et~al\mbox{.}}{2016}]%
        {dalmia2016columbia}
\bibfield{author}{\bibinfo{person}{Vaidehi Dalmia}, \bibinfo{person}{Hongyi
  Liu}, {and} \bibinfo{person}{Shih-Fu Chang}.}
  \bibinfo{year}{2016}\natexlab{}.
\newblock \showarticletitle{Columbia mvso image sentiment dataset}.
\newblock \bibinfo{journal}{\emph{arXiv preprint arXiv:1611.04455}}
  (\bibinfo{year}{2016}).
\newblock


\bibitem[\protect\citeauthoryear{Dan-Glauser and Scherer}{Dan-Glauser and
  Scherer}{2011}]%
        {dan2011geneva}
\bibfield{author}{\bibinfo{person}{Elise~S Dan-Glauser} {and}
  \bibinfo{person}{Klaus~R Scherer}.} \bibinfo{year}{2011}\natexlab{}.
\newblock \showarticletitle{The Geneva affective picture database (GAPED): a
  new 730-picture database focusing on valence and normative significance}.
\newblock \bibinfo{journal}{\emph{Behavior Research Methods}}
  \bibinfo{volume}{43}, \bibinfo{number}{2} (\bibinfo{year}{2011}),
  \bibinfo{pages}{468--477}.
\newblock


\bibitem[\protect\citeauthoryear{Dellandr{\'e}a, Chen, Baveye, Sj{\"o}berg,
  Chamaret, et~al\mbox{.}}{Dellandr{\'e}a et~al\mbox{.}}{2016}]%
        {dellandrea2016mediaeval}
\bibfield{author}{\bibinfo{person}{Emmanuel Dellandr{\'e}a},
  \bibinfo{person}{Liming Chen}, \bibinfo{person}{Yoann Baveye},
  \bibinfo{person}{Mats~Viktor Sj{\"o}berg}, \bibinfo{person}{Christel
  Chamaret}, {et~al\mbox{.}}} \bibinfo{year}{2016}\natexlab{}.
\newblock \showarticletitle{The mediaeval 2016 emotional impact of movies
  task}. In \bibinfo{booktitle}{\emph{CEUR Workshop Proceedings}}.
\newblock


\bibitem[\protect\citeauthoryear{Dellandr{\'{e}}a, Huigsloot, Chen, Baveye, and
  Sj{\"{o}}berg}{Dellandr{\'{e}}a et~al\mbox{.}}{2017}]%
        {DBLP:conf/mediaeval/DellandreaH0BS17}
\bibfield{author}{\bibinfo{person}{Emmanuel Dellandr{\'{e}}a},
  \bibinfo{person}{Martijn Huigsloot}, \bibinfo{person}{Liming Chen},
  \bibinfo{person}{Yoann Baveye}, {and} \bibinfo{person}{Mats Sj{\"{o}}berg}.}
  \bibinfo{year}{2017}\natexlab{}.
\newblock \showarticletitle{The MediaEval 2017 Emotional Impact of Movies
  Task}. In \bibinfo{booktitle}{\emph{Working Notes Proceedings of the
  MediaEval 2017 Workshop}}.
\newblock


\bibitem[\protect\citeauthoryear{Dellandr{\'{e}}a, Huigsloot, Chen, Baveye,
  Xiao, and Sj{\"{o}}berg}{Dellandr{\'{e}}a et~al\mbox{.}}{2018}]%
        {DBLP:conf/mediaeval/DellandreaH0BXS18}
\bibfield{author}{\bibinfo{person}{Emmanuel Dellandr{\'{e}}a},
  \bibinfo{person}{Martijn Huigsloot}, \bibinfo{person}{Liming Chen},
  \bibinfo{person}{Yoann Baveye}, \bibinfo{person}{Zhongzhe Xiao}, {and}
  \bibinfo{person}{Mats Sj{\"{o}}berg}.} \bibinfo{year}{2018}\natexlab{}.
\newblock \showarticletitle{The MediaEval 2018 Emotional Impact of Movies
  Task}. In \bibinfo{booktitle}{\emph{Working Notes Proceedings of the
  MediaEval 2018 Workshop}}.
\newblock


\bibitem[\protect\citeauthoryear{Dellandr{\'e}a, Huigsloot, Chen, Baveye, Xiao,
  and Sj{\"o}berg}{Dellandr{\'e}a et~al\mbox{.}}{2019}]%
        {dellandrea2019datasets}
\bibfield{author}{\bibinfo{person}{Emmanuel Dellandr{\'e}a},
  \bibinfo{person}{Martijn Huigsloot}, \bibinfo{person}{Liming Chen},
  \bibinfo{person}{Yoann Baveye}, \bibinfo{person}{Zhongzhe Xiao}, {and}
  \bibinfo{person}{Mats Sj{\"o}berg}.} \bibinfo{year}{2019}\natexlab{}.
\newblock \showarticletitle{Datasets column: predicting the emotional impact of
  movies}.
\newblock \bibinfo{journal}{\emph{ACM SIGMultimedia Records}}
  \bibinfo{volume}{10}, \bibinfo{number}{4} (\bibinfo{year}{2019}),
  \bibinfo{pages}{6}.
\newblock


\bibitem[\protect\citeauthoryear{Devlin, Chang, Lee, and Toutanova}{Devlin
  et~al\mbox{.}}{2019}]%
        {devlin-etal-2019-bert}
\bibfield{author}{\bibinfo{person}{Jacob Devlin}, \bibinfo{person}{Ming-Wei
  Chang}, \bibinfo{person}{Kenton Lee}, {and} \bibinfo{person}{Kristina
  Toutanova}.} \bibinfo{year}{2019}\natexlab{}.
\newblock \showarticletitle{{BERT}: Pre-training of Deep Bidirectional
  Transformers for Language Understanding}. In \bibinfo{booktitle}{\emph{Annual
  Conference of the North American Chapter of the Association for Computational
  Linguistics}}. \bibinfo{pages}{4171--4186}.
\newblock


\bibitem[\protect\citeauthoryear{Downie, Laurier, and Ehmann}{Downie
  et~al\mbox{.}}{2008}]%
        {mirex07}
\bibfield{author}{\bibinfo{person}{XHJS Downie}, \bibinfo{person}{Cyril
  Laurier}, {and} \bibinfo{person}{MBAF Ehmann}.}
  \bibinfo{year}{2008}\natexlab{}.
\newblock \showarticletitle{The 2007 MIREX audio mood classification task:
  Lessons learned}. In \bibinfo{booktitle}{\emph{International Society for
  Music Information Retrieval Conference}}. \bibinfo{pages}{462--467}.
\newblock


\bibitem[\protect\citeauthoryear{Eerola and Vuoskoski}{Eerola and
  Vuoskoski}{2011}]%
        {Eerola2011}
\bibfield{author}{\bibinfo{person}{Tuomas Eerola} {and}
  \bibinfo{person}{Jonna~K Vuoskoski}.} \bibinfo{year}{2011}\natexlab{}.
\newblock \showarticletitle{A comparison of the discrete and dimensional models
  of emotion in music}.
\newblock \bibinfo{journal}{\emph{Psychology of Music}} \bibinfo{volume}{39},
  \bibinfo{number}{1} (\bibinfo{year}{2011}), \bibinfo{pages}{18--49}.
\newblock


\bibitem[\protect\citeauthoryear{Ekman}{Ekman}{1992}]%
        {ekman1992argument}
\bibfield{author}{\bibinfo{person}{Paul Ekman}.}
  \bibinfo{year}{1992}\natexlab{}.
\newblock \showarticletitle{An argument for basic emotions}.
\newblock \bibinfo{journal}{\emph{Cognition \& Emotion}} \bibinfo{volume}{6},
  \bibinfo{number}{3-4} (\bibinfo{year}{1992}), \bibinfo{pages}{169--200}.
\newblock


\bibitem[\protect\citeauthoryear{El~Ayadi, Kamel, and Karray}{El~Ayadi
  et~al\mbox{.}}{2011}]%
        {el2011survey}
\bibfield{author}{\bibinfo{person}{Moataz El~Ayadi}, \bibinfo{person}{Mohamed~S
  Kamel}, {and} \bibinfo{person}{Fakhri Karray}.}
  \bibinfo{year}{2011}\natexlab{}.
\newblock \showarticletitle{Survey on speech emotion recognition: Features,
  classification schemes, and databases}.
\newblock \bibinfo{journal}{\emph{Pattern Recognition}} \bibinfo{volume}{44},
  \bibinfo{number}{3} (\bibinfo{year}{2011}), \bibinfo{pages}{572--587}.
\newblock


\bibitem[\protect\citeauthoryear{Eyben, Weninger, Gross, and Schuller}{Eyben
  et~al\mbox{.}}{2013}]%
        {eyben2013}
\bibfield{author}{\bibinfo{person}{Florian Eyben}, \bibinfo{person}{Felix
  Weninger}, \bibinfo{person}{Florian Gross}, {and}
  \bibinfo{person}{Bj{\"{o}}rn Schuller}.} \bibinfo{year}{2013}\natexlab{}.
\newblock \showarticletitle{Recent developments in openSMILE, the munich
  open-source multimedia feature extractor}. In \bibinfo{booktitle}{\emph{ACM
  International Conference on Multimedia}}. \bibinfo{pages}{835--838}.
\newblock
\showISBNx{9781450324045}


\bibitem[\protect\citeauthoryear{Eyben, W\"{o}llmer, and Schuller}{Eyben
  et~al\mbox{.}}{2010}]%
        {eyben2010}
\bibfield{author}{\bibinfo{person}{Florian Eyben}, \bibinfo{person}{Martin
  W\"{o}llmer}, {and} \bibinfo{person}{Bj\"{o}rn Schuller}.}
  \bibinfo{year}{2010}\natexlab{}.
\newblock \showarticletitle{OpenSMILE: The Munich Versatile and Fast
  Open-source Audio Feature Extractor}. In \bibinfo{booktitle}{\emph{ACM
  International Conference on Multimedia}}. \bibinfo{pages}{1459--1462}.
\newblock


\bibitem[\protect\citeauthoryear{G{\"{o}}nen and Alpaydın}{G{\"{o}}nen and
  Alpaydın}{2011}]%
        {Gonen2011}
\bibfield{author}{\bibinfo{person}{Mehmet G{\"{o}}nen} {and}
  \bibinfo{person}{Ethem Alpaydın}.} \bibinfo{year}{2011}\natexlab{}.
\newblock \showarticletitle{Multiple Kernel Learning Algorithms}.
\newblock \bibinfo{journal}{\emph{Journal of Machine Learning Research}}
  \bibinfo{volume}{12}, \bibinfo{number}{Jul} (\bibinfo{year}{2011}),
  \bibinfo{pages}{2211--2268}.
\newblock


\bibitem[\protect\citeauthoryear{Goodfellow, Pouget-Abadie, Mirza, Xu,
  Warde-Farley, Ozair, Courville, and Bengio}{Goodfellow et~al\mbox{.}}{2014}]%
        {goodfellow2014generative}
\bibfield{author}{\bibinfo{person}{Ian Goodfellow}, \bibinfo{person}{Jean
  Pouget-Abadie}, \bibinfo{person}{Mehdi Mirza}, \bibinfo{person}{Bing Xu},
  \bibinfo{person}{David Warde-Farley}, \bibinfo{person}{Sherjil Ozair},
  \bibinfo{person}{Aaron Courville}, {and} \bibinfo{person}{Yoshua Bengio}.}
  \bibinfo{year}{2014}\natexlab{}.
\newblock \showarticletitle{Generative adversarial nets}. In
  \bibinfo{booktitle}{\emph{Advances in Neural Information Processing
  Systems}}. \bibinfo{pages}{2672--2680}.
\newblock


\bibitem[\protect\citeauthoryear{Gui, Zhong, and Ming}{Gui
  et~al\mbox{.}}{2018}]%
        {gui2018implicit}
\bibfield{author}{\bibinfo{person}{Dongdong Gui}, \bibinfo{person}{Sheng-hua
  Zhong}, {and} \bibinfo{person}{Zhong Ming}.} \bibinfo{year}{2018}\natexlab{}.
\newblock \showarticletitle{Implicit Affective Video Tagging Using Pupillary
  Response}. In \bibinfo{booktitle}{\emph{International Conference on
  Multimedia Modeling}}. \bibinfo{pages}{165--176}.
\newblock


\bibitem[\protect\citeauthoryear{Guo, Song, Zhang, Ma, Luo, et~al\mbox{.}}{Guo
  et~al\mbox{.}}{2019}]%
        {guo2019affective}
\bibfield{author}{\bibinfo{person}{Jie Guo}, \bibinfo{person}{Bin Song},
  \bibinfo{person}{Peng Zhang}, \bibinfo{person}{Mengdi Ma},
  \bibinfo{person}{Wenwen Luo}, {et~al\mbox{.}}}
  \bibinfo{year}{2019}\natexlab{}.
\newblock \showarticletitle{Affective video content analysis based on
  multimodal data fusion in heterogeneous networks}.
\newblock \bibinfo{journal}{\emph{Information Fusion}}  \bibinfo{volume}{51}
  (\bibinfo{year}{2019}), \bibinfo{pages}{224--232}.
\newblock


\bibitem[\protect\citeauthoryear{Gupta, Khomami~Abadi, C{\'a}rdenes~Cabr{\'e},
  Morreale, Falk, and Sebe}{Gupta et~al\mbox{.}}{2016}]%
        {gupta2016quality}
\bibfield{author}{\bibinfo{person}{Rishabh Gupta}, \bibinfo{person}{Mojtaba
  Khomami~Abadi}, \bibinfo{person}{Jes{\'u}s~Alejandro C{\'a}rdenes~Cabr{\'e}},
  \bibinfo{person}{Fabio Morreale}, \bibinfo{person}{Tiago~H Falk}, {and}
  \bibinfo{person}{Nicu Sebe}.} \bibinfo{year}{2016}\natexlab{}.
\newblock \showarticletitle{A quality adaptive multimodal affect recognition
  system for user-centric multimedia indexing}. In
  \bibinfo{booktitle}{\emph{ACM International Conference on Multimedia
  Retrieval}}. \bibinfo{pages}{317--320}.
\newblock


\bibitem[\protect\citeauthoryear{Han, Ji, Hu, Guo, and Liu}{Han
  et~al\mbox{.}}{2015}]%
        {han2015arousal}
\bibfield{author}{\bibinfo{person}{Junwei Han}, \bibinfo{person}{Xiang Ji},
  \bibinfo{person}{Xintao Hu}, \bibinfo{person}{Lei Guo}, {and}
  \bibinfo{person}{Tianming Liu}.} \bibinfo{year}{2015}\natexlab{}.
\newblock \showarticletitle{Arousal recognition using audio-visual features and
  FMRI-based brain response}.
\newblock \bibinfo{journal}{\emph{IEEE Transactions on Affective Computing}}
  \bibinfo{volume}{6}, \bibinfo{number}{4} (\bibinfo{year}{2015}),
  \bibinfo{pages}{337--347}.
\newblock


\bibitem[\protect\citeauthoryear{Han, Zhang, Cheng, Liu, and Xu}{Han
  et~al\mbox{.}}{2018}]%
        {han2018advanced}
\bibfield{author}{\bibinfo{person}{Junwei Han}, \bibinfo{person}{Dingwen
  Zhang}, \bibinfo{person}{Gong Cheng}, \bibinfo{person}{Nian Liu}, {and}
  \bibinfo{person}{Dong Xu}.} \bibinfo{year}{2018}\natexlab{}.
\newblock \showarticletitle{Advanced deep-learning techniques for salient and
  category-specific object detection: a survey}.
\newblock \bibinfo{journal}{\emph{IEEE Signal Processing Magazine}}
  \bibinfo{volume}{35}, \bibinfo{number}{1} (\bibinfo{year}{2018}),
  \bibinfo{pages}{84--100}.
\newblock


\bibitem[\protect\citeauthoryear{Hanjalic}{Hanjalic}{2006}]%
        {hanjalic2006extracting}
\bibfield{author}{\bibinfo{person}{Alan Hanjalic}.}
  \bibinfo{year}{2006}\natexlab{}.
\newblock \showarticletitle{Extracting moods from pictures and sounds: Towards
  truly personalized TV}.
\newblock \bibinfo{journal}{\emph{IEEE Signal Processing Magazine}}
  \bibinfo{volume}{23}, \bibinfo{number}{2} (\bibinfo{year}{2006}),
  \bibinfo{pages}{90--100}.
\newblock


\bibitem[\protect\citeauthoryear{Hanjalic and Xu}{Hanjalic and Xu}{2005}]%
        {hanjalic2005affective}
\bibfield{author}{\bibinfo{person}{Alan Hanjalic} {and} \bibinfo{person}{Li-Qun
  Xu}.} \bibinfo{year}{2005}\natexlab{}.
\newblock \showarticletitle{Affective video content representation and
  modeling}.
\newblock \bibinfo{journal}{\emph{IEEE Transactions on Multimedia}}
  \bibinfo{volume}{7}, \bibinfo{number}{1} (\bibinfo{year}{2005}),
  \bibinfo{pages}{143--154}.
\newblock


\bibitem[\protect\citeauthoryear{Hansen and Hasan}{Hansen and Hasan}{2015}]%
        {hansen2015speaker}
\bibfield{author}{\bibinfo{person}{John~HL Hansen} {and}
  \bibinfo{person}{Taufiq Hasan}.} \bibinfo{year}{2015}\natexlab{}.
\newblock \showarticletitle{Speaker recognition by machines and humans: A
  tutorial review}.
\newblock \bibinfo{journal}{\emph{IEEE Signal Processing Magazine}}
  \bibinfo{volume}{32}, \bibinfo{number}{6} (\bibinfo{year}{2015}),
  \bibinfo{pages}{74--99}.
\newblock


\bibitem[\protect\citeauthoryear{Herath, Harandi, and Porikli}{Herath
  et~al\mbox{.}}{2017}]%
        {herath2017going}
\bibfield{author}{\bibinfo{person}{Samitha Herath}, \bibinfo{person}{Mehrtash
  Harandi}, {and} \bibinfo{person}{Fatih Porikli}.}
  \bibinfo{year}{2017}\natexlab{}.
\newblock \showarticletitle{Going deeper into action recognition: A survey}.
\newblock \bibinfo{journal}{\emph{Image and Vision Computing}}
  \bibinfo{volume}{60} (\bibinfo{year}{2017}), \bibinfo{pages}{4--21}.
\newblock


\bibitem[\protect\citeauthoryear{Hu, Ding, Li, Wang, Gao, Wang, and Maybank}{Hu
  et~al\mbox{.}}{2016}]%
        {hu2016multi}
\bibfield{author}{\bibinfo{person}{Weiming Hu}, \bibinfo{person}{Xinmiao Ding},
  \bibinfo{person}{Bing Li}, \bibinfo{person}{Jianchao Wang},
  \bibinfo{person}{Yan Gao}, \bibinfo{person}{Fangshi Wang}, {and}
  \bibinfo{person}{Stephen Maybank}.} \bibinfo{year}{2016}\natexlab{}.
\newblock \showarticletitle{Multi-perspective cost-sensitive context-aware
  multi-instance sparse coding and its application to sensitive video
  recognition}.
\newblock \bibinfo{journal}{\emph{IEEE Transactions on Multimedia}}
  \bibinfo{volume}{18}, \bibinfo{number}{1} (\bibinfo{year}{2016}),
  \bibinfo{pages}{76--89}.
\newblock


\bibitem[\protect\citeauthoryear{Hu and Downie}{Hu and Downie}{2007}]%
        {hu2007exploring}
\bibfield{author}{\bibinfo{person}{Xiao Hu} {and} \bibinfo{person}{J~Stephen
  Downie}.} \bibinfo{year}{2007}\natexlab{}.
\newblock \showarticletitle{Exploring Mood Metadata: Relationships with Genre,
  Artist and Usage Metadata.}. In \bibinfo{booktitle}{\emph{International
  Society for Music Information Retrieval Conference}}.
  \bibinfo{pages}{67--72}.
\newblock


\bibitem[\protect\citeauthoryear{Huang, Li, Tao, Lian, and Yi}{Huang
  et~al\mbox{.}}{2018}]%
        {huang2018end}
\bibfield{author}{\bibinfo{person}{Jian Huang}, \bibinfo{person}{Ya Li},
  \bibinfo{person}{Jianhua Tao}, \bibinfo{person}{Zheng Lian}, {and}
  \bibinfo{person}{Jiangyan Yi}.} \bibinfo{year}{2018}\natexlab{}.
\newblock \showarticletitle{End-to-End Continuous Emotion Recognition from
  Video Using 3D Convlstm Networks}. In \bibinfo{booktitle}{\emph{IEEE
  International Conference on Acoustics, Speech and Signal Processing}}.
  \bibinfo{pages}{6837--6841}.
\newblock


\bibitem[\protect\citeauthoryear{Inskip, Macfarlane, and Rafferty}{Inskip
  et~al\mbox{.}}{2012}]%
        {inskip2012}
\bibfield{author}{\bibinfo{person}{Charles Inskip}, \bibinfo{person}{Andy
  Macfarlane}, {and} \bibinfo{person}{Pauline Rafferty}.}
  \bibinfo{year}{2012}\natexlab{}.
\newblock \showarticletitle{Towards the disintermediation of creative music
  search: analysing queries to determine important facets}.
\newblock \bibinfo{journal}{\emph{International Journal on Digital Libraries}}
  \bibinfo{volume}{12}, \bibinfo{number}{2-3} (\bibinfo{year}{2012}),
  \bibinfo{pages}{137--147}.
\newblock


\bibitem[\protect\citeauthoryear{Jansen, Zhang, Sobel, and Chowdury}{Jansen
  et~al\mbox{.}}{2009}]%
        {jansen2009twitter}
\bibfield{author}{\bibinfo{person}{Bernard~J Jansen}, \bibinfo{person}{Mimi
  Zhang}, \bibinfo{person}{Kate Sobel}, {and} \bibinfo{person}{Abdur
  Chowdury}.} \bibinfo{year}{2009}\natexlab{}.
\newblock \showarticletitle{Twitter power: Tweets as electronic word of mouth}.
\newblock \bibinfo{journal}{\emph{Journal of the Association for Information
  Science and Technology}} \bibinfo{volume}{60}, \bibinfo{number}{11}
  (\bibinfo{year}{2009}), \bibinfo{pages}{2169--2188}.
\newblock


\bibitem[\protect\citeauthoryear{{Jeffrey H. Kahn, Ren{\'{e}}e M. Tobin} and
  Anderson}{{Jeffrey H. Kahn, Ren{\'{e}}e M. Tobin} and Anderson}{2007}]%
        {liwc2007}
\bibfield{author}{\bibinfo{person}{Audra E.~Massey {Jeffrey H. Kahn,
  Ren{\'{e}}e M. Tobin}} {and} \bibinfo{person}{Jennifer~A. Anderson}.}
  \bibinfo{year}{2007}\natexlab{}.
\newblock \showarticletitle{Measuring Emotional Expression with the Linguistic
  Inquiry and Word Count}.
\newblock \bibinfo{journal}{\emph{JSTOR: The American Journal of Psychology}}
  \bibinfo{volume}{120}, \bibinfo{number}{2} (\bibinfo{year}{2007}),
  \bibinfo{pages}{263--286}.
\newblock


\bibitem[\protect\citeauthoryear{Jiang, Xu, and Xue}{Jiang
  et~al\mbox{.}}{2014}]%
        {jiang2014predicting}
\bibfield{author}{\bibinfo{person}{Yu-Gang Jiang}, \bibinfo{person}{Baohan Xu},
  {and} \bibinfo{person}{Xiangyang Xue}.} \bibinfo{year}{2014}\natexlab{}.
\newblock \showarticletitle{Predicting emotions in user-generated videos}. In
  \bibinfo{booktitle}{\emph{Twenty-Eighth AAAI Conference on Artificial
  Intelligence}}.
\newblock


\bibitem[\protect\citeauthoryear{Joshi, Datta, Fedorovskaya, Luong, Wang, Li,
  and Luo}{Joshi et~al\mbox{.}}{2011}]%
        {joshi:emotion-survey}
\bibfield{author}{\bibinfo{person}{Dhiraj Joshi}, \bibinfo{person}{Ritendra
  Datta}, \bibinfo{person}{Elena Fedorovskaya}, \bibinfo{person}{Quang-Tuan
  Luong}, \bibinfo{person}{James~Z. Wang}, \bibinfo{person}{Jia Li}, {and}
  \bibinfo{person}{Jiebo Luo}.} \bibinfo{year}{2011}\natexlab{}.
\newblock \showarticletitle{Aesthetics and emotions in images}.
\newblock \bibinfo{journal}{\emph{IEEE Signal Processing Magazine}}
  \bibinfo{volume}{28}, \bibinfo{number}{5} (\bibinfo{year}{2011}),
  \bibinfo{pages}{94--115}.
\newblock


\bibitem[\protect\citeauthoryear{Jou, Chen, Pappas, Redi, Topkara, and
  Chang}{Jou et~al\mbox{.}}{2015}]%
        {Jou:MVSO}
\bibfield{author}{\bibinfo{person}{Brendan Jou}, \bibinfo{person}{Tao Chen},
  \bibinfo{person}{Nikolaos Pappas}, \bibinfo{person}{Miriam Redi},
  \bibinfo{person}{Mercan Topkara}, {and} \bibinfo{person}{Shih-Fu Chang}.}
  \bibinfo{year}{2015}\natexlab{}.
\newblock \showarticletitle{Visual Affect Around the World: {A} Large-scale
  Multilingual Visual Sentiment Ontology}. In \bibinfo{booktitle}{\emph{ACM
  International Conference on Multimedia}}. \bibinfo{pages}{159--168}.
\newblock


\bibitem[\protect\citeauthoryear{Jou, Yuying~Qian, and Chang}{Jou
  et~al\mbox{.}}{2016}]%
        {senticart:icmr16}
\bibfield{author}{\bibinfo{person}{Brendan Jou}, \bibinfo{person}{Margaret
  Yuying~Qian}, {and} \bibinfo{person}{Shih-Fu Chang}.}
  \bibinfo{year}{2016}\natexlab{}.
\newblock \showarticletitle{{SentiCart}: {C}artography and
  Geo-contextualization for Multilingual Visual Sentiment}. In
  \bibinfo{booktitle}{\emph{ACM International Conference on Multimedia
  Retrieval}}. \bibinfo{pages}{389--392}.
\newblock


\bibitem[\protect\citeauthoryear{Khorrami, Le~Paine, Brady, Dagli, and
  Huang}{Khorrami et~al\mbox{.}}{2016}]%
        {khorrami2016deep}
\bibfield{author}{\bibinfo{person}{Pooya Khorrami}, \bibinfo{person}{Tom
  Le~Paine}, \bibinfo{person}{Kevin Brady}, \bibinfo{person}{Charlie Dagli},
  {and} \bibinfo{person}{Thomas~S Huang}.} \bibinfo{year}{2016}\natexlab{}.
\newblock \showarticletitle{How deep neural networks can improve emotion
  recognition on video data}. In \bibinfo{booktitle}{\emph{IEEE International
  Conference on Image Processing}}. \bibinfo{pages}{619--623}.
\newblock


\bibitem[\protect\citeauthoryear{Kim, Schmidt, Migneco, Morton, Richardson,
  Scott, Speck, and Turnbull}{Kim et~al\mbox{.}}{2010}]%
        {Kim2010}
\bibfield{author}{\bibinfo{person}{Youngmoo~E Kim}, \bibinfo{person}{Erik~M
  Schmidt}, \bibinfo{person}{Raymond Migneco}, \bibinfo{person}{Brandon~G
  Morton}, \bibinfo{person}{Patrick Richardson}, \bibinfo{person}{Jeffrey
  Scott}, \bibinfo{person}{Jacquelin~A Speck}, {and} \bibinfo{person}{Douglas
  Turnbull}.} \bibinfo{year}{2010}\natexlab{}.
\newblock \showarticletitle{Music emotion recognition: A state of the art
  review}. In \bibinfo{booktitle}{\emph{International Society for Music
  Information Retrieval Conference}}, Vol.~\bibinfo{volume}{86}.
  \bibinfo{pages}{937--952}.
\newblock


\bibitem[\protect\citeauthoryear{Koelstra, Muhl, Soleymani, Lee, Yazdani,
  Ebrahimi, Pun, Nijholt, and Patras}{Koelstra et~al\mbox{.}}{2012}]%
        {koelstra2012tac}
\bibfield{author}{\bibinfo{person}{Sander Koelstra}, \bibinfo{person}{Christian
  Muhl}, \bibinfo{person}{Mohammad Soleymani}, \bibinfo{person}{Jong-Seok Lee},
  \bibinfo{person}{Ashkan Yazdani}, \bibinfo{person}{Touradj Ebrahimi},
  \bibinfo{person}{Thierry Pun}, \bibinfo{person}{Anton Nijholt}, {and}
  \bibinfo{person}{Ioannis Patras}.} \bibinfo{year}{2012}\natexlab{}.
\newblock \showarticletitle{Deap: A database for emotion analysis using
  physiological signals}.
\newblock \bibinfo{journal}{\emph{IEEE Transactions on Affective Computing}}
  \bibinfo{volume}{3}, \bibinfo{number}{1} (\bibinfo{year}{2012}),
  \bibinfo{pages}{18--31}.
\newblock


\bibitem[\protect\citeauthoryear{Lang, Bradley, and Cuthbert}{Lang
  et~al\mbox{.}}{1997}]%
        {lang1997international}
\bibfield{author}{\bibinfo{person}{Peter~J Lang}, \bibinfo{person}{Margaret~M
  Bradley}, {and} \bibinfo{person}{Bruce~N Cuthbert}.}
  \bibinfo{year}{1997}\natexlab{}.
\newblock \showarticletitle{International affective picture system (IAPS):
  Technical manual and affective ratings}.
\newblock \bibinfo{journal}{\emph{NIMH Center for the Study of Emotion and
  Attention}} (\bibinfo{year}{1997}), \bibinfo{pages}{39--58}.
\newblock


\bibitem[\protect\citeauthoryear{Larson, Soleymani, Gravier, Ionescu, and
  Jones}{Larson et~al\mbox{.}}{2017}]%
        {mediaeval}
\bibfield{author}{\bibinfo{person}{Martha Larson}, \bibinfo{person}{Mohammad
  Soleymani}, \bibinfo{person}{Guillaume Gravier}, \bibinfo{person}{Bogdan
  Ionescu}, {and} \bibinfo{person}{Gareth~JF Jones}.}
  \bibinfo{year}{2017}\natexlab{}.
\newblock \showarticletitle{The benchmarking initiative for multimedia
  evaluation: MediaEval 2016}.
\newblock \bibinfo{journal}{\emph{IEEE MultiMedia}} \bibinfo{volume}{24},
  \bibinfo{number}{1} (\bibinfo{year}{2017}), \bibinfo{pages}{93--96}.
\newblock


\bibitem[\protect\citeauthoryear{Lartillot, Toiviainen, and Eerola}{Lartillot
  et~al\mbox{.}}{2008}]%
        {mirtoolbox}
\bibfield{author}{\bibinfo{person}{Olivier Lartillot}, \bibinfo{person}{Petri
  Toiviainen}, {and} \bibinfo{person}{Tuomas Eerola}.}
  \bibinfo{year}{2008}\natexlab{}.
\newblock \showarticletitle{A Matlab Toolbox for Music Information Retrieval}.
  In \bibinfo{booktitle}{\emph{Data Analysis, Machine Learning and
  Applications}}. \bibinfo{pages}{261--268}.
\newblock


\bibitem[\protect\citeauthoryear{Laurier, Herrera, Mandel, and Ellis}{Laurier
  et~al\mbox{.}}{2007}]%
        {laurier07mirex}
\bibfield{author}{\bibinfo{person}{Cyril Laurier}, \bibinfo{person}{Perfecto
  Herrera}, \bibinfo{person}{M Mandel}, {and} \bibinfo{person}{D Ellis}.}
  \bibinfo{year}{2007}\natexlab{}.
\newblock \showarticletitle{Audio music mood classification using support
  vector machine}. In \bibinfo{booktitle}{\emph{MIREX task on Audio Mood
  Classification}}. \bibinfo{pages}{2--4}.
\newblock


\bibitem[\protect\citeauthoryear{Lee and Park}{Lee and Park}{2011}]%
        {lee2011fuzzy}
\bibfield{author}{\bibinfo{person}{Joonwhoan Lee} {and}
  \bibinfo{person}{EunJong Park}.} \bibinfo{year}{2011}\natexlab{}.
\newblock \showarticletitle{Fuzzy similarity-based emotional classification of
  color images}.
\newblock \bibinfo{journal}{\emph{IEEE Transactions on Multimedia}}
  \bibinfo{volume}{13}, \bibinfo{number}{5} (\bibinfo{year}{2011}),
  \bibinfo{pages}{1031--1039}.
\newblock


\bibitem[\protect\citeauthoryear{Li, Xiong, Hu, and Ding}{Li
  et~al\mbox{.}}{2012}]%
        {li2012context}
\bibfield{author}{\bibinfo{person}{Bing Li}, \bibinfo{person}{Weihua Xiong},
  \bibinfo{person}{Weiming Hu}, {and} \bibinfo{person}{Xinmiao Ding}.}
  \bibinfo{year}{2012}\natexlab{}.
\newblock \showarticletitle{Context-aware affective images classification based
  on bilayer sparse representation}. In \bibinfo{booktitle}{\emph{ACM
  International Conference on Multimedia}}. \bibinfo{pages}{721--724}.
\newblock


\bibitem[\protect\citeauthoryear{Li, Bailenson, Pines, Greenleaf, and
  Williams}{Li et~al\mbox{.}}{2017}]%
        {Li_VR}
\bibfield{author}{\bibinfo{person}{Benjamin~J. Li}, \bibinfo{person}{Jeremy~N.
  Bailenson}, \bibinfo{person}{Adam Pines}, \bibinfo{person}{Walter~J.
  Greenleaf}, {and} \bibinfo{person}{Leanne~M. Williams}.}
  \bibinfo{year}{2017}\natexlab{}.
\newblock \showarticletitle{A Public Database of Immersive VR Videos with
  Corresponding Ratings of Arousal, Valence, and Correlations between Head
  Movements and Self Report Measures}.
\newblock \bibinfo{journal}{\emph{Frontiers in Psychology}}
  \bibinfo{volume}{8} (\bibinfo{year}{2017}), \bibinfo{pages}{2116}.
\newblock


\bibitem[\protect\citeauthoryear{Lin, RoyChowdhury, and Maji}{Lin
  et~al\mbox{.}}{2015}]%
        {Lin_2015_ICCV}
\bibfield{author}{\bibinfo{person}{Tsung-Yu Lin}, \bibinfo{person}{Aruni
  RoyChowdhury}, {and} \bibinfo{person}{Subhransu Maji}.}
  \bibinfo{year}{2015}\natexlab{}.
\newblock \showarticletitle{Bilinear CNN Models for Fine-Grained Visual
  Recognition}. In \bibinfo{booktitle}{\emph{IEEE International Conference on
  Computer Vision}}. \bibinfo{pages}{1449--1457}.
\newblock


\bibitem[\protect\citeauthoryear{Lu, Suryanarayan, Adams~Jr, Li, Newman, and
  Wang}{Lu et~al\mbox{.}}{2012}]%
        {lu2012shape}
\bibfield{author}{\bibinfo{person}{Xin Lu}, \bibinfo{person}{Poonam
  Suryanarayan}, \bibinfo{person}{Reginald~B Adams~Jr}, \bibinfo{person}{Jia
  Li}, \bibinfo{person}{Michelle~G Newman}, {and} \bibinfo{person}{James~Z
  Wang}.} \bibinfo{year}{2012}\natexlab{}.
\newblock \showarticletitle{On shape and the computability of emotions}. In
  \bibinfo{booktitle}{\emph{ACM International Conference on Multimedia}}.
  \bibinfo{pages}{229--238}.
\newblock


\bibitem[\protect\citeauthoryear{Machajdik and Hanbury}{Machajdik and
  Hanbury}{2010}]%
        {machajdik2010affective}
\bibfield{author}{\bibinfo{person}{Jana Machajdik} {and} \bibinfo{person}{Allan
  Hanbury}.} \bibinfo{year}{2010}\natexlab{}.
\newblock \showarticletitle{Affective image classification using features
  inspired by psychology and art theory}. In \bibinfo{booktitle}{\emph{ACM
  International Conference on Multimedia}}. \bibinfo{pages}{83--92}.
\newblock


\bibitem[\protect\citeauthoryear{Malandrakis, Potamianos, Evangelopoulos, and
  Zlatintsi}{Malandrakis et~al\mbox{.}}{2011}]%
        {5946961}
\bibfield{author}{\bibinfo{person}{Nikos Malandrakis},
  \bibinfo{person}{Alexandros Potamianos}, \bibinfo{person}{Georgios
  Evangelopoulos}, {and} \bibinfo{person}{Athanasia Zlatintsi}.}
  \bibinfo{year}{2011}\natexlab{}.
\newblock \showarticletitle{A supervised approach to movie emotion tracking}.
  In \bibinfo{booktitle}{\emph{IEEE International Conference on Acoustics,
  Speech and Signal Processing}}. \bibinfo{pages}{2376--2379}.
\newblock


\bibitem[\protect\citeauthoryear{Malheiro, Panda, Gomes, and Paiva}{Malheiro
  et~al\mbox{.}}{2016}]%
        {7536113}
\bibfield{author}{\bibinfo{person}{Ricardo Malheiro}, \bibinfo{person}{Renato
  Panda}, \bibinfo{person}{Paulo Gomes}, {and} \bibinfo{person}{Rui~Pedro
  Paiva}.} \bibinfo{year}{2016}\natexlab{}.
\newblock \showarticletitle{Emotionally-relevant features for classification
  and regression of music lyrics}.
\newblock \bibinfo{journal}{\emph{IEEE Transactions on Affective Computing}}
  \bibinfo{volume}{9}, \bibinfo{number}{2} (\bibinfo{year}{2016}),
  \bibinfo{pages}{240--254}.
\newblock


\bibitem[\protect\citeauthoryear{McDuff and Soleymani}{McDuff and
  Soleymani}{2017}]%
        {mcduff2017large}
\bibfield{author}{\bibinfo{person}{Daniel McDuff} {and}
  \bibinfo{person}{Mohammad Soleymani}.} \bibinfo{year}{2017}\natexlab{}.
\newblock \showarticletitle{Large-scale affective content analysis: Combining
  media content features and facial reactions}. In
  \bibinfo{booktitle}{\emph{IEEE International Conference on Automatic Face \&
  Gesture Recognition}}. \bibinfo{pages}{339--345}.
\newblock


\bibitem[\protect\citeauthoryear{McFee, Raffel, Liang, Ellis, McVicar,
  Battenberg, and Nieto}{McFee et~al\mbox{.}}{2015}]%
        {mcfee2015librosa}
\bibfield{author}{\bibinfo{person}{Brian McFee}, \bibinfo{person}{Colin
  Raffel}, \bibinfo{person}{Dawen Liang}, \bibinfo{person}{Daniel~PW Ellis},
  \bibinfo{person}{Matt McVicar}, \bibinfo{person}{Eric Battenberg}, {and}
  \bibinfo{person}{Oriol Nieto}.} \bibinfo{year}{2015}\natexlab{}.
\newblock \showarticletitle{librosa: Audio and music signal analysis in
  python}. In \bibinfo{booktitle}{\emph{Proceedings of the Python in Science
  Conferences}}, Vol.~\bibinfo{volume}{8}. \bibinfo{pages}{18--25}.
\newblock


\bibitem[\protect\citeauthoryear{Merler, Mac, Joshi, Nguyen, Hammer, Kent,
  Xiong, Do, Smith, and Feris}{Merler et~al\mbox{.}}{2018}]%
        {merler:highlights}
\bibfield{author}{\bibinfo{person}{Michele Merler},
  \bibinfo{person}{Khoi-Nguyen~C. Mac}, \bibinfo{person}{Dhiraj Joshi},
  \bibinfo{person}{Quoc-Bao Nguyen}, \bibinfo{person}{Stephen Hammer},
  \bibinfo{person}{John Kent}, \bibinfo{person}{Jinjun Xiong},
  \bibinfo{person}{Minh~N. Do}, \bibinfo{person}{John~R. Smith}, {and}
  \bibinfo{person}{Rogerio~S. Feris}.} \bibinfo{year}{2018}\natexlab{}.
\newblock \showarticletitle{Automatic curation of sports highlights using
  multimodal excitement features}.
\newblock \bibinfo{journal}{\emph{IEEE Transactions on Multimedia}}
  \bibinfo{volume}{21}, \bibinfo{number}{5} (\bibinfo{year}{2018}),
  \bibinfo{pages}{1147--1160}.
\newblock


\bibitem[\protect\citeauthoryear{Mikels, Fredrickson, Larkin, Lindberg, Maglio,
  and Reuter-Lorenz}{Mikels et~al\mbox{.}}{2005}]%
        {mikels2005emotional}
\bibfield{author}{\bibinfo{person}{Joseph~A Mikels}, \bibinfo{person}{Barbara~L
  Fredrickson}, \bibinfo{person}{Gregory~R Larkin}, \bibinfo{person}{Casey~M
  Lindberg}, \bibinfo{person}{Sam~J Maglio}, {and} \bibinfo{person}{Patricia~A
  Reuter-Lorenz}.} \bibinfo{year}{2005}\natexlab{}.
\newblock \showarticletitle{Emotional category data on images from the
  International Affective Picture System}.
\newblock \bibinfo{journal}{\emph{Behavior Research Methods}}
  \bibinfo{volume}{37}, \bibinfo{number}{4} (\bibinfo{year}{2005}),
  \bibinfo{pages}{626--630}.
\newblock


\bibitem[\protect\citeauthoryear{Mikolov, Sutskever, Chen, Corrado, and
  Dean}{Mikolov et~al\mbox{.}}{2013}]%
        {word2vec}
\bibfield{author}{\bibinfo{person}{Tomas Mikolov}, \bibinfo{person}{Ilya
  Sutskever}, \bibinfo{person}{Kai Chen}, \bibinfo{person}{Greg~S Corrado},
  {and} \bibinfo{person}{Jeff Dean}.} \bibinfo{year}{2013}\natexlab{}.
\newblock \showarticletitle{Distributed Representations of Words and Phrases
  and their Compositionality}.
\newblock In \bibinfo{booktitle}{\emph{Advances in Neural Information
  Processing Systems 26}}. \bibinfo{pages}{3111--3119}.
\newblock


\bibitem[\protect\citeauthoryear{Mo, Niu, Su, and Das}{Mo
  et~al\mbox{.}}{2018}]%
        {mo2018novel}
\bibfield{author}{\bibinfo{person}{Shasha Mo}, \bibinfo{person}{Jianwei Niu},
  \bibinfo{person}{Yiming Su}, {and} \bibinfo{person}{Sajal~K Das}.}
  \bibinfo{year}{2018}\natexlab{}.
\newblock \showarticletitle{A novel feature set for video emotion recognition}.
\newblock \bibinfo{journal}{\emph{Neurocomputing}}  \bibinfo{volume}{291}
  (\bibinfo{year}{2018}), \bibinfo{pages}{11--20}.
\newblock


\bibitem[\protect\citeauthoryear{Morreale, Masu, De~Angeli,
  et~al\mbox{.}}{Morreale et~al\mbox{.}}{2013}]%
        {morreale2013robin}
\bibfield{author}{\bibinfo{person}{Fabio Morreale}, \bibinfo{person}{Raul
  Masu}, \bibinfo{person}{Antonella De~Angeli}, {et~al\mbox{.}}}
  \bibinfo{year}{2013}\natexlab{}.
\newblock \showarticletitle{Robin: an algorithmic composer for interactive
  scenarios}.
\newblock \bibinfo{journal}{\emph{Sound and Music Computing Conference}}
  \bibinfo{volume}{2013} (\bibinfo{year}{2013}), \bibinfo{pages}{207--212}.
\newblock


\bibitem[\protect\citeauthoryear{Muszynski, Tian, Lai, Moore, Kostoulas,
  Lombardo, Pun, and Chanel}{Muszynski et~al\mbox{.}}{2019}]%
        {muszynski2019recognizing}
\bibfield{author}{\bibinfo{person}{Michal Muszynski}, \bibinfo{person}{Leimin
  Tian}, \bibinfo{person}{Catherine Lai}, \bibinfo{person}{Johanna Moore},
  \bibinfo{person}{Theodoros Kostoulas}, \bibinfo{person}{Patrizia Lombardo},
  \bibinfo{person}{Thierry Pun}, {and} \bibinfo{person}{Guillaume Chanel}.}
  \bibinfo{year}{2019}\natexlab{}.
\newblock \showarticletitle{Recognizing Induced Emotions of Movie Audiences
  From Multimodal Information}.
\newblock \bibinfo{journal}{\emph{IEEE Transactions on Affective Computing}}
  (\bibinfo{year}{2019}).
\newblock


\bibitem[\protect\citeauthoryear{Nemati and Naghsh-Nilchi}{Nemati and
  Naghsh-Nilchi}{2016}]%
        {nemati2016incorporating}
\bibfield{author}{\bibinfo{person}{Shahla Nemati} {and}
  \bibinfo{person}{Ahmad~Reza Naghsh-Nilchi}.} \bibinfo{year}{2016}\natexlab{}.
\newblock \showarticletitle{Incorporating social media comments in affective
  video retrieval}.
\newblock \bibinfo{journal}{\emph{Journal of Information Science}}
  \bibinfo{volume}{42}, \bibinfo{number}{4} (\bibinfo{year}{2016}),
  \bibinfo{pages}{524--538}.
\newblock


\bibitem[\protect\citeauthoryear{Nemati and Naghsh-Nilchi}{Nemati and
  Naghsh-Nilchi}{2017}]%
        {nemati2017evidential}
\bibfield{author}{\bibinfo{person}{Shahla Nemati} {and}
  \bibinfo{person}{Ahmad~Reza Naghsh-Nilchi}.} \bibinfo{year}{2017}\natexlab{}.
\newblock \showarticletitle{An evidential data fusion method for affective
  music video retrieval}.
\newblock \bibinfo{journal}{\emph{Intelligent Data Analysis}}
  \bibinfo{volume}{21}, \bibinfo{number}{2} (\bibinfo{year}{2017}),
  \bibinfo{pages}{427--441}.
\newblock


\bibitem[\protect\citeauthoryear{Nicolaou, Gunes, and Pantic}{Nicolaou
  et~al\mbox{.}}{2011}]%
        {Nicolaou2011}
\bibfield{author}{\bibinfo{person}{Mihalis~A Nicolaou}, \bibinfo{person}{Hatice
  Gunes}, {and} \bibinfo{person}{Maja Pantic}.}
  \bibinfo{year}{2011}\natexlab{}.
\newblock \showarticletitle{Continuous Prediction of Spontaneous Affect from
  Multiple Cues and Modalities in Valence-Arousal Space}.
\newblock \bibinfo{journal}{\emph{IEEE Transactions on Affective Computing}}
  \bibinfo{volume}{2}, \bibinfo{number}{2} (\bibinfo{year}{2011}),
  \bibinfo{pages}{92--105}.
\newblock


\bibitem[\protect\citeauthoryear{Niu, Su, Mo, and Zhu}{Niu
  et~al\mbox{.}}{2017a}]%
        {niu2017novel}
\bibfield{author}{\bibinfo{person}{Jianwei Niu}, \bibinfo{person}{Yiming Su},
  \bibinfo{person}{Shasha Mo}, {and} \bibinfo{person}{Zeyu Zhu}.}
  \bibinfo{year}{2017}\natexlab{a}.
\newblock \showarticletitle{A Novel Affective Visualization System for Videos
  Based on Acoustic and Visual Features}. In
  \bibinfo{booktitle}{\emph{International Conference on Multimedia Modeling}}.
  \bibinfo{pages}{15--27}.
\newblock


\bibitem[\protect\citeauthoryear{Niu, Wang, Su, and Guo}{Niu
  et~al\mbox{.}}{2017b}]%
        {Niu2017TemporalFV}
\bibfield{author}{\bibinfo{person}{Jianwei Niu}, \bibinfo{person}{Shihao Wang},
  \bibinfo{person}{Yiming Su}, {and} \bibinfo{person}{Song Guo}.}
  \bibinfo{year}{2017}\natexlab{b}.
\newblock \showarticletitle{Temporal Factor-Aware Video Affective Analysis and
  Recommendation for Cyber-Based Social Media}.
\newblock \bibinfo{journal}{\emph{IEEE Transactions on Emerging Topics in
  Computing}}  \bibinfo{volume}{5} (\bibinfo{year}{2017}),
  \bibinfo{pages}{412--424}.
\newblock


\bibitem[\protect\citeauthoryear{Niu, Zhao, and Aziz}{Niu
  et~al\mbox{.}}{2016}]%
        {niu2016novel}
\bibfield{author}{\bibinfo{person}{Jianwei Niu}, \bibinfo{person}{Xiaoke Zhao},
  {and} \bibinfo{person}{Muhammad Ali~Abdul Aziz}.}
  \bibinfo{year}{2016}\natexlab{}.
\newblock \showarticletitle{A novel affect-based model of similarity measure of
  videos}.
\newblock \bibinfo{journal}{\emph{Neurocomputing}}  \bibinfo{volume}{173}
  (\bibinfo{year}{2016}), \bibinfo{pages}{339--345}.
\newblock


\bibitem[\protect\citeauthoryear{Ortony, Clore, and Collins}{Ortony
  et~al\mbox{.}}{1988}]%
        {ortony88emotion}
\bibfield{author}{\bibinfo{person}{Andrew Ortony}, \bibinfo{person}{Gerald~L.
  Clore}, {and} \bibinfo{person}{Allan Collins}.}
  \bibinfo{year}{1988}\natexlab{}.
\newblock \bibinfo{booktitle}{\emph{{The Cognitive Structure of Emotions}}}.
\newblock \bibinfo{publisher}{{Cambridge University Press}}.
\newblock
\showISBNx{0521353645}


\bibitem[\protect\citeauthoryear{Panda}{Panda}{2019}]%
        {panda_thesis}
\bibfield{author}{\bibinfo{person}{Renato Eduardo~Silva Panda}.}
  \bibinfo{year}{2019}\natexlab{}.
\newblock \emph{\bibinfo{title}{Emotion-based Analysis and Classification of
  Audio Music}}.
\newblock \bibinfo{thesistype}{Ph.D. Dissertation}. \bibinfo{school}{Univesity
  of Coimbra}, \bibinfo{address}{Coimbra, Portugal}.
\newblock


\bibitem[\protect\citeauthoryear{PANDEYA and Joonwhoan}{PANDEYA and
  Joonwhoan}{2019}]%
        {pandeya2019music}
\bibfield{author}{\bibinfo{person}{Yagya~Raj PANDEYA} {and}
  \bibinfo{person}{LEE Joonwhoan}.} \bibinfo{year}{2019}\natexlab{}.
\newblock \showarticletitle{Music-Video Emotion Analysis Using Late Fusion of
  Multimodal}.
\newblock \bibinfo{journal}{\emph{DEStech Transactions on Computer Science and
  Engineering}} \bibinfo{number}{iteee} (\bibinfo{year}{2019}).
\newblock


\bibitem[\protect\citeauthoryear{Pang, Lee, et~al\mbox{.}}{Pang
  et~al\mbox{.}}{2008}]%
        {pang2008opinion}
\bibfield{author}{\bibinfo{person}{Bo Pang}, \bibinfo{person}{Lillian Lee},
  {et~al\mbox{.}}} \bibinfo{year}{2008}\natexlab{}.
\newblock \showarticletitle{Opinion mining and sentiment analysis}.
\newblock \bibinfo{journal}{\emph{Foundations and Trends{\textregistered} in
  Information Retrieval}} \bibinfo{volume}{2}, \bibinfo{number}{1--2}
  (\bibinfo{year}{2008}), \bibinfo{pages}{1--135}.
\newblock


\bibitem[\protect\citeauthoryear{Pang and Ngo}{Pang and Ngo}{2015}]%
        {pang2015mutlimodal}
\bibfield{author}{\bibinfo{person}{Lei Pang} {and} \bibinfo{person}{Chong-Wah
  Ngo}.} \bibinfo{year}{2015}\natexlab{}.
\newblock \showarticletitle{Mutlimodal learning with deep boltzmann machine for
  emotion prediction in user generated videos}. In
  \bibinfo{booktitle}{\emph{ACM International Conference on Multimedia
  Retrieval}}. \bibinfo{pages}{619--622}.
\newblock


\bibitem[\protect\citeauthoryear{Pang, Zhu, and Ngo}{Pang
  et~al\mbox{.}}{2015}]%
        {Pang2015}
\bibfield{author}{\bibinfo{person}{Lei Pang}, \bibinfo{person}{Shiai Zhu},
  {and} \bibinfo{person}{Chong~Wah Ngo}.} \bibinfo{year}{2015}\natexlab{}.
\newblock \showarticletitle{Deep Multimodal Learning for Affective Analysis and
  Retrieval}.
\newblock \bibinfo{journal}{\emph{IEEE Transactions on Multimedia}}
  \bibinfo{volume}{17}, \bibinfo{number}{11} (\bibinfo{year}{2015}),
  \bibinfo{pages}{2008--2020}.
\newblock


\bibitem[\protect\citeauthoryear{Patterson and Hays}{Patterson and
  Hays}{2012}]%
        {patterson2012sun}
\bibfield{author}{\bibinfo{person}{Genevieve Patterson} {and}
  \bibinfo{person}{James Hays}.} \bibinfo{year}{2012}\natexlab{}.
\newblock \showarticletitle{Sun attribute database: Discovering, annotating,
  and recognizing scene attributes}. In \bibinfo{booktitle}{\emph{IEEE
  Conference on Computer Vision and Pattern Recognition}}.
  \bibinfo{pages}{2751--2758}.
\newblock


\bibitem[\protect\citeauthoryear{Peng, Sadovnik, Gallagher, and Chen}{Peng
  et~al\mbox{.}}{2015}]%
        {peng2015mixed}
\bibfield{author}{\bibinfo{person}{Kuan-Chuan Peng}, \bibinfo{person}{Amir
  Sadovnik}, \bibinfo{person}{Andrew Gallagher}, {and} \bibinfo{person}{Tsuhan
  Chen}.} \bibinfo{year}{2015}\natexlab{}.
\newblock \showarticletitle{A Mixed Bag of Emotions: Model, Predict, and
  Transfer Emotion Distributions}. In \bibinfo{booktitle}{\emph{IEEE Conference
  on Computer Vision and Pattern Recognition}}. \bibinfo{pages}{860--868}.
\newblock


\bibitem[\protect\citeauthoryear{Pennington, Socher, and Manning}{Pennington
  et~al\mbox{.}}{2014}]%
        {pennington2014glove}
\bibfield{author}{\bibinfo{person}{Jeffrey Pennington},
  \bibinfo{person}{Richard Socher}, {and} \bibinfo{person}{Christopher
  Manning}.} \bibinfo{year}{2014}\natexlab{}.
\newblock \showarticletitle{Glove: Global vectors for word representation}. In
  \bibinfo{booktitle}{\emph{Conference on Empirical Methods in Natural Language
  Processing}}. \bibinfo{pages}{1532--1543}.
\newblock


\bibitem[\protect\citeauthoryear{Plutchik}{Plutchik}{1980}]%
        {plutchik1980emotion}
\bibfield{author}{\bibinfo{person}{Robert Plutchik}.}
  \bibinfo{year}{1980}\natexlab{}.
\newblock \bibinfo{booktitle}{\emph{Emotion: A psychoevolutionary synthesis}}.
\newblock \bibinfo{publisher}{Harpercollins College Division}.
\newblock


\bibitem[\protect\citeauthoryear{Poria, Hazarika, Majumder, Naik, Cambria, and
  Mihalcea}{Poria et~al\mbox{.}}{2019}]%
        {poria-etal-2019-meld}
\bibfield{author}{\bibinfo{person}{Soujanya Poria}, \bibinfo{person}{Devamanyu
  Hazarika}, \bibinfo{person}{Navonil Majumder}, \bibinfo{person}{Gautam Naik},
  \bibinfo{person}{Erik Cambria}, {and} \bibinfo{person}{Rada Mihalcea}.}
  \bibinfo{year}{2019}\natexlab{}.
\newblock \showarticletitle{{MELD}: A Multimodal Multi-Party Dataset for
  Emotion Recognition in Conversations}. In \bibinfo{booktitle}{\emph{Annual
  Meeting of the Association for Computational Linguistics}}.
  \bibinfo{pages}{527--536}.
\newblock


\bibitem[\protect\citeauthoryear{Rajagopalan, Morency, Baltrus̆aitis, and
  Goecke}{Rajagopalan et~al\mbox{.}}{2016}]%
        {Rajagopalan2016}
\bibfield{author}{\bibinfo{person}{Shyam~Sundar Rajagopalan},
  \bibinfo{person}{Louis-Philippe Morency}, \bibinfo{person}{Tadas
  Baltrus̆aitis}, {and} \bibinfo{person}{Roland Goecke}.}
  \bibinfo{year}{2016}\natexlab{}.
\newblock \showarticletitle{Extending Long Short-Term Memory for Multi-View
  Structured Learning}. In \bibinfo{booktitle}{\emph{European Conference on
  Computer Vision}}. \bibinfo{pages}{338--353}.
\newblock


\bibitem[\protect\citeauthoryear{Rao, Xu, Liu, Wang, and Burnett}{Rao
  et~al\mbox{.}}{2016b}]%
        {rao2016multi}
\bibfield{author}{\bibinfo{person}{Tianrong Rao}, \bibinfo{person}{Min Xu},
  \bibinfo{person}{Huiying Liu}, \bibinfo{person}{Jinqiao Wang}, {and}
  \bibinfo{person}{Ian Burnett}.} \bibinfo{year}{2016}\natexlab{b}.
\newblock \showarticletitle{Multi-scale blocks based image emotion
  classification using multiple instance learning}. In
  \bibinfo{booktitle}{\emph{IEEE International Conference on Image
  Processing}}. \bibinfo{pages}{634--638}.
\newblock


\bibitem[\protect\citeauthoryear{Rao, Xu, and Xu}{Rao et~al\mbox{.}}{2016a}]%
        {rao2016learning}
\bibfield{author}{\bibinfo{person}{Tianrong Rao}, \bibinfo{person}{Min Xu},
  {and} \bibinfo{person}{Dong Xu}.} \bibinfo{year}{2016}\natexlab{a}.
\newblock \showarticletitle{Learning multi-level deep representations for image
  emotion classification}.
\newblock \bibinfo{journal}{\emph{arXiv preprint arXiv:1611.07145}}
  (\bibinfo{year}{2016}).
\newblock


\bibitem[\protect\citeauthoryear{Sander, Grandjean, and Scherer}{Sander
  et~al\mbox{.}}{2005}]%
        {citeulike:3014462}
\bibfield{author}{\bibinfo{person}{David Sander}, \bibinfo{person}{Didier
  Grandjean}, {and} \bibinfo{person}{Klaus~R. Scherer}.}
  \bibinfo{year}{2005}\natexlab{}.
\newblock \showarticletitle{A systems approach to appraisal mechanisms in
  emotion}.
\newblock \bibinfo{journal}{\emph{Neural Networks}} \bibinfo{volume}{18},
  \bibinfo{number}{4} (\bibinfo{year}{2005}), \bibinfo{pages}{317--352}.
\newblock


\bibitem[\protect\citeauthoryear{Sartori, Culibrk, Yan, and Sebe}{Sartori
  et~al\mbox{.}}{2015}]%
        {sartori2015s}
\bibfield{author}{\bibinfo{person}{Andreza Sartori}, \bibinfo{person}{Dubravko
  Culibrk}, \bibinfo{person}{Yan Yan}, {and} \bibinfo{person}{Nicu Sebe}.}
  \bibinfo{year}{2015}\natexlab{}.
\newblock \showarticletitle{Who's afraid of itten: Using the art theory of
  color combination to analyze emotions in abstract paintings}. In
  \bibinfo{booktitle}{\emph{ACM International Conference on Multimedia}}.
  \bibinfo{pages}{311--320}.
\newblock


\bibitem[\protect\citeauthoryear{Schlosberg}{Schlosberg}{1954}]%
        {schlosberg1954three}
\bibfield{author}{\bibinfo{person}{Harold Schlosberg}.}
  \bibinfo{year}{1954}\natexlab{}.
\newblock \showarticletitle{Three dimensions of emotion}.
\newblock \bibinfo{journal}{\emph{Psychological Review}} \bibinfo{volume}{61},
  \bibinfo{number}{2} (\bibinfo{year}{1954}), \bibinfo{pages}{81}.
\newblock


\bibitem[\protect\citeauthoryear{She, Yang, Cheng, Lai, Rosin, and Wang}{She
  et~al\mbox{.}}{2019}]%
        {she2019wscnet}
\bibfield{author}{\bibinfo{person}{Dongyu She}, \bibinfo{person}{Jufeng Yang},
  \bibinfo{person}{Ming-Ming Cheng}, \bibinfo{person}{Yu-Kun Lai},
  \bibinfo{person}{Paul~L Rosin}, {and} \bibinfo{person}{Liang Wang}.}
  \bibinfo{year}{2019}\natexlab{}.
\newblock \showarticletitle{WSCNet: Weakly supervised coupled networks for
  visual sentiment classification and detection}.
\newblock \bibinfo{journal}{\emph{IEEE Transactions on Multimedia}}
  (\bibinfo{year}{2019}).
\newblock


\bibitem[\protect\citeauthoryear{Shen, Jia, Nie, Feng, Zhang, Hu, Chua, and
  Zhu}{Shen et~al\mbox{.}}{2017}]%
        {shen2017depression}
\bibfield{author}{\bibinfo{person}{Guangyao Shen}, \bibinfo{person}{Jia Jia},
  \bibinfo{person}{Liqiang Nie}, \bibinfo{person}{Fuli Feng},
  \bibinfo{person}{Cunjun Zhang}, \bibinfo{person}{Tianrui Hu},
  \bibinfo{person}{Tat-Seng Chua}, {and} \bibinfo{person}{Wenwu Zhu}.}
  \bibinfo{year}{2017}\natexlab{}.
\newblock \showarticletitle{Depression Detection via Harvesting Social Media: A
  Multimodal Dictionary Learning Solution.}. In
  \bibinfo{booktitle}{\emph{International Joint Conference on Artificial
  Intelligence}}. \bibinfo{pages}{3838--3844}.
\newblock


\bibitem[\protect\citeauthoryear{Shukla}{Shukla}{2018}]%
        {shukla2018multimodal}
\bibfield{author}{\bibinfo{person}{Abhinav Shukla}.}
  \bibinfo{year}{2018}\natexlab{}.
\newblock \emph{\bibinfo{title}{Multimodal Emotion Recognition from
  Advertisements with Application to Computational Advertising}}.
\newblock \bibinfo{thesistype}{Ph.D. Dissertation}.
  \bibinfo{school}{International Institute of Information Technology
  Hyderabad}.
\newblock


\bibitem[\protect\citeauthoryear{Shukla, Gullapuram, Katti, Yadati,
  Kankanhalli, and Subramanian}{Shukla et~al\mbox{.}}{2017a}]%
        {shukla2017affect}
\bibfield{author}{\bibinfo{person}{Abhinav Shukla},
  \bibinfo{person}{Shruti~Shriya Gullapuram}, \bibinfo{person}{Harish Katti},
  \bibinfo{person}{Karthik Yadati}, \bibinfo{person}{Mohan Kankanhalli}, {and}
  \bibinfo{person}{Ramanathan Subramanian}.} \bibinfo{year}{2017}\natexlab{a}.
\newblock \showarticletitle{Affect recognition in ads with application to
  computational advertising}. In \bibinfo{booktitle}{\emph{ACM International
  Conference on Multimedia}}. \bibinfo{pages}{1148--1156}.
\newblock


\bibitem[\protect\citeauthoryear{Shukla, Gullapuram, Katti, Yadati,
  Kankanhalli, and Subramanian}{Shukla et~al\mbox{.}}{2017b}]%
        {shukla2017evaluating}
\bibfield{author}{\bibinfo{person}{Abhinav Shukla},
  \bibinfo{person}{Shruti~Shriya Gullapuram}, \bibinfo{person}{Harish Katti},
  \bibinfo{person}{Karthik Yadati}, \bibinfo{person}{Mohan Kankanhalli}, {and}
  \bibinfo{person}{Ramanathan Subramanian}.} \bibinfo{year}{2017}\natexlab{b}.
\newblock \showarticletitle{Evaluating content-centric vs. user-centric ad
  affect recognition}. In \bibinfo{booktitle}{\emph{ACM International
  Conference on Multimodal Interaction}}. \bibinfo{pages}{402--410}.
\newblock


\bibitem[\protect\citeauthoryear{Shukla, Katti, Kankanhalli, and
  Subramanian}{Shukla et~al\mbox{.}}{2018}]%
        {shukla2018looking}
\bibfield{author}{\bibinfo{person}{Abhinav Shukla}, \bibinfo{person}{Harish
  Katti}, \bibinfo{person}{Mohan Kankanhalli}, {and}
  \bibinfo{person}{Ramanathan Subramanian}.} \bibinfo{year}{2018}\natexlab{}.
\newblock \showarticletitle{Looking Beyond a Clever Narrative: Visual Context
  and Attention are Primary Drivers of Affect in Video Advertisements}. In
  \bibinfo{booktitle}{\emph{ACM International Conference on Multimodal
  Interaction}}. \bibinfo{pages}{210--219}.
\newblock


\bibitem[\protect\citeauthoryear{Sivaprasad, Joshi, Agrawal, and
  Pedanekar}{Sivaprasad et~al\mbox{.}}{2018}]%
        {sivaprasad2018multimodal}
\bibfield{author}{\bibinfo{person}{Sarath Sivaprasad},
  \bibinfo{person}{Tanmayee Joshi}, \bibinfo{person}{Rishabh Agrawal}, {and}
  \bibinfo{person}{Niranjan Pedanekar}.} \bibinfo{year}{2018}\natexlab{}.
\newblock \showarticletitle{Multimodal Continuous Prediction of Emotions in
  Movies using Long Short-Term Memory Networks}. In
  \bibinfo{booktitle}{\emph{ACM International Conference on Multimedia
  Retrieval}}. \bibinfo{pages}{413--419}.
\newblock


\bibitem[\protect\citeauthoryear{Sj{\"o}berg, Baveye, Wang, Quang, Ionescu,
  Dellandr{\'e}a, Schedl, Demarty, and Chen}{Sj{\"o}berg et~al\mbox{.}}{2015}]%
        {sjoberg2015mediaeval}
\bibfield{author}{\bibinfo{person}{Mats Sj{\"o}berg}, \bibinfo{person}{Yoann
  Baveye}, \bibinfo{person}{Hanli Wang}, \bibinfo{person}{Vu~Lam Quang},
  \bibinfo{person}{Bogdan Ionescu}, \bibinfo{person}{Emmanuel Dellandr{\'e}a},
  \bibinfo{person}{Markus Schedl}, \bibinfo{person}{Claire-H{\'e}l{\`e}ne
  Demarty}, {and} \bibinfo{person}{Liming Chen}.}
  \bibinfo{year}{2015}\natexlab{}.
\newblock \showarticletitle{The MediaEval 2015 Affective Impact of Movies
  Task}. In \bibinfo{booktitle}{\emph{MediaEval}}.
\newblock


\bibitem[\protect\citeauthoryear{Smith, Joshi, Huet, Hsu, and Cota}{Smith
  et~al\mbox{.}}{2017}]%
        {Smith:augmenting}
\bibfield{author}{\bibinfo{person}{John~R. Smith}, \bibinfo{person}{Dhiraj
  Joshi}, \bibinfo{person}{Benoit Huet}, \bibinfo{person}{Winston Hsu}, {and}
  \bibinfo{person}{Jozef Cota}.} \bibinfo{year}{2017}\natexlab{}.
\newblock \showarticletitle{Harnessing A.I. for augmenting creativity:
  Application to movie trailer creation}. In \bibinfo{booktitle}{\emph{ACM
  International Conference on Multimedia}}. \bibinfo{pages}{1799--1808}.
\newblock


\bibitem[\protect\citeauthoryear{Soleymani}{Soleymani}{2015}]%
        {Soleymani2015}
\bibfield{author}{\bibinfo{person}{Mohammad Soleymani}.}
  \bibinfo{year}{2015}\natexlab{}.
\newblock \showarticletitle{The quest for visual interest}. In
  \bibinfo{booktitle}{\emph{ACM International Conference on Multimedia}}.
  \bibinfo{pages}{919--922}.
\newblock


\bibitem[\protect\citeauthoryear{Soleymani, Caro, Schmidt, Sha, and
  Yang}{Soleymani et~al\mbox{.}}{2013}]%
        {Soleymani1000songs}
\bibfield{author}{\bibinfo{person}{Mohammad Soleymani},
  \bibinfo{person}{Micheal~N. Caro}, \bibinfo{person}{Erik~M. Schmidt},
  \bibinfo{person}{Cheng-Ya Sha}, {and} \bibinfo{person}{Yi-Hsuan Yang}.}
  \bibinfo{year}{2013}\natexlab{}.
\newblock \showarticletitle{1000 Songs for Emotional Analysis of Music}. In
  \bibinfo{booktitle}{\emph{ACM International Workshop on Crowdsourcing for
  Multimedia}}. \bibinfo{pages}{1--6}.
\newblock


\bibitem[\protect\citeauthoryear{Soleymani, Garcia, Jou, Schuller, Chang, and
  Pantic}{Soleymani et~al\mbox{.}}{2017}]%
        {soleymani2017survey}
\bibfield{author}{\bibinfo{person}{Mohammad Soleymani}, \bibinfo{person}{David
  Garcia}, \bibinfo{person}{Brendan Jou}, \bibinfo{person}{Bj{\"o}rn Schuller},
  \bibinfo{person}{Shih-Fu Chang}, {and} \bibinfo{person}{Maja Pantic}.}
  \bibinfo{year}{2017}\natexlab{}.
\newblock \showarticletitle{A survey of multimodal sentiment analysis}.
\newblock \bibinfo{journal}{\emph{Image and Vision Computing}}
  \bibinfo{volume}{65} (\bibinfo{year}{2017}), \bibinfo{pages}{3--14}.
\newblock


\bibitem[\protect\citeauthoryear{Soleymani, Kierkels, Chanel, and
  Pun}{Soleymani et~al\mbox{.}}{2009}]%
        {soleymani2009}
\bibfield{author}{\bibinfo{person}{Mohammad Soleymani},
  \bibinfo{person}{Joep~JM Kierkels}, \bibinfo{person}{Guillaume Chanel}, {and}
  \bibinfo{person}{Thierry Pun}.} \bibinfo{year}{2009}\natexlab{}.
\newblock \showarticletitle{A bayesian framework for video affective
  representation}. In \bibinfo{booktitle}{\emph{International Conference on
  Affective Computing and Intelligent Interaction and Workshops}}.
  \bibinfo{pages}{1--7}.
\newblock


\bibitem[\protect\citeauthoryear{Soleymani, Larson, Pun, and
  Hanjalic}{Soleymani et~al\mbox{.}}{2014}]%
        {soleymani2014corpus}
\bibfield{author}{\bibinfo{person}{Mohammad Soleymani}, \bibinfo{person}{Martha
  Larson}, \bibinfo{person}{Thierry Pun}, {and} \bibinfo{person}{Alan
  Hanjalic}.} \bibinfo{year}{2014}\natexlab{}.
\newblock \showarticletitle{Corpus development for affective video indexing}.
\newblock \bibinfo{journal}{\emph{IEEE Transactions on Multimedia}}
  \bibinfo{volume}{16}, \bibinfo{number}{4} (\bibinfo{year}{2014}),
  \bibinfo{pages}{1075--1089}.
\newblock


\bibitem[\protect\citeauthoryear{Soleymani, Lichtenauer, Pun, and
  Pantic}{Soleymani et~al\mbox{.}}{2012}]%
        {soleymani2012multimodal}
\bibfield{author}{\bibinfo{person}{Mohammad Soleymani}, \bibinfo{person}{Jeroen
  Lichtenauer}, \bibinfo{person}{Thierry Pun}, {and} \bibinfo{person}{Maja
  Pantic}.} \bibinfo{year}{2012}\natexlab{}.
\newblock \showarticletitle{A multimodal database for affect recognition and
  implicit tagging}.
\newblock \bibinfo{journal}{\emph{IEEE Transactions on Affective Computing}}
  \bibinfo{volume}{3}, \bibinfo{number}{1} (\bibinfo{year}{2012}),
  \bibinfo{pages}{42--55}.
\newblock


\bibitem[\protect\citeauthoryear{Song and Soleymani}{Song and
  Soleymani}{2019}]%
        {Song_2019_CVPR}
\bibfield{author}{\bibinfo{person}{Yale Song} {and} \bibinfo{person}{Mohammad
  Soleymani}.} \bibinfo{year}{2019}\natexlab{}.
\newblock \showarticletitle{Polysemous Visual-Semantic Embedding for
  Cross-Modal Retrieval}. In \bibinfo{booktitle}{\emph{IEEE Conference on
  Computer Vision and Pattern Recognition}}. \bibinfo{pages}{1979--1988}.
\newblock


\bibitem[\protect\citeauthoryear{Speck, Schmidt, Morton, and Kim}{Speck
  et~al\mbox{.}}{2011}]%
        {speck11ismir}
\bibfield{author}{\bibinfo{person}{Jacquelin~A Speck}, \bibinfo{person}{Erik~M
  Schmidt}, \bibinfo{person}{Brandon~G Morton}, {and}
  \bibinfo{person}{Youngmoo~E Kim}.} \bibinfo{year}{2011}\natexlab{}.
\newblock \showarticletitle{A Comparative Study of Collaborative vs.
  Traditional Musical Mood Annotation}. In
  \bibinfo{booktitle}{\emph{International Society for Music Information
  Retrieval Conference}}, Vol.~\bibinfo{volume}{104}.
  \bibinfo{pages}{549--554}.
\newblock


\bibitem[\protect\citeauthoryear{Staiano and Guerini}{Staiano and
  Guerini}{2014}]%
        {staiano2014depeche}
\bibfield{author}{\bibinfo{person}{Jacopo Staiano} {and} \bibinfo{person}{Marco
  Guerini}.} \bibinfo{year}{2014}\natexlab{}.
\newblock \showarticletitle{Depeche Mood: a Lexicon for Emotion Analysis from
  Crowd Annotated News}. In \bibinfo{booktitle}{\emph{Annual Meeting of the
  Association for Computational Linguistics}}. \bibinfo{pages}{427--433}.
\newblock


\bibitem[\protect\citeauthoryear{Subramanian, Wache, Abadi, Vieriu, Winkler,
  and Sebe}{Subramanian et~al\mbox{.}}{2018}]%
        {subramanian2018ascertain}
\bibfield{author}{\bibinfo{person}{Ramanathan Subramanian},
  \bibinfo{person}{Julia Wache}, \bibinfo{person}{Mojtaba~Khomami Abadi},
  \bibinfo{person}{Radu~L Vieriu}, \bibinfo{person}{Stefan Winkler}, {and}
  \bibinfo{person}{Nicu Sebe}.} \bibinfo{year}{2018}\natexlab{}.
\newblock \showarticletitle{ASCERTAIN: Emotion and personality recognition
  using commercial sensors}.
\newblock \bibinfo{journal}{\emph{IEEE Transactions on Affective Computing}}
  \bibinfo{volume}{9}, \bibinfo{number}{2} (\bibinfo{year}{2018}),
  \bibinfo{pages}{147--160}.
\newblock


\bibitem[\protect\citeauthoryear{Sun, Yu, Huang, and Hu}{Sun
  et~al\mbox{.}}{2009}]%
        {sun2009improved}
\bibfield{author}{\bibinfo{person}{Kai Sun}, \bibinfo{person}{Junqing Yu},
  \bibinfo{person}{Yue Huang}, {and} \bibinfo{person}{Xiaoqiang Hu}.}
  \bibinfo{year}{2009}\natexlab{}.
\newblock \showarticletitle{An improved valence-arousal emotion space for video
  affective content representation and recognition}. In
  \bibinfo{booktitle}{\emph{IEEE International Conference on Multimedia and
  Expo}}. \bibinfo{pages}{566--569}.
\newblock


\bibitem[\protect\citeauthoryear{Tarvainen, Laaksonen, and Takala}{Tarvainen
  et~al\mbox{.}}{2018}]%
        {Tarvainen2018}
\bibfield{author}{\bibinfo{person}{Jussi Tarvainen}, \bibinfo{person}{Jorma
  Laaksonen}, {and} \bibinfo{person}{Tapio Takala}.}
  \bibinfo{year}{2018}\natexlab{}.
\newblock \showarticletitle{Film mood and its quantitative determinants in
  different types of scenes}.
\newblock \bibinfo{journal}{\emph{IEEE Transactions on Affective Computing}}
  (\bibinfo{year}{2018}).
\newblock


\bibitem[\protect\citeauthoryear{Teixeira, Yamasaki, and Aizawa}{Teixeira
  et~al\mbox{.}}{2012}]%
        {Teixeira2012}
\bibfield{author}{\bibinfo{person}{Ren{\'e} Marcelino~Abritta Teixeira},
  \bibinfo{person}{Toshihiko Yamasaki}, {and} \bibinfo{person}{Kiyoharu
  Aizawa}.} \bibinfo{year}{2012}\natexlab{}.
\newblock \showarticletitle{Determination of emotional content of video clips
  by low-level audiovisual features}.
\newblock \bibinfo{journal}{\emph{Multimedia Tools and Applications}}
  \bibinfo{volume}{61}, \bibinfo{number}{1} (\bibinfo{year}{2012}),
  \bibinfo{pages}{21--49}.
\newblock


\bibitem[\protect\citeauthoryear{Torralba and Efros}{Torralba and
  Efros}{2011}]%
        {torralba2011unbiased}
\bibfield{author}{\bibinfo{person}{Antonio Torralba} {and}
  \bibinfo{person}{Alexei~A Efros}.} \bibinfo{year}{2011}\natexlab{}.
\newblock \showarticletitle{Unbiased look at dataset bias}. In
  \bibinfo{booktitle}{\emph{IEEE Conference on Computer Vision and Pattern
  Recognition}}. \bibinfo{pages}{1521--1528}.
\newblock


\bibitem[\protect\citeauthoryear{Turnbull, Barrington, Torres, and
  Lanckriet}{Turnbull et~al\mbox{.}}{2007}]%
        {turnbull2007towards}
\bibfield{author}{\bibinfo{person}{Douglas Turnbull}, \bibinfo{person}{Luke
  Barrington}, \bibinfo{person}{David Torres}, {and} \bibinfo{person}{Gert
  Lanckriet}.} \bibinfo{year}{2007}\natexlab{}.
\newblock \showarticletitle{Towards musical query-by-semantic-description using
  the cal500 data set}. In \bibinfo{booktitle}{\emph{International ACM SIGIR
  Conference on Research and Development in Information Retrieval}}.
  \bibinfo{pages}{439--446}.
\newblock


\bibitem[\protect\citeauthoryear{Tzanetakis and Cook}{Tzanetakis and
  Cook}{2000}]%
        {tzanetakis2000marsyas}
\bibfield{author}{\bibinfo{person}{George Tzanetakis} {and}
  \bibinfo{person}{Perry Cook}.} \bibinfo{year}{2000}\natexlab{}.
\newblock \showarticletitle{Marsyas: A framework for audio analysis}.
\newblock \bibinfo{journal}{\emph{Organised Sound}} \bibinfo{volume}{4},
  \bibinfo{number}{3} (\bibinfo{year}{2000}), \bibinfo{pages}{169--175}.
\newblock


\bibitem[\protect\citeauthoryear{Wang, Chen, and Ji}{Wang
  et~al\mbox{.}}{2019}]%
        {wang2017content}
\bibfield{author}{\bibinfo{person}{Shangfei Wang}, \bibinfo{person}{Shiyu
  Chen}, {and} \bibinfo{person}{Qiang Ji}.} \bibinfo{year}{2019}\natexlab{}.
\newblock \showarticletitle{Content-based video emotion tagging augmented by
  users' multiple physiological responses}.
\newblock \bibinfo{journal}{\emph{IEEE Transactions on Affective Computing}}
  \bibinfo{volume}{10}, \bibinfo{number}{2} (\bibinfo{year}{2019}),
  \bibinfo{pages}{155--166}.
\newblock


\bibitem[\protect\citeauthoryear{Wang and Ji}{Wang and Ji}{2015}]%
        {wang2015video}
\bibfield{author}{\bibinfo{person}{Shangfei Wang} {and} \bibinfo{person}{Qiang
  Ji}.} \bibinfo{year}{2015}\natexlab{}.
\newblock \showarticletitle{Video affective content analysis: a survey of
  state-of-the-art methods}.
\newblock \bibinfo{journal}{\emph{IEEE Transactions on Affective Computing}}
  \bibinfo{volume}{6}, \bibinfo{number}{4} (\bibinfo{year}{2015}),
  \bibinfo{pages}{410--430}.
\newblock


\bibitem[\protect\citeauthoryear{Wang, Zhu, Yue, and Ji}{Wang
  et~al\mbox{.}}{2015}]%
        {Wang2015Emotion}
\bibfield{author}{\bibinfo{person}{Shangfei Wang}, \bibinfo{person}{Yachen
  Zhu}, \bibinfo{person}{Lihua Yue}, {and} \bibinfo{person}{Qiang Ji}.}
  \bibinfo{year}{2015}\natexlab{}.
\newblock \showarticletitle{Emotion recognition with the help of privileged
  information}.
\newblock \bibinfo{journal}{\emph{IEEE Transactions on Autonomous Mental
  Development}} \bibinfo{volume}{7}, \bibinfo{number}{3}
  (\bibinfo{year}{2015}), \bibinfo{pages}{189--200}.
\newblock


\bibitem[\protect\citeauthoryear{Wang, Jia, Yin, and Cai}{Wang
  et~al\mbox{.}}{2013}]%
        {wang2013interpretable}
\bibfield{author}{\bibinfo{person}{Xiaohui Wang}, \bibinfo{person}{Jia Jia},
  \bibinfo{person}{Jiaming Yin}, {and} \bibinfo{person}{Lianhong Cai}.}
  \bibinfo{year}{2013}\natexlab{}.
\newblock \showarticletitle{Interpretable aesthetic features for affective
  image classification}. In \bibinfo{booktitle}{\emph{IEEE International
  Conference on Image Processing}}. \bibinfo{pages}{3230--3234}.
\newblock


\bibitem[\protect\citeauthoryear{Warriner, Kuperman, and Brysbaert}{Warriner
  et~al\mbox{.}}{2013}]%
        {warriner2013norms}
\bibfield{author}{\bibinfo{person}{Amy~Beth Warriner}, \bibinfo{person}{Victor
  Kuperman}, {and} \bibinfo{person}{Marc Brysbaert}.}
  \bibinfo{year}{2013}\natexlab{}.
\newblock \showarticletitle{Norms of valence, arousal, and dominance for 13,915
  English lemmas}.
\newblock \bibinfo{journal}{\emph{Behavior Research Methods}}
  \bibinfo{volume}{45}, \bibinfo{number}{4} (\bibinfo{year}{2013}),
  \bibinfo{pages}{1191--1207}.
\newblock


\bibitem[\protect\citeauthoryear{Xing, Zhang, Zhang, Zhang, Wu, Shi, Yu, and
  Zhang}{Xing et~al\mbox{.}}{2019}]%
        {xing2019exploiting}
\bibfield{author}{\bibinfo{person}{Baixi Xing}, \bibinfo{person}{Hui Zhang},
  \bibinfo{person}{Kejun Zhang}, \bibinfo{person}{Lekai Zhang},
  \bibinfo{person}{Xinda Wu}, \bibinfo{person}{Xiaoying Shi},
  \bibinfo{person}{Shanghai Yu}, {and} \bibinfo{person}{Sanyuan Zhang}.}
  \bibinfo{year}{2019}\natexlab{}.
\newblock \showarticletitle{Exploiting EEG Signals and Audiovisual Feature
  Fusion for Video Emotion Recognition}.
\newblock \bibinfo{journal}{\emph{IEEE Access}}  \bibinfo{volume}{7}
  (\bibinfo{year}{2019}), \bibinfo{pages}{59844--59861}.
\newblock


\bibitem[\protect\citeauthoryear{Xu, Fu, Jiang, Li, and Sigal}{Xu
  et~al\mbox{.}}{2016}]%
        {xu2016heterogeneous}
\bibfield{author}{\bibinfo{person}{Baohan Xu}, \bibinfo{person}{Yanwei Fu},
  \bibinfo{person}{Yu-Gang Jiang}, \bibinfo{person}{Boyang Li}, {and}
  \bibinfo{person}{Leonid Sigal}.} \bibinfo{year}{2016}\natexlab{}.
\newblock \showarticletitle{Heterogeneous knowledge transfer in video emotion
  recognition, attribution and summarization}.
\newblock \bibinfo{journal}{\emph{IEEE Transactions on Affective Computing}}
  \bibinfo{volume}{9}, \bibinfo{number}{2} (\bibinfo{year}{2016}),
  \bibinfo{pages}{255--270}.
\newblock


\bibitem[\protect\citeauthoryear{Xu, Cetintas, Lee, and Li}{Xu
  et~al\mbox{.}}{2014}]%
        {xu5731visual}
\bibfield{author}{\bibinfo{person}{Can Xu}, \bibinfo{person}{Suleyman
  Cetintas}, \bibinfo{person}{Kuang-Chih Lee}, {and} \bibinfo{person}{Li-Jia
  Li}.} \bibinfo{year}{2014}\natexlab{}.
\newblock \showarticletitle{Visual Sentiment Prediction with Deep Convolutional
  Neural Networks}.
\newblock \bibinfo{journal}{\emph{arXiv preprint arXiv:1411.5731}}
  (\bibinfo{year}{2014}).
\newblock


\bibitem[\protect\citeauthoryear{Yang, She, Lai, and Yang}{Yang
  et~al\mbox{.}}{2018b}]%
        {yang2018retrieving}
\bibfield{author}{\bibinfo{person}{Jufeng Yang}, \bibinfo{person}{Dongyu She},
  \bibinfo{person}{Yukun Lai}, {and} \bibinfo{person}{Ming-Hsuan Yang}.}
  \bibinfo{year}{2018}\natexlab{b}.
\newblock \showarticletitle{Retrieving and classifying affective Images via
  deep metric learning}. In \bibinfo{booktitle}{\emph{AAAI Conference on
  Artificial Intelligence}}.
\newblock


\bibitem[\protect\citeauthoryear{Yang, She, Lai, Rosin, and Yang}{Yang
  et~al\mbox{.}}{2018c}]%
        {yang2018weakly}
\bibfield{author}{\bibinfo{person}{Jufeng Yang}, \bibinfo{person}{Dongyu She},
  \bibinfo{person}{Yu-Kun Lai}, \bibinfo{person}{Paul~L Rosin}, {and}
  \bibinfo{person}{Ming-Hsuan Yang}.} \bibinfo{year}{2018}\natexlab{c}.
\newblock \showarticletitle{Weakly supervised coupled networks for visual
  sentiment analysis}. In \bibinfo{booktitle}{\emph{IEEE conference on computer
  vision and pattern recognition}}. \bibinfo{pages}{7584--7592}.
\newblock


\bibitem[\protect\citeauthoryear{Yang, She, and Sun}{Yang
  et~al\mbox{.}}{2017a}]%
        {yang2017joint}
\bibfield{author}{\bibinfo{person}{Jufeng Yang}, \bibinfo{person}{Dongyu She},
  {and} \bibinfo{person}{Ming Sun}.} \bibinfo{year}{2017}\natexlab{a}.
\newblock \showarticletitle{Joint image emotion classification and distribution
  learning via deep convolutional neural network}. In
  \bibinfo{booktitle}{\emph{International Joint Conference on Artificial
  Intelligence}}. \bibinfo{pages}{3266--3272}.
\newblock


\bibitem[\protect\citeauthoryear{Yang, Sun, and Sun}{Yang
  et~al\mbox{.}}{2017b}]%
        {yang2017learning}
\bibfield{author}{\bibinfo{person}{Jufeng Yang}, \bibinfo{person}{Ming Sun},
  {and} \bibinfo{person}{Xiaoxiao Sun}.} \bibinfo{year}{2017}\natexlab{b}.
\newblock \showarticletitle{Learning Visual Sentiment Distributions via
  Augmented Conditional Probability Neural Network}. In
  \bibinfo{booktitle}{\emph{AAAI Conference on Artificial Intelligence}}.
  \bibinfo{pages}{224--230}.
\newblock


\bibitem[\protect\citeauthoryear{Yang, Liu, and Metaxas}{Yang
  et~al\mbox{.}}{2010}]%
        {yang2010exploring}
\bibfield{author}{\bibinfo{person}{Peng Yang}, \bibinfo{person}{Qingshan Liu},
  {and} \bibinfo{person}{Dimitris~N Metaxas}.} \bibinfo{year}{2010}\natexlab{}.
\newblock \showarticletitle{Exploring facial expressions with compositional
  features}. In \bibinfo{booktitle}{\emph{IEEE Conference on Computer Vision
  and Pattern Recognition}}. \bibinfo{pages}{2638--2644}.
\newblock


\bibitem[\protect\citeauthoryear{Yang, Dong, and Li}{Yang
  et~al\mbox{.}}{2018a}]%
        {Yang2018}
\bibfield{author}{\bibinfo{person}{Xinyu Yang}, \bibinfo{person}{Yizhuo Dong},
  {and} \bibinfo{person}{Juan Li}.} \bibinfo{year}{2018}\natexlab{a}.
\newblock \showarticletitle{Review of data features-based music emotion
  recognition methods}.
\newblock \bibinfo{journal}{\emph{Multimedia Systems}} \bibinfo{volume}{24},
  \bibinfo{number}{4} (\bibinfo{year}{2018}), \bibinfo{pages}{365--389}.
\newblock


\bibitem[\protect\citeauthoryear{Yang, Jia, Zhang, Wu, Chen, Li, Xing, and
  Tang}{Yang et~al\mbox{.}}{2014}]%
        {yang2014your}
\bibfield{author}{\bibinfo{person}{Yang Yang}, \bibinfo{person}{Jia Jia},
  \bibinfo{person}{Shumei Zhang}, \bibinfo{person}{Boya Wu},
  \bibinfo{person}{Qicong Chen}, \bibinfo{person}{Juanzi Li},
  \bibinfo{person}{Chunxiao Xing}, {and} \bibinfo{person}{Jie Tang}.}
  \bibinfo{year}{2014}\natexlab{}.
\newblock \showarticletitle{How Do Your Friends on Social Media Disclose Your
  Emotions?}. In \bibinfo{booktitle}{\emph{AAAI Conference on Artificial
  Intelligence}}. \bibinfo{pages}{306--312}.
\newblock


\bibitem[\protect\citeauthoryear{Yang and Chen}{Yang and Chen}{2012}]%
        {yang2012machine}
\bibfield{author}{\bibinfo{person}{Yi-Hsuan Yang} {and}
  \bibinfo{person}{Homer~H Chen}.} \bibinfo{year}{2012}\natexlab{}.
\newblock \showarticletitle{Machine recognition of music emotion: A review}.
\newblock \bibinfo{journal}{\emph{ACM Transactions on Intelligent Systems and
  Technology}} \bibinfo{volume}{3}, \bibinfo{number}{3} (\bibinfo{year}{2012}),
  \bibinfo{pages}{40}.
\newblock


\bibitem[\protect\citeauthoryear{Yao, She, Zhao, Liang, Lai, and Yang}{Yao
  et~al\mbox{.}}{2019}]%
        {yao2019attention}
\bibfield{author}{\bibinfo{person}{Xingxu Yao}, \bibinfo{person}{Dongyu She},
  \bibinfo{person}{Sicheng Zhao}, \bibinfo{person}{Jie Liang},
  \bibinfo{person}{Yu-Kun Lai}, {and} \bibinfo{person}{Jufeng Yang}.}
  \bibinfo{year}{2019}\natexlab{}.
\newblock \showarticletitle{Attention-aware Polarity Sensitive Embedding for
  Affective Image Retrieval}. In \bibinfo{booktitle}{\emph{IEEE International
  Conference on Computer Vision}}.
\newblock


\bibitem[\protect\citeauthoryear{Yi and Wang}{Yi and Wang}{2018}]%
        {yi2018multi}
\bibfield{author}{\bibinfo{person}{Yun Yi} {and} \bibinfo{person}{Hanli Wang}.}
  \bibinfo{year}{2018}\natexlab{}.
\newblock \showarticletitle{Multi-modal learning for affective content analysis
  in movies}.
\newblock \bibinfo{journal}{\emph{Multimedia Tools and Applications}}
  (\bibinfo{year}{2018}), \bibinfo{pages}{1--20}.
\newblock


\bibitem[\protect\citeauthoryear{You, Luo, Jin, and Yang}{You
  et~al\mbox{.}}{2015}]%
        {you2015robust}
\bibfield{author}{\bibinfo{person}{Quanzeng You}, \bibinfo{person}{Jiebo Luo},
  \bibinfo{person}{Hailin Jin}, {and} \bibinfo{person}{Jianchao Yang}.}
  \bibinfo{year}{2015}\natexlab{}.
\newblock \showarticletitle{Robust Image Sentiment Analysis Using Progressively
  Trained and Domain Transferred Deep Networks.}. In
  \bibinfo{booktitle}{\emph{AAAI Conference on Artificial Intelligence}}.
  \bibinfo{pages}{381--388}.
\newblock


\bibitem[\protect\citeauthoryear{You, Luo, Jin, and Yang}{You
  et~al\mbox{.}}{2016}]%
        {you2016building}
\bibfield{author}{\bibinfo{person}{Quanzeng You}, \bibinfo{person}{Jiebo Luo},
  \bibinfo{person}{Hailin Jin}, {and} \bibinfo{person}{Jianchao Yang}.}
  \bibinfo{year}{2016}\natexlab{}.
\newblock \showarticletitle{Building a large scale dataset for image emotion
  recognition: The fine print and the benchmark}. In
  \bibinfo{booktitle}{\emph{AAAI Conference on Artificial Intelligence}}.
  \bibinfo{pages}{308--314}.
\newblock


\bibitem[\protect\citeauthoryear{Yuan, Mcdonough, You, and Luo}{Yuan
  et~al\mbox{.}}{2013}]%
        {yuan2013sentribute}
\bibfield{author}{\bibinfo{person}{Jianbo Yuan}, \bibinfo{person}{Sean
  Mcdonough}, \bibinfo{person}{Quanzeng You}, {and} \bibinfo{person}{Jiebo
  Luo}.} \bibinfo{year}{2013}\natexlab{}.
\newblock \showarticletitle{Sentribute: image sentiment analysis from a
  mid-level perspective}. In \bibinfo{booktitle}{\emph{ACM International
  Workshop on Issues of Sentiment Discovery and Opinion Mining}}.
  \bibinfo{pages}{10}.
\newblock


\bibitem[\protect\citeauthoryear{Zadeh, Chen, Poria, Cambria, and
  Morency}{Zadeh et~al\mbox{.}}{2017}]%
        {zadeh2017tensor}
\bibfield{author}{\bibinfo{person}{Amir Zadeh}, \bibinfo{person}{Minghai Chen},
  \bibinfo{person}{Soujanya Poria}, \bibinfo{person}{Erik Cambria}, {and}
  \bibinfo{person}{Louis-Philippe Morency}.} \bibinfo{year}{2017}\natexlab{}.
\newblock \showarticletitle{Tensor Fusion Network for Multimodal Sentiment
  Analysis}. In \bibinfo{booktitle}{\emph{Conference on Empirical Methods in
  Natural Language Processing}}. \bibinfo{pages}{1103--1114}.
\newblock


\bibitem[\protect\citeauthoryear{Zadeh, Liang, Poria, Vij, Cambria, and
  Morency}{Zadeh et~al\mbox{.}}{2018}]%
        {zadeh2018multi}
\bibfield{author}{\bibinfo{person}{Amir Zadeh}, \bibinfo{person}{Paul~Pu
  Liang}, \bibinfo{person}{Soujanya Poria}, \bibinfo{person}{Prateek Vij},
  \bibinfo{person}{Erik Cambria}, {and} \bibinfo{person}{Louis-Philippe
  Morency}.} \bibinfo{year}{2018}\natexlab{}.
\newblock \showarticletitle{Multi-attention recurrent network for human
  communication comprehension}. In \bibinfo{booktitle}{\emph{AAAI Conference on
  Artificial Intelligence}}.
\newblock


\bibitem[\protect\citeauthoryear{Zentner, Grandjean, and Scherer}{Zentner
  et~al\mbox{.}}{2008}]%
        {Zentner2008}
\bibfield{author}{\bibinfo{person}{Marcel Zentner}, \bibinfo{person}{Didier
  Grandjean}, {and} \bibinfo{person}{Klaus~R Scherer}.}
  \bibinfo{year}{2008}\natexlab{}.
\newblock \showarticletitle{Emotions evoked by the sound of music:
  characterization, classification, and measurement.}
\newblock \bibinfo{journal}{\emph{Emotion}} \bibinfo{volume}{8},
  \bibinfo{number}{4} (\bibinfo{year}{2008}), \bibinfo{pages}{494--521}.
\newblock


\bibitem[\protect\citeauthoryear{Zhan, She, Zhao, Cheng, and Yang}{Zhan
  et~al\mbox{.}}{2019}]%
        {zhan2019zero}
\bibfield{author}{\bibinfo{person}{Chi Zhan}, \bibinfo{person}{Dongyu She},
  \bibinfo{person}{Sicheng Zhao}, \bibinfo{person}{Ming-Ming Cheng}, {and}
  \bibinfo{person}{Jufeng Yang}.} \bibinfo{year}{2019}\natexlab{}.
\newblock \showarticletitle{Zero-Shot Emotion Recognition via Affective
  Structural Embedding}. In \bibinfo{booktitle}{\emph{IEEE International
  Conference on Computer Vision}}.
\newblock


\bibitem[\protect\citeauthoryear{Zhang, Zhang, Li, Yang, and Sun}{Zhang
  et~al\mbox{.}}{2018}]%
        {Zhang2018PMEmo}
\bibfield{author}{\bibinfo{person}{Kejun Zhang}, \bibinfo{person}{Hui Zhang},
  \bibinfo{person}{Simeng Li}, \bibinfo{person}{Changyuan Yang}, {and}
  \bibinfo{person}{Lingyun Sun}.} \bibinfo{year}{2018}\natexlab{}.
\newblock \showarticletitle{The PMEmo Dataset for Music Emotion Recognition}.
  In \bibinfo{booktitle}{\emph{ACM International Conference on Multimedia
  Retrieval}}. \bibinfo{pages}{135--142}.
\newblock


\bibitem[\protect\citeauthoryear{Zhang, Qin, Ji, Zhao, Huang, and Luo}{Zhang
  et~al\mbox{.}}{2016}]%
        {zhang2016exploring}
\bibfield{author}{\bibinfo{person}{Yanhao Zhang}, \bibinfo{person}{Lei Qin},
  \bibinfo{person}{Rongrong Ji}, \bibinfo{person}{Sicheng Zhao},
  \bibinfo{person}{Qingming Huang}, {and} \bibinfo{person}{Jiebo Luo}.}
  \bibinfo{year}{2016}\natexlab{}.
\newblock \showarticletitle{Exploring coherent motion patterns via structured
  trajectory learning for crowd mood modeling}.
\newblock \bibinfo{journal}{\emph{IEEE Transactions on Circuits and Systems for
  Video technology}} \bibinfo{volume}{27}, \bibinfo{number}{3}
  (\bibinfo{year}{2016}), \bibinfo{pages}{635--648}.
\newblock


\bibitem[\protect\citeauthoryear{Zhao, Ding, Gao, and Han}{Zhao
  et~al\mbox{.}}{2017a}]%
        {zhao2017approximating}
\bibfield{author}{\bibinfo{person}{Sicheng Zhao}, \bibinfo{person}{Guiguang
  Ding}, \bibinfo{person}{Yue Gao}, {and} \bibinfo{person}{Jungong Han}.}
  \bibinfo{year}{2017}\natexlab{a}.
\newblock \showarticletitle{Approximating Discrete Probability Distribution of
  Image Emotions by Multi-Modal Features Fusion}. In
  \bibinfo{booktitle}{\emph{International Joint Conference on Artificial
  Intelligence}}. \bibinfo{pages}{4669--4675}.
\newblock


\bibitem[\protect\citeauthoryear{Zhao, Ding, Gao, and Han}{Zhao
  et~al\mbox{.}}{2017b}]%
        {zhao2017learning}
\bibfield{author}{\bibinfo{person}{Sicheng Zhao}, \bibinfo{person}{Guiguang
  Ding}, \bibinfo{person}{Yue Gao}, {and} \bibinfo{person}{Jungong Han}.}
  \bibinfo{year}{2017}\natexlab{b}.
\newblock \showarticletitle{Learning Visual Emotion Distributions via
  Multi-Modal Features Fusion}. In \bibinfo{booktitle}{\emph{ACM International
  Conference on Multimedia}}. \bibinfo{pages}{369--377}.
\newblock


\bibitem[\protect\citeauthoryear{Zhao, Ding, Gao, Zhao, Tang, Han, Yao, and
  Huang}{Zhao et~al\mbox{.}}{2018a}]%
        {zhao2018discrete}
\bibfield{author}{\bibinfo{person}{Sicheng Zhao}, \bibinfo{person}{Guiguang
  Ding}, \bibinfo{person}{Yue Gao}, \bibinfo{person}{Xin Zhao},
  \bibinfo{person}{Youbao Tang}, \bibinfo{person}{Jungong Han},
  \bibinfo{person}{Hongxun Yao}, {and} \bibinfo{person}{Qingming Huang}.}
  \bibinfo{year}{2018}\natexlab{a}.
\newblock \showarticletitle{Discrete Probability Distribution Prediction of
  Image Emotions With Shared Sparse Learning}.
\newblock \bibinfo{journal}{\emph{IEEE Transactions on Affective Computing}}
  (\bibinfo{year}{2018}).
\newblock


\bibitem[\protect\citeauthoryear{Zhao, Ding, Han, and Gao}{Zhao
  et~al\mbox{.}}{2018b}]%
        {zhao2018personality}
\bibfield{author}{\bibinfo{person}{Sicheng Zhao}, \bibinfo{person}{Guiguang
  Ding}, \bibinfo{person}{Jungong Han}, {and} \bibinfo{person}{Yue Gao}.}
  \bibinfo{year}{2018}\natexlab{b}.
\newblock \showarticletitle{Personality-Aware Personalized Emotion Recognition
  from Physiological Signals}. In \bibinfo{booktitle}{\emph{International Joint
  Conference on Artificial Intelligence}}.
\newblock


\bibitem[\protect\citeauthoryear{Zhao, Ding, Huang, Chua, Schuller, and
  Keutzer}{Zhao et~al\mbox{.}}{2018c}]%
        {zhao2018affective}
\bibfield{author}{\bibinfo{person}{Sicheng Zhao}, \bibinfo{person}{Guiguang
  Ding}, \bibinfo{person}{Qingming Huang}, \bibinfo{person}{Tat-Seng Chua},
  \bibinfo{person}{Bj{\"o}rn~W Schuller}, {and} \bibinfo{person}{Kurt
  Keutzer}.} \bibinfo{year}{2018}\natexlab{c}.
\newblock \showarticletitle{Affective Image Content Analysis: A Comprehensive
  Survey}. In \bibinfo{booktitle}{\emph{International Joint Conference on
  Artificial Intelligence}}. \bibinfo{pages}{5534--5541}.
\newblock


\bibitem[\protect\citeauthoryear{Zhao, Gao, Jiang, Yao, Chua, and Sun}{Zhao
  et~al\mbox{.}}{2014a}]%
        {zhao2014exploring}
\bibfield{author}{\bibinfo{person}{Sicheng Zhao}, \bibinfo{person}{Yue Gao},
  \bibinfo{person}{Xiaolei Jiang}, \bibinfo{person}{Hongxun Yao},
  \bibinfo{person}{Tat-Seng Chua}, {and} \bibinfo{person}{Xiaoshuai Sun}.}
  \bibinfo{year}{2014}\natexlab{a}.
\newblock \showarticletitle{Exploring principles-of-art features for image
  emotion recognition}. In \bibinfo{booktitle}{\emph{ACM International
  Conference on Multimedia}}. \bibinfo{pages}{47--56}.
\newblock


\bibitem[\protect\citeauthoryear{Zhao, Gholaminejad, Ding, Gao, Han, and
  Keutzer}{Zhao et~al\mbox{.}}{2019a}]%
        {zhao2019personalized}
\bibfield{author}{\bibinfo{person}{Sicheng Zhao}, \bibinfo{person}{Amir
  Gholaminejad}, \bibinfo{person}{Guiguang Ding}, \bibinfo{person}{Yue Gao},
  \bibinfo{person}{Jungong Han}, {and} \bibinfo{person}{Kurt Keutzer}.}
  \bibinfo{year}{2019}\natexlab{a}.
\newblock \showarticletitle{Personalized emotion recognition by
  personality-aware high-order learning of physiological signals}.
\newblock \bibinfo{journal}{\emph{ACM Transactions on Multimedia Computing,
  Communications, and Applications}} \bibinfo{volume}{15}, \bibinfo{number}{1s}
  (\bibinfo{year}{2019}), \bibinfo{pages}{14}.
\newblock


\bibitem[\protect\citeauthoryear{Zhao, Jia, Chen, Li, Ding, and Keutzer}{Zhao
  et~al\mbox{.}}{2019b}]%
        {zhao2019pdanet}
\bibfield{author}{\bibinfo{person}{Sicheng Zhao}, \bibinfo{person}{Zizhou Jia},
  \bibinfo{person}{Hui Chen}, \bibinfo{person}{Leida Li},
  \bibinfo{person}{Guiguang Ding}, {and} \bibinfo{person}{Kurt Keutzer}.}
  \bibinfo{year}{2019}\natexlab{b}.
\newblock \showarticletitle{PDANet: Polarity-consistent Deep Attention Network
  for Fine-grained Visual Emotion Regression}. In \bibinfo{booktitle}{\emph{ACM
  International Conference on Multimedia}}.
\newblock


\bibitem[\protect\citeauthoryear{Zhao, Lin, Xu, Zhao, Guo, Krishna, Ding, and
  Keutzer}{Zhao et~al\mbox{.}}{2019c}]%
        {zhao2019cycleemotiongan}
\bibfield{author}{\bibinfo{person}{Sicheng Zhao}, \bibinfo{person}{Chuang Lin},
  \bibinfo{person}{Pengfei Xu}, \bibinfo{person}{Sendong Zhao},
  \bibinfo{person}{Yuchen Guo}, \bibinfo{person}{Ravi Krishna},
  \bibinfo{person}{Guiguang Ding}, {and} \bibinfo{person}{Kurt Keutzer}.}
  \bibinfo{year}{2019}\natexlab{c}.
\newblock \showarticletitle{CycleEmotionGAN: Emotional Semantic Consistency
  Preserved CycleGAN for Adapting Image Emotions}. In
  \bibinfo{booktitle}{\emph{AAAI Conference on Artificial Intelligence}}.
  \bibinfo{pages}{2620--2627}.
\newblock


\bibitem[\protect\citeauthoryear{Zhao, Yao, Gao, Ding, and Chua}{Zhao
  et~al\mbox{.}}{2018d}]%
        {zhao2018predicting}
\bibfield{author}{\bibinfo{person}{Sicheng Zhao}, \bibinfo{person}{Hongxun
  Yao}, \bibinfo{person}{Yue Gao}, \bibinfo{person}{Guiguang Ding}, {and}
  \bibinfo{person}{Tat-Seng Chua}.} \bibinfo{year}{2018}\natexlab{d}.
\newblock \showarticletitle{Predicting personalized image emotion perceptions
  in social networks}.
\newblock \bibinfo{journal}{\emph{IEEE Transactions on Affective Computing}}
  \bibinfo{volume}{9}, \bibinfo{number}{4} (\bibinfo{year}{2018}),
  \bibinfo{pages}{526--540}.
\newblock


\bibitem[\protect\citeauthoryear{Zhao, Yao, Gao, Ji, and Ding}{Zhao
  et~al\mbox{.}}{2017c}]%
        {zhao2017continuous}
\bibfield{author}{\bibinfo{person}{Sicheng Zhao}, \bibinfo{person}{Hongxun
  Yao}, \bibinfo{person}{Yue Gao}, \bibinfo{person}{Rongrong Ji}, {and}
  \bibinfo{person}{Guiguang Ding}.} \bibinfo{year}{2017}\natexlab{c}.
\newblock \showarticletitle{Continuous Probability Distribution Prediction of
  Image Emotions via Multi-Task Shared Sparse Regression}.
\newblock \bibinfo{journal}{\emph{IEEE Transactions on Multimedia}}
  \bibinfo{volume}{19}, \bibinfo{number}{3} (\bibinfo{year}{2017}),
  \bibinfo{pages}{632--645}.
\newblock


\bibitem[\protect\citeauthoryear{Zhao, Yao, Gao, Ji, Xie, Jiang, and Chua}{Zhao
  et~al\mbox{.}}{2016}]%
        {zhao2016predicting}
\bibfield{author}{\bibinfo{person}{Sicheng Zhao}, \bibinfo{person}{Hongxun
  Yao}, \bibinfo{person}{Yue Gao}, \bibinfo{person}{Rongrong Ji},
  \bibinfo{person}{Wenlong Xie}, \bibinfo{person}{Xiaolei Jiang}, {and}
  \bibinfo{person}{Tat-Seng Chua}.} \bibinfo{year}{2016}\natexlab{}.
\newblock \showarticletitle{Predicting personalized emotion perceptions of
  social images}. In \bibinfo{booktitle}{\emph{ACM International Conference on
  Multimedia}}. \bibinfo{pages}{1385--1394}.
\newblock


\bibitem[\protect\citeauthoryear{Zhao, Yao, Jiang, and Sun}{Zhao
  et~al\mbox{.}}{2015}]%
        {zhao2015predicting}
\bibfield{author}{\bibinfo{person}{Sicheng Zhao}, \bibinfo{person}{Hongxun
  Yao}, \bibinfo{person}{Xiaolei Jiang}, {and} \bibinfo{person}{Xiaoshuai
  Sun}.} \bibinfo{year}{2015}\natexlab{}.
\newblock \showarticletitle{Predicting discrete probability distribution of
  image emotions}. In \bibinfo{booktitle}{\emph{IEEE International Conference
  on Image Processing}}. \bibinfo{pages}{2459--2463}.
\newblock


\bibitem[\protect\citeauthoryear{Zhao, Yao, Sun, Jiang, and Xu}{Zhao
  et~al\mbox{.}}{2013}]%
        {zhao2013flexible}
\bibfield{author}{\bibinfo{person}{Sicheng Zhao}, \bibinfo{person}{Hongxun
  Yao}, \bibinfo{person}{Xiaoshuai Sun}, \bibinfo{person}{Xiaolei Jiang}, {and}
  \bibinfo{person}{Pengfei Xu}.} \bibinfo{year}{2013}\natexlab{}.
\newblock \showarticletitle{Flexible presentation of videos based on affective
  content analysis}. In \bibinfo{booktitle}{\emph{International Conference on
  Multimedia Modeling}}. \bibinfo{pages}{368--379}.
\newblock


\bibitem[\protect\citeauthoryear{Zhao, Yao, Yang, and Zhang}{Zhao
  et~al\mbox{.}}{2014b}]%
        {zhao2014affective}
\bibfield{author}{\bibinfo{person}{Sicheng Zhao}, \bibinfo{person}{Hongxun
  Yao}, \bibinfo{person}{You Yang}, {and} \bibinfo{person}{Yanhao Zhang}.}
  \bibinfo{year}{2014}\natexlab{b}.
\newblock \showarticletitle{Affective image retrieval via multi-graph
  learning}. In \bibinfo{booktitle}{\emph{ACM International Conference on
  Multimedia}}. \bibinfo{pages}{1025--1028}.
\newblock


\bibitem[\protect\citeauthoryear{Zhao, Zhao, Ding, and Keutzer}{Zhao
  et~al\mbox{.}}{2018e}]%
        {zhao2018emotiongan}
\bibfield{author}{\bibinfo{person}{Sicheng Zhao}, \bibinfo{person}{Xin Zhao},
  \bibinfo{person}{Guiguang Ding}, {and} \bibinfo{person}{Kurt Keutzer}.}
  \bibinfo{year}{2018}\natexlab{e}.
\newblock \showarticletitle{EmotionGAN: unsupervised domain adaptation for
  learning discrete probability distributions of image emotions}. In
  \bibinfo{booktitle}{\emph{ACM International Conference on Multimedia}}.
  \bibinfo{pages}{1319--1327}.
\newblock


\bibitem[\protect\citeauthoryear{Zhong, Wu, and Jiang}{Zhong
  et~al\mbox{.}}{2019}]%
        {zhong2019video}
\bibfield{author}{\bibinfo{person}{Sheng-hua Zhong}, \bibinfo{person}{Jiaxin
  Wu}, {and} \bibinfo{person}{Jianmin Jiang}.} \bibinfo{year}{2019}\natexlab{}.
\newblock \showarticletitle{Video summarization via spatio-temporal deep
  architecture}.
\newblock \bibinfo{journal}{\emph{Neurocomputing}}  \bibinfo{volume}{332}
  (\bibinfo{year}{2019}), \bibinfo{pages}{224--235}.
\newblock


\bibitem[\protect\citeauthoryear{Zhu, Park, Isola, and Efros}{Zhu
  et~al\mbox{.}}{2017b}]%
        {zhu2017unpaired}
\bibfield{author}{\bibinfo{person}{Jun-Yan Zhu}, \bibinfo{person}{Taesung
  Park}, \bibinfo{person}{Phillip Isola}, {and} \bibinfo{person}{Alexei~A
  Efros}.} \bibinfo{year}{2017}\natexlab{b}.
\newblock \showarticletitle{Unpaired Image-To-Image Translation Using
  Cycle-Consistent Adversarial Networks}. In \bibinfo{booktitle}{\emph{IEEE
  International Conference on Computer Vision}}. \bibinfo{pages}{2223--2232}.
\newblock


\bibitem[\protect\citeauthoryear{Zhu, Li, Zhang, Rao, Xu, Huang, and Xu}{Zhu
  et~al\mbox{.}}{2017a}]%
        {zhu2017dependency}
\bibfield{author}{\bibinfo{person}{Xinge Zhu}, \bibinfo{person}{Liang Li},
  \bibinfo{person}{Weigang Zhang}, \bibinfo{person}{Tianrong Rao},
  \bibinfo{person}{Min Xu}, \bibinfo{person}{Qingming Huang}, {and}
  \bibinfo{person}{Dong Xu}.} \bibinfo{year}{2017}\natexlab{a}.
\newblock \showarticletitle{Dependency exploitation: a unified CNN-RNN approach
  for visual emotion recognition}. In \bibinfo{booktitle}{\emph{International
  Joint Conference on Artificial Intelligence}}. \bibinfo{pages}{3595--3601}.
\newblock


\bibitem[\protect\citeauthoryear{Zhu, Jiang, Peng, and Zhong}{Zhu
  et~al\mbox{.}}{2016}]%
        {zhu2016video}
\bibfield{author}{\bibinfo{person}{Yingying Zhu}, \bibinfo{person}{Zhengbo
  Jiang}, \bibinfo{person}{Jianfeng Peng}, {and} \bibinfo{person}{Sheng-hua
  Zhong}.} \bibinfo{year}{2016}\natexlab{}.
\newblock \showarticletitle{Video affective content analysis based on
  protagonist via convolutional neural network}. In
  \bibinfo{booktitle}{\emph{Pacific Rim Conference on Multimedia}}.
  \bibinfo{pages}{170--180}.
\newblock


\bibitem[\protect\citeauthoryear{Zhu, Tong, Jiang, Zhong, and Tian}{Zhu
  et~al\mbox{.}}{2019}]%
        {zhu2019hybrid}
\bibfield{author}{\bibinfo{person}{Yingying Zhu}, \bibinfo{person}{Min Tong},
  \bibinfo{person}{Zhengbo Jiang}, \bibinfo{person}{Shenghua Zhong}, {and}
  \bibinfo{person}{Qi Tian}.} \bibinfo{year}{2019}\natexlab{}.
\newblock \showarticletitle{Hybrid feature-based analysis of video's affective
  content using protagonist detection}.
\newblock \bibinfo{journal}{\emph{Expert Systems with Applications}}
  \bibinfo{volume}{128} (\bibinfo{year}{2019}), \bibinfo{pages}{316--326}.
\newblock


\bibitem[\protect\citeauthoryear{Zlatintsi, Koutras, Evangelopoulos,
  Malandrakis, Efthymiou, Pastra, Potamianos, and Maragos}{Zlatintsi
  et~al\mbox{.}}{2017}]%
        {Zlatintsi2017}
\bibfield{author}{\bibinfo{person}{Athanasia Zlatintsi},
  \bibinfo{person}{Petros Koutras}, \bibinfo{person}{Georgios Evangelopoulos},
  \bibinfo{person}{Nikolaos Malandrakis}, \bibinfo{person}{Niki Efthymiou},
  \bibinfo{person}{Katerina Pastra}, \bibinfo{person}{Alexandros Potamianos},
  {and} \bibinfo{person}{Petros Maragos}.} \bibinfo{year}{2017}\natexlab{}.
\newblock \showarticletitle{COGNIMUSE: a multimodal video database annotated
  with saliency, events, semantics and emotion with application to
  summarization}.
\newblock \bibinfo{journal}{\emph{EURASIP Journal on Image and Video
  Processing}} \bibinfo{volume}{2017}, \bibinfo{number}{1}
  (\bibinfo{year}{2017}), \bibinfo{pages}{54}.
\newblock


\end{thebibliography}

%
% If your work has an appendix, this is the place to put it.

\end{document}